\documentclass[12pt]{article}
\pdfoutput=1
\usepackage{putex}
\usepackage{comment}
\usepackage{graphicx}
\usepackage{caption}
\usepackage{amsmath}
\usepackage{array}
\usepackage{subcaption}
\usepackage{epstopdf}
\usepackage{enumerate}
\usepackage{cite}
\usepackage{youngtab}
\usepackage{tensor}
\usepackage{slashed}
\usepackage[export]{adjustbox}
\usepackage[aligntableaux=center]{ytableau}
\usepackage[utf8]{inputenc}
\usepackage[
      colorlinks=true,
      linkcolor=blue,
      urlcolor=blue,
      filecolor=black,
      citecolor=red,
      ]{hyperref}
\newcommand{\RNum}[1]{\uppercase\expandafter{\romannumeral #1\relax}}

\newcommand {\be} {\begin {equation}}
\newcommand {\ee} {\end {equation}}

\newcommand {\bes} {\begin {equation*}}
\newcommand {\ees} {\end {equation*}}

\newcommand{\beq}{\begin{equation}}
\newcommand{\eeq}{\end{equation}}

\def\ie{\begin{equation}\begin{aligned}}
\def\fe{\end{aligned}\end{equation}}

\numberwithin{equation}{section}

\def\<{\langle}
\def\>{\rangle}
\def \eps {\epsilon}

\begin{document}

%\preprint{PUPT-}

\institution{PU}{Department of Physics, Princeton University, Princeton, NJ 08544, USA}

\title{CFT in AdS and boundary RG flows}

\authors{Simone Giombi and Himanshu Khanchandani}

\abstract{Using the fact that flat space with a boundary is related by a Weyl transformation to anti-de Sitter (AdS) space, one may 
study observables in boundary conformal field theory (BCFT) by placing a CFT in AdS. In addition to correlation functions 
of local operators, a quantity of interest is the free energy of the CFT computed on the AdS space with hyperbolic ball metric, 
i.e. with a spherical boundary. It is natural to expect that the AdS free energy can be used to define a quantity that decreases under 
boundary renormalization group flows. We test this idea by discussing in detail the case of the large $N$ critical $O(N)$ model in general dimension 
$d$, as well as its perturbative descriptions in the epsilon-expansion. Using the AdS approach, we recover the various known boundary critical behaviors of the model, and we compute the free energy for each boundary fixed point, 
finding results which are consistent with the conjectured $F$-theorem in a continuous range of dimensions. Finally, we also use the AdS 
setup to compute correlation functions and extract some of the BCFT data. In particular, we show that using the bulk equations of motion,
in conjunction with crossing symmetry, gives an efficient way to constrain bulk two-point functions and extract anomalous dimensions of boundary operators.}

\maketitle

\tableofcontents

\section{Introduction and summary}

A boundary conformal field theory (BCFT) defined on the flat half-space may be related via a Weyl transformation to the same conformal field theory defined in anti-de Sitter space. Indeed, the flat metric on the half-space with coordinates $(y,\mathbf{x})$, $y>0$ can be written as 
\begin{equation}
    ds^2 = dy ^2 + d \mathbf{x}^2 = y^2 ds^2_{AdS_d}, \ \ \mathbf{x} = (x_1, ... , x_{d-1})\,,
\end{equation}
where $ds^2_{AdS_d}$ is the standard Poincar\'e metric
\begin{equation}
ds^2_{AdS_d} = \frac{1}{y^2} \left(dy^2+d \mathbf{x}^2\right)\,.
\label{AdS-Poincare}
\end{equation}
The correlation functions of the BCFT can be then translated to correlation functions in AdS by performing the required Weyl rescaling of the operators. For instance, for a scalar operator of dimension $\Delta$, we have $\mathcal{O}_{AdS} = y^{\Delta} \mathcal{O}_{\textrm{half-space}}$. This implies, in particular, that the BCFT one-point functions in AdS are simply constant. In this paper, we use the AdS approach to study various properties of boundary conformal field theories. We will see that several aspects of a BCFT appear naturally when the CFT is defined on AdS, and the technical machinery developed in the AdS/CFT literature can be used to extract new results about the BCFT data. The connection between CFT on AdS and the BCFT problem has been noted before several times in the literature, see e.g. \cite{Paulos:2016fap, Carmi:2018qzm, Herzog:2019bom, Herzog:2020lel}, and \cite{Doyon:2004fv, Aharony:2010ay, Aharony:2015hix} for earlier related work. The more general idea of studing quantum field theory in AdS background appeared a long time ago in \cite{CALLAN1990366}. 

In a BCFT, it is possible to add relevant perturbations localized on the boundary, which may then drive non-trivial boundary RG flows connecting boundary critical points. Under such boundary flows, the bulk theory stays conformal and the bulk OPE data remains unaffected, but the boundary data changes. There has been considerable progress on studying quantities that are argued to be monotonic under boundary RG flows (and more general flows in defect CFT) \cite{PhysRevLett.67.161, Yamaguchi:2002pa, Friedan:2003yc, Takayanagi:2011zk, Fujita:2011fp, Nozaki:2012qd, Gaiotto:2014gha, Estes:2014hka, Jensen:2015swa, Casini:2016fgb, Kobayashi:2018lil, Casini:2018nym}. For $d = 3$ Euclidean BCFT, a proof was given in \cite{Jensen:2015swa} that the coefficient of the Euler density term in the boundary trace anomaly decreases under a boundary RG flow. Such anomaly coefficient may be extracted from the logarithmic term in either the 3d hemisphere \cite{Jensen:2015swa} or round ball \cite{Nozaki:2012qd} free energy.   
%They demonstrated their theorem for the example of a free scalar by computing hemisphere partition function. 
In $d = 4$, the free energy on a hemisphere \cite{Gaiotto:2014gha} (suitably normalized by the round 4-sphere free energy) was proposed and checked in free and perturbative examples to decrease under boundary RG flows. The boundary trace anomaly is also related to the entanglement entropy in the presence of a boundary which has been discussed in \cite{Fursaev:2013mxa, Fursaev:2016inw, Seminara:2017hhh, Berthiere:2016ott, Berthiere:2019lks, Loveridge:2020xhl}. 

It is natural to expect that in general $d$ the free energy of the BCFT defined on a space with spherical boundary may be used to define a suitable quantity that decreases under boundary RG flows. When the CFT is placed in AdS space, such a free energy can be defined by using the hyperbolic ball coordinates of AdS, so that the boundary is a sphere and the problem is conformally related to the BCFT on the round ball. Extending the idea of the generalized $F$-theorem proposed in \cite{Giombi:2014xxa} for the case of CFT with no boundaries, it is then a plausible conjecture that the quantity 
\begin{equation} \label{ConjectureDefinition}
\tilde{F} = - \sin \left( \frac{\pi (d - 1)}{2} \right) F_{{\rm AdS}_d}\,,
\end{equation}
where $F_{{\rm AdS}_d}$ is the free energy of the CFT on the hyperbolic space with sphere boundary, decreases under boundary RG flows in general $d$.\footnote{A proposal to use AdS space to define a candidate $c$-function for {\it bulk} RG flows was made in \cite{CAPPELLI1991616}.} A similar conjecture was presented in \cite{Kobayashi:2018lil}, but our main point here is the suggestion of using the AdS background to compute the free energy of the BCFT. In odd $d$, i.e. even-dimensional boundary, there is no bulk conformal anomaly, but the free energy $F_{{\rm AdS}_d}$ has a logarithmic divergence coming from the regularized volume of hyperbolic space \cite{Diaz:2007an, Casini:2011kv}. The coefficient of the logarithmic divergence is related to one of the boundary conformal anomaly coefficients (the one that does not vanish for round sphere boundary). When working in dimensional regularization, this logarithmic divergence appears as a pole, which is cancelled by the sine factor in (\ref{ConjectureDefinition}). Thus, in odd $d$, the quantity $\tilde F$ captures the boundary anomaly coefficient. In even $d$, the regularized volume of hyperbolic space is finite, but the free energy $F_{{\rm AdS}_d}$ has UV logarithmic divergences related to the bulk conformal anomaly.\footnote{The conformal anomaly in even $d$ in the presence of a boundary includes, in addition to the bulk terms, various boundary terms, see \cite{Herzog:2015ioa} where the case of $d=4$ was worked out in generality. But for the case of round sphere boundary, the only surviving boundary term should be the topological term completing the bulk Euler density. The combination of bulk Euler density and corresponding boundary term is proportional to the $a$-anomaly coefficient, which is fixed by short-distance bulk physics.} Since this is fixed by short distance physics in the bulk, it is not expected to change under boundary RG flows. Hence, the difference of $\tilde F$ between UV and IR should be a finite quantity in even $d$, and it still makes sense to ask for positivity of $\tilde F^{UV}-\tilde F^{IR}$. One can verify this explicitly in the case of free fields, as we show below, but should be true more generally. 

After some warm-up calculations in free field theory in section \ref{SectionGeneralRemarks}, we will compute the quantity (\ref{ConjectureDefinition}) in interacting BCFT and verify that the boundary $F$-theorem for $\tilde F$ holds for boundary RG flows. Our primary example in this paper will be the critical $O(N)$ model. In the case of the free $O(N)$ model, there are two possible boundary conditions, Neumann or Dirichlet, for each of the fundamental fields. One may flow from Neumann to Dirichlet by adding a boundary mass term, and it is easy to verify that $\tilde F^{\rm Neumann}>\tilde F^{\rm Dirichlet}$ for all $d$.\footnote{Calculations of the free energy for free conformal fields in hyperbolic space as well as round ball, and their relation to conformal anomalies, were carried out previously in \cite{Rodriguez-Gomez:2017kxf, Rodriguez-Gomez:2017aca}.} But when we add interactions in the bulk, there is a much richer phase structure on the boundary. The theory of boundary fixed points in the $O(N)$ model has been studied in great detail in the literature before \cite{Die86a, Die97, cardy_1996}. It can be studied perturbatively near $4$ dimensions by means of an $\epsilon$ expansion \cite{Diehl:1981zz, McAvity:1993ue, McAvity:1995zd}, in general $d$ by means of a large $N$ expansion \cite{PhysRevLett.38.1046, 10.1143/PTP.70.1226, McAvity:1995zd}, or using bootstrap techniques \cite{Liendo:2012hy, Gliozzi:2015qsa, Bissi:2018mcq, Kaviraj:2018tfd}. The $O(N)$ model defined in AdS has also received some attention on its own right \cite{Bertan:2018afl, Carmi:2018qzm}. We briefly review some of the main features here and explain some of the phenomena from a large $N$ perspective. 

\begin{figure} 
\centering
\includegraphics[scale=0.8]{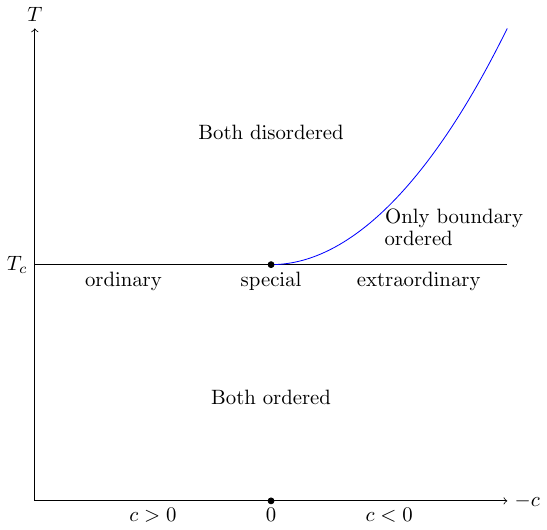}
\caption{Phase diagram of the $O(N)$ model as we vary the boundary interaction strength $c$ \cite{Die97}, defined such that $c=0$ corresponds to tuning to the special transition. The blue line describes the boundary critical temperature when it is above the bulk critical temperature, and immediately below this line one has a phase where there is ordering only on the boundary and not in the bulk.}
\label{FigurePhaseDiagram}
\end{figure}
In terms of a lattice system of spins interacting with a ferromagnetic nearest-neighbor interaction, the boundary has a lower coordination number than bulk. So we expect ordering and magnetization on the boundary to be driven by the bulk, and hence the boundary should undergo the phase transition at the same temperature as the bulk. This is what typically happens and it is referred to as the ``ordinary transition". However, in the presence of sufficiently strong interactions at the boundary, the boundary can undergo a phase transition at a temperature higher than the bulk. As we lower the temperature and reach the bulk critical temperature (with the boundary already ordered), we have the so-called ``extraordinary" transition. In this case, the $O(N)$ symmetry is broken and the fundamental field has a non-zero one-point function. Finally, there is a critical value of the boundary interaction strength at which the bulk and boundary critical temperatures become equal. This corresponds to the so-called ``special transition".  We reproduce the well-known phase diagram in figure \ref{FigurePhaseDiagram} showing different phases of the system at different values of the surface interaction strength $c$. 

\begin{figure} 
\centering
\includegraphics[scale=1]{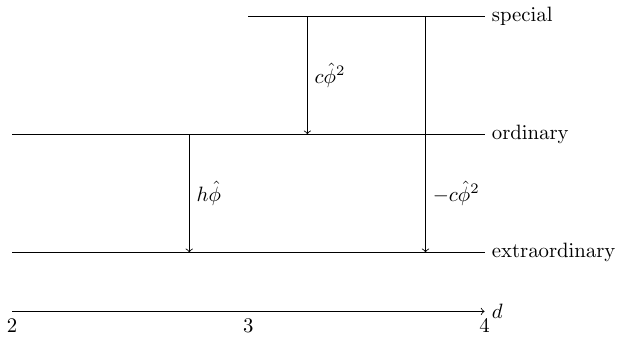}
\caption{Surface RG flow in the large $N$ $O(N)$ model between $2 < d < 4$.}
 \label{FigBoundaryFixedPoints}
\end{figure}

In order to describe these phases and the RG flows between them more explicitly, let us turn to the field theory description. Near four dimensions, the critical behavior of the $O(N)$ model can be described by the Wilson-Fisher fixed point of the scalar field theory with quartic interactions.
The bulk action on the flat half-space is\footnote{We will always assume that the bulk is critical, so any other mass terms have been tuned to zero.} 
%To study the problem in the presence of boundary, we can map the theory to AdS and use the the following action
\begin{equation} \label{ActionPhi4-flat}
S = \int d^{d-1}\mathbf{x}dy  \left( \frac{1}{2} (\partial_{\mu} \phi^I)^2  +\frac{\lambda}{4} (\phi^I \phi^I)^2  \right)\,.
\end{equation}
This model has a perturbative IR fixed point in $d=4-\epsilon$ at a certain critical coupling $\lambda=\lambda_*$. This is determined by bulk physics, and it is fixed by the renormalization of the theory in the usual flat space without boundary.  
%Notice that this AdS action is actually well suited to describe the symmetry breaking phase corresponding to extraordinary transition, where $\phi^I$ gets a one-point function: it corresponds to the minima of the potential. 
As it is well-known, the large $N$ expansion of the critical theory may be developed by performing a Hubbard-Stratonovich transformation, which yields the action in terms of the auxiliary field $\sigma(x)$ 
\begin{equation} \label{ActionLargeN-flat}
S = \int d^{d-1}\mathbf{x}dy  \left( \frac{1}{2} (\partial_{\mu} \phi^I)^2  + \frac{1}{2} \sigma \phi^I \phi^I\right)\,.
\end{equation}
In this action we omitted the $\sigma^2 / 4 \lambda$ term, which can be dropped in the critical limit (see e.g. \cite{Fei:2014yja, Giombi:2016ejx} for reviews). Note that the $\sigma$ operator, with bulk dimension $2+O(1/N)$,  plays the role of $\phi^2$ at the interacting fixed point. Let us assume that we are at the bulk critical point, and further tune the boundary interactions so that we reach the special transition point. In the field theory description, this corresponds to tuning the boundary mass term $\hat\phi^2 = \phi^2(\mathbf{x},0)$, which is relevant at the special transition (in addition, the operator $\hat\phi^I = \phi^I(\mathbf{x},0)$ is also relevant. Operators with a hat will denote operators in the boundary spectrum throughout this paper). Then, we can flow out of the special transition by adding the relevant boundary interaction $c\hat\phi^2$. For $c>0$, this drives the system to the ordinary transition, where the $O(N)$ symmetry is unbroken. The case of $c<0$ corresponds instead to flowing to the extraordinary transition, which favours a non-zero vev for $\phi$.  As we will see later, in the large $N$ description of the special transition, there is a boundary operator induced by $\sigma$ with dimension $2$ at large $N$: this plays the role of the boundary mass term driving the flow. One can see that this operator is relevant on the boundary only for $d > 3$, which is consistent with the fact that $d = 3$ is the lower critical dimension for the special transition for $N>1$ \cite{Die97} (as we will review below, from the large $N$ approach one finds that the dimension of the leading boundary operator induced by $\phi^I$ goes to zero as $d\rightarrow 3$). Note that another possible relevant interaction we can add on the boundary is $ h \hat{\phi}^I$, which is like adding a surface magnetic field. This also has the effect of ordering the boundary and drives the system to the extraordinary transition.\footnote{To be precise, the fixed point that is reached by $h \hat{\phi}^I $ flow is referred to as the ``normal transition" in some of the literature, but normal and extraordinary transition belong to the same universality class \cite{Die97} and we will not distinguish between them.} In the case of the ordinary transition, we will see below that the leading boundary operator induced by $\phi$ is relevant and has dimension $d - 2$ at large $N$. This operator can be used to drive a flow from ordinary to extraordinary transition, and such flow exists also in the range $2<d<3$. To summarize, assuming we are at bulk criticality, there are three distinct boundary critical behaviors: the special transition, which has two relevant boundary operators; the ordinary transition, with a single relevant boundary operator; and the extraordinary transition, which has no relevant boundary operators. We show all the three fixed points in figure \ref{FigBoundaryFixedPoints}. We will compute the free energy for all these three boundary critical points, both at large $N$ and using $\epsilon$-expansion, and find that $\tilde{F}$ is highest for the special transition, followed by ordinary and then by extraordinary transition, in agreement with the conjectured boundary $F$-theorem. 

Near $4$ dimensions, all the fixed points described above should match with the possible boundary conditions of the model in eq. \eqref{ActionPhi4-flat}. Indeed, it is well-known that the special transition corresponds to perturbing the free theory with Neumann boundary conditions for all the $N$ fields, and the ordinary transition to perturbing the free theory with Dirichlet boundary conditions. The flow from special to ordinary transition described above corresponds in the free field limit to the familiar fact that we can flow from Neumann to Dirichlet boundary conditions by adding a boundary mass term. On the other hand, the extraordinary transition involves giving a non-zero one-point function to one of the $\phi^I$ (and having Dirichlet boundary condition for the remaining $N-1$ fields). We will see below that this phase has a natural realization in the AdS description: it simply corresponds to the non-trivial minimum of the scalar potential of the theory in AdS, which arises due to the negative conformal coupling term to the AdS curvature, see eq. \eqref{ActionPhi4}. 

As was recently noted in \cite{Diehl:2020rfx}, and as we also observe in subsection \ref{SectionBulkEOM}, the description in terms of simple Neumann and Dirichlet boundary conditions is really only appropriate in the vicinity of $d=4$ where the CFT is nearly free. In general, the boundary critical behaviors of the model may have realizations in terms of different boundary conditions in different perturbative descriptions of the same underlying BCFT, and we will see some explicit examples of this in the paper (see figure \ref{SurfaceFlowsInteracting}). 
%As we explain in \ref{SectionBulkEOM}, to do perturbation theory away from $d = 4$, we need to look at the boundary spectrum to recognize that the fixed point arises by perturbing the Neumann boundary condition for free fields.  

Recall that near $2$ dimension, the critical properties of the $O(N)$ model can also be described by the non-linear sigma model (NL$\sigma$M). In the flat half-space, the action is
%which we may place in AdS as 
\begin{equation} \label{Action2d-flat}
S = \int d^{d-1}\mathbf{x}dy \left( \frac{1}{2} (\partial_{\mu} \phi^I)^2 + \sigma \left( \phi^I \phi^I  - \frac{1}{t^2} \right)\right).
\end{equation}
In the presence of the boundary, after solving the constraint, we can assign either Neumann or Dirichlet boundary conditions to the $N - 1$ unconstrained fields. We will check below, using our calculations of the free energy and anomalous dimensions in section \ref{SectionBoundarySpectrum}, that the Neumann case matches onto the ordinary transition while the Dirichlet case matches onto the extraordinary transition in the large $N$ theory. Note that the Neumann/Dirichlet boundary conditions near $d=2$ are not correlated with the boundary conditions in the Wilson-Fisher description near $d=4$.   

The large $N$ $O(N)$ model can be formally continued above $d = 4$, and near $6$ dimensions, it is described by the $6 - \epsilon$ expansion in a cubic theory with $N+1$ fields \cite{Fei:2014yja, Fei:2014xta}. When mapped to the AdS background, the action reads
\begin{equation} \label{Action6d-flat}
    S = \int d^{d-1}\mathbf{x}dy \bigg[ \frac{1}{2} (\partial_{\mu} \phi^I)^2+ \frac{1}{2} (\partial_{\mu} \sigma)^2 + \frac{g_1}{2} \sigma \phi^I \phi^I + \frac{g_2}{6} \sigma^3 \bigg].
\end{equation}
For large enough $N$, the model has a perturbatively unitary fixed point with real couplings \cite{Fei:2014yja, Fei:2014xta}.\footnote{Non-perturbatively, instantons generates exponentially suppressed imaginary parts for any $N$ \cite{Giombi:2019upv}. In this paper we will focus on perturbation theory and use the cubic model (\ref{Action6d}) as a useful check of the large $N$ results.}
We will see that the continuation of the special transition above $d=4$ matches onto the critical point of the cubic model with Dirichlet boundary conditions, while the continuation of the ordinary transition matches onto a phase where $\sigma$ acquires a vev. The case of Neumann boundary conditions in the $d=6-\epsilon$ theory matches instead with an additional phase of the large $N$ theory which appears above $d=5$, where the leading boundary operator in the fundamental of $O(N)$ has dimension $d-4+O(1/N)$. The extraordinary transition may also be formally continued to $d>4$, though it becomes non-unitary as we explain below; it matches onto a phase of the cubic model where both $\phi^I$ and $\sigma$ get a non-zero one-point function.

The rest of this paper is organized as follows: in section \ref{SectionGeneralRemarks}, we compute the AdS free energy in free theories and in conformal perturbation theory, and spell out the connection to trace anomalies in $d=3$. In section \ref{SectionO(N)} we study the $O(N)$ model in AdS, focusing on the large $N$ expansion, and describe the different boundary critical behaviors of the model. We calculate the corresponding values of the AdS free energy and verify consistency with the conjectured $F$-theorem. We also make explicit comparisons between the large $N$ and the various $\epsilon$ expansions near even $d$. 
In section \ref{SectionBoundarySpectrum}, we give more details on the BCFT spectrum in these models. We suggest that using the equations of motion obeyed by the bulk fields gives a convenient way to extract the anomalous dimensions of boundary operators. This is essentially an application to the case of BCFT of the idea described in \cite{Rychkov:2015naa}. For the bulk two-point function, it is particularly convenient to do this calculation in the AdS setup, because the correlation function is then just a function of the chordal distance. Moreover, the equation of motion operator takes a simple form and the boundary conformal blocks are the eigenfunctions of this operator. Using this idea, we reproduce in a straightforward manner the $\epsilon$ expansion results for Wilson-Fisher fixed point previously obtained in \cite{Bissi:2018mcq, Kaviraj:2018tfd}. At large $N$, we combine this idea with the BCFT crossing equation to get the $1/N$ correction to the anomalous dimension of the leading boundary operator and OPE coefficients of subleading boundary operators in the case of the ordinary transition.\footnote{The anomalous dimension was originally obtained some time ago by using different methods \cite{10.1143/PTP.70.1226}} This can be thought of as a version of analytic bootstrap for BCFT. We then go on to calculate some examples of boundary four-point functions using Witten diagrams in AdS, and obtain the boundary data appearing in the conformal block decomposition of the four-point function. In section \ref{sec-Concl} we make some concluding remarks and comment on possible future directions.

\section{AdS free energy and boundary RG flows: simple examples} \label{SectionGeneralRemarks}
In preparation to the calculations in the interacting $O(N)$ model, in this section we compute the AdS free energy in simple free field theory examples, and check consistency with the conjectured boundary $F$-theorem in terms of the quantity defined in (\ref{ConjectureDefinition}). We also briefly discuss the case of weakly relevant boundary flows, and elaborate on the relation of the free energy to the trace anomaly coefficients, focusing on the $d=3$ case. 

As explained in the introduction, to calculate the free energy we consider the case in which the boundary of AdS is a round sphere, in other words we will be computing the free energy on a hyperbolic ball. The metric may be obtained, for instance, from the Poincar\'e metric (\ref{AdS-Poincare}), by the following stereographic projection (throughout this paper, the index $i$ runs from $1$ to $d - 1$, while the index $\mu$ runs from $1$ to $d$)
\begin{equation} \label{BallCoordinates}
    x_i = \frac{2 u \Omega_{i + 1}}{ 1 + u^2 - 2 u \Omega_1}, \ \ \ \ y = \frac{1 - u^2}{ 1 + u^2 - 2 u \Omega_1} 
\end{equation}
where $(\Omega_1, ... \Omega_d)$ are the coordinates on the $d - 1$ sphere with $|\Omega_i|^2 = 1$. This gives the hyperbolic ball metric
\begin{equation}
    ds^2 = \frac{4}{(1-u^2)^2}\left(d u ^2 + u ^2 d \Omega^2_{d -1}\right)\,.
\end{equation}
Then  conjecture is that under a RG flow driven by a relevant boundary operator, $\tilde{F}$ computed on AdS$_d$ with sphere boundary decreases. From now on, whenever we write $F$ or $\tilde{F}$ it should be understood to be computed on AdS. 

\subsection{Neumann to Dirichlet flow in free field theory}
The simplest example that we can study is the case of a conformally coupled scalar on AdS, which can be obtained via a Weyl transformation from a free massless scalar on half-space. The action in AdS reads
\begin{equation}
S = \int d^d x \sqrt{g} \left( \frac{1}{2} 
(\partial_{\mu} \phi)^2 - \frac{d (d - 2)}{8} \phi^2 \right) , 
\end{equation}
where the ``mass" term comes from the conformal coupling to the AdS curvature (the Ricci scalar is ${\cal R}=-d(d-1)/R^2$, and we have set the radius to one for convenience). Using the usual AdS/CFT mass-dimension relation, we can get the conformal dimension of the boundary operator induced by $\phi$
\begin{equation} \label{AdS/CFTMassDimension}
    \hat{\Delta} (\hat{\Delta} - (d - 1)) = - \frac{d (d - 2)}{4} \implies \hat{\Delta}_{+} = \frac{d}{2} , \quad \hat{\Delta}_{-} = \frac{d}{2} -1. 
\end{equation}
The case of $\hat{\Delta} = d/2 - 1$ corresponds to the Neumann boundary condition while $d/2$ to Dirichlet boundary condition. 

A relevant boundary mass term, $c \hat{\phi}(\mathbf{x})^2 = c\phi(\mathbf{x},0)^2$ triggers a RG flow from Neumann (UV) to Dirichlet (IR) boundary conditions. The free energy can be computed by calculating determinants on AdS, since the action is quadratic (for later use, we consider slightly more general massive case)
\begin{equation} \label{FreeEnergyMassiveFree}
\begin{split}
    F (m^2) &=\frac{1}{2} \textrm{tr} \log \bigg( -\nabla^2 + m^2  - \frac{d (d - 2)}{4} \bigg) \\
    &= \frac{  \textrm{Vol}(H^d)}{ 2 (4 \pi)^{d/2} \Gamma(\frac{d}{2})} \int_{-\infty}^{\infty} d \nu \frac{\Gamma( i \nu + \frac{d - 1}{2}) \Gamma(- i \nu + \frac{d - 1}{2})}{\Gamma(i \nu) \Gamma(- i \nu)} \log \bigg( \nu^2 + m^2 + \frac{1}{4}\bigg)
\end{split}
\end{equation}
where we used the fact that the eigenvalues of Laplacian on $AdS_d$  are $\nu^2 + \frac{(d-1)^2}{4} $ with spectral density \cite{doi:10.1063/1.530850, Bytsenko:1994bc}
\begin{equation}
\mu(\nu) =  \frac{  \ \textrm{Vol}(H^d)}{(4 \pi)^{d/2} \Gamma(\frac{d}{2})}\frac{\Gamma( i \nu + \frac{d - 1}{2}) \Gamma(- i \nu + \frac{d - 1}{2})}{\Gamma(i \nu) \Gamma(- i \nu)} .
\end{equation}
We can also write the free energy in terms of the boundary dimension $\hat{\Delta}$
\begin{equation}
\begin{split}
     & F (\hat{\Delta}) =    \frac{ \textrm{Vol}(H^d)}{2 (4 \pi)^{d/2} \Gamma(\frac{d}{2})} \int_{-\infty}^{\infty} d \nu \frac{\Gamma( i \nu + \frac{d - 1}{2}) \Gamma(- i \nu + \frac{d - 1}{2})}{\Gamma(i \nu) \Gamma(- i \nu)}  \log \bigg( \nu^2 + \bigg(\hat{\Delta} - \frac{d -1}{2} \bigg)^2 \bigg) 
\\
 &= - \frac{\partial}{\partial \alpha}   
\left[\frac{  \textrm{Vol}(H^d)}{2 (4 \pi)^{d/2} \Gamma(\frac{d}{2})}  \int_{-\infty}^{\infty} d \nu \frac{\Gamma( i \nu + \frac{d - 1}{2}) \Gamma(- i \nu + \frac{d - 1}{2})}{\Gamma(i \nu) \Gamma(- i \nu)}  \frac{1}{ \left( \nu^2 + \left(\hat{\Delta} - \frac{d -1}{2} \right)^2 \right)^{\alpha}} \right] \Bigg|_{\alpha \rightarrow 0}
\label{FDelta-integral}
\end{split}      
\end{equation}
where the second line is equivalent to using the standard spectral zeta function regularization. 
This integral can be explicitly computed in three dimensions and gives the result
\begin{equation} \label{FreeEnergy3dFreeMassive}
F (\hat{\Delta}) = - \frac{\textrm{Vol}(H^3)}{12 \pi} (\hat{\Delta} - 1)^3. 
\end{equation} 
As defined in the introduction in eq. \eqref{ConjectureDefinition}, the quantity that is conjectured to decrease under a boundary RG flow is $\tilde{F}$. Using the regularized volume of the Hyperbolic space \cite{Diaz:2007an, Casini:2011kv} 
$\textrm{Vol}(H^d) = \pi^{\frac{d - 1}{2}} \Gamma(\frac{1 - d}{2})$, we get 
\begin{equation}
\tilde{ F}^N = \tilde{ F} \left(\hat{\Delta} = \frac{1}{2} \right) = \frac{\pi}{96}, \hspace{1cm} \tilde{ F}^D = \tilde{ F} \left( \hat{\Delta} = \frac{3}{2} \right) = -\frac{\pi}{96}.
\end{equation}
So $\tilde{F}^N > \tilde{F}^D$ in agreement with the expected boundary $F$-theorem. The quantity $\tilde F$ is related to one of the boundary anomaly coefficient, as we review in section \ref{SectionTraceAnomaly} below in the $d=3$ case, and hence the inequality for the AdS free energy is in accordance with what was proved in \cite{Jensen:2015swa}. We can also evaluate the integral in $d = 5$, where it gives 
\begin{equation}\label{FreeEnergy5dFreeMassive}
\begin{split}
F (\hat{\Delta}) &=  \frac{\textrm{Vol}(H^5)}{360 \pi^2} (\hat{\Delta} - 2)^3 (7 + 3 \ \hat{\Delta} (\hat{\Delta} - 4)) \\
\implies \tilde{ F}^N &= \tilde{ F} \left(\hat{\Delta} = \frac{3}{2} \right) =  \frac{17 \pi}{23040}, \hspace{1cm} \tilde{ F}^D = \tilde{ F} \left( \hat{\Delta} = \frac{5}{2} \right) = -\frac{17 \pi}{23040}. 
\end{split}
\end{equation}

\begin{figure} 
\centering
\includegraphics[scale=1]{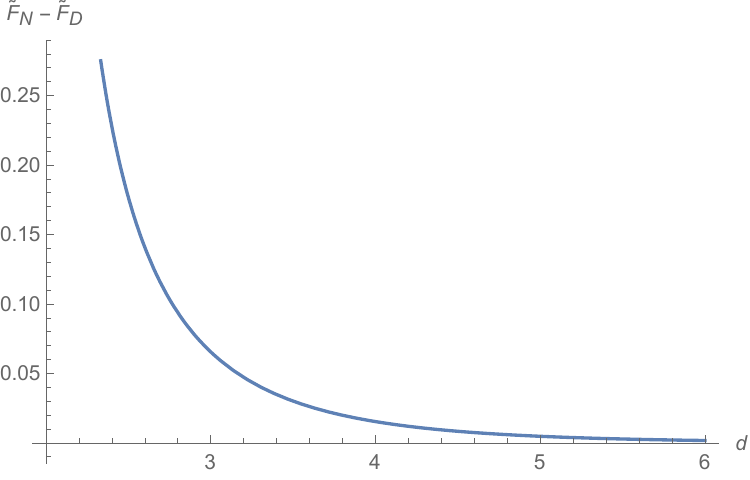} 
\caption{Free energy difference between the Neumann and Dirichlet boundary fixed points of the free bulk theory.}
\label{FigFreeEnergyDiffFree}
\end{figure}

For general dimensions, this integral is not easy to perform, but there is a shortcut if we are just interested in the free energy difference between the two boundary conditions. We can think of the flow from Neumann to Dirichlet boundary conditions as analogous to a ``double trace" flow in the $d - 1$ dimensional CFT on the boundary driven by operator $c \hat{\phi}^2$. Under such a flow, the dimension of the boundary operator flows from $\hat{\Delta}$ to $d -1 - \hat{\Delta}$. So we can use the general result for the free energy change under a double trace flow driven by the square of a primary scalar operator of dimension $\hat{\Delta}$ \cite{Diaz:2007an, Giombi:2014xxa}
\begin{equation}
  F_{d-1-\hat{\Delta}} - F_{\hat{\Delta}} =-\frac{1}{\sin(\frac{\pi (d- 1)}{2}) \Gamma(d)} \int_{0}^{\hat{\Delta} - \frac{d -1}{2}} d u \ u \ \sin \pi u \ \Gamma\bigg(\frac{d -1}{2} + u \bigg) \Gamma \bigg(\frac{d -1}{2} - u \bigg).  
\end{equation}
This implies that the free energy change under the flow from Neumann to Dirichlet is
\begin{equation} \label{DoubleTraceFEDiff}
\begin{split}
    \delta F = F^{N} - F^{D} &= F \left( \hat{\Delta} = \frac{d}{2} - 1 \right) - F\left(\hat{\Delta} = \frac{d}{2} \right)  \\
    & = -\frac{1}{\sin(\frac{\pi (d- 1)}{2}) \Gamma(d)} \int_{0}^{\frac{1}{2}} d u \ u \ \sin \pi u \ \Gamma\bigg(\frac{d -1}{2} + u \bigg) \Gamma \bigg(\frac{d -1}{2} - u \bigg)
\end{split}
\end{equation}
%where the subscript means that we have conformally coupled scalars. 
It is easy to check that this result agrees with our calculation in $d = 3, 5$ done above. It also agrees with a calculation of the free energy difference between Neumann and Dirichlet performed on a hemisphere in \cite{Herzog:2019bom}. In $d = 4$, (\ref{DoubleTraceFEDiff}) gives
\begin{equation}
    \delta \tilde{F} = \frac{\zeta(3)}{8 \pi^2}\,,
\end{equation}
while in $d = 6$, we find 
\begin{equation}
    \delta \tilde{F} = \frac{\pi^2 \zeta(3) + 3 \zeta(5)}{96 \pi^6}.
\end{equation}
These results agree with the hemisphere free energy calculation done in $d = 4$ in \cite{Gaiotto:2014gha} and later generalized to even dimensions in \cite{Dowker:2014rva, Dowker:2017cqe, Dowker:2017hpm}. For general $d$, in terms of $\tilde F$, we may write 
\begin{equation}
\delta \tilde{ F} = \tilde{F}^{N} - \tilde{F}^{D} =   \frac{1}{\Gamma(d)} \int_{0}^{\frac{1}{2}} d u \ u \ \sin \pi u \ \Gamma\bigg(\frac{d -1}{2} + u \bigg) \Gamma \bigg(\frac{d -1}{2} - u \bigg).
\label{delFND}
\end{equation}
This can be checked to be positive numerically in all $d > 2$,\footnote{In the limit $d\rightarrow 2$, one can see that (\ref{delFND}) diverges logarithmically. This is due to the fact that for Neumann boundary condition the free scalar develops a zero mode in $d=2$. The zero mode should be separated out and treated carefully, but we will not discuss this case in detail in the paper.} in accordance with $\tilde{F}^{UV}>\tilde{F}^{IR}$ for continuous $d$. We plot this quantity in fig. \ref{FigFreeEnergyDiffFree}. Note that in even $d$ the free energies $F^N$ and $F^D$ have UV divergences related to the bulk conformal anomaly, but they cancel when taking the difference leaving a well-defined, finite quantity.\footnote{In the standard spectral zeta function approach, the bulk UV divergence is captured by $\zeta(0)$ \cite{Hawking:1976ja} (equivalently in the heat kernel approach, it is captured by the Seeley coefficient $b_d$). This can be computed from the second line in (\ref{FDelta-integral}) by evaluating the integral by analytic continuation in $\alpha$, and setting $\alpha=0$ at the end (without taking the derivative in front). The result is easily seen to be independent on the boundary condition, as expected since it is related to short-distance physics in the bulk. For instance, in $d=4$ one finds $\zeta(0)=-1/90$ for both $\hat\Delta=1$ and $\hat\Delta=2$, see e.g. \cite{Giombi:2013fka}.} 

\subsection{Weakly relevant boundary flows}
Another simple situation where we can discuss the boundary $F$-theorem using the AdS free energy is in conformal perturbation theory. Since the boundary is a sphere, for small perturbations localized on the boundary, the statement is similar to the statement of generalized $F$-theorem for the sphere free energy in field theories without a boundary \cite{Giombi:2014xxa, Fei:2015oha} (even though in the BCFT case the boundary theory is not a local CFT by itself, this is not important for the calculation of $F$ in conformal perturbation theory). It was shown in \cite{Fei:2015oha} that when a CFT is perturbed by a weakly relevant operator, the universal part of the free energy decreases. The same argument basically goes through when we put the CFT in AdS and perturb the boundary by a weakly relevant boundary operator, with just the replacement $d \rightarrow d - 1$. We give a very brief review of the argument here and refer the reader to \cite{Fei:2015oha} for more details. 

Consider a CFT in $AdS_d$ perturbed by a weakly relevant operator localized on the boundary
\begin{equation}
S = S_{CFT_0} \ + \ g_b \int_{S^{d - 1}} O
\end{equation}
where $g_b$ is the bare coupling constant and $O$ has bare dimension $\hat{\Delta} = d - 1 - \epsilon$ with small $\epsilon$. The correlation functions of $O$ can be written in terms of the chordal distance $s(\mathbf{x},\mathbf{y})$ on the boundary sphere
\begin{equation}
\begin{split}
&\langle O(\mathbf{x}) O(\mathbf{y})\rangle_0 = \frac{\mathcal{C}_2}{s(\mathbf{x},\mathbf{y})^{2 \hat{\Delta}}}, \hspace{1cm} \langle O(\mathbf{x}) O(\mathbf{y}) O(\mathbf{z}) \rangle_0 = \frac{\mathcal{C}_3}{s(\mathbf{x},\mathbf{y})^{\Delta} s(\mathbf{y},\mathbf{z})^{ \Delta} s(\mathbf{x},\mathbf{z})^{\hat{\Delta}} }, \\
&s(\mathbf{x},\mathbf{y}) = \frac{2 |\mathbf{x} - \mathbf{y}|}{(1 + \mathbf{x}^2)^{1/2} (1 + \mathbf{y}^2)^{1/2} }.
\end{split}
\end{equation}
It is possible to find an IR fixed point using conformal perturbation theory at the renormalized coupling
\begin{equation}
g = g_* = \frac{\Gamma\left(\frac{d - 1}{2} \right) \mathcal{C}_3}{\pi^{\frac{d - 1}{2}} \mathcal{C}_2} \epsilon + O(\epsilon^2).
\end{equation}
The change in the AdS free energy under the flow can be calculated as \cite{Fei:2015oha} 
\begin{equation}
\delta F = F^{UV} - F^{IR} = \frac{g_b^2}{2} \int \int_{S^{d - 1}} \langle O O \rangle - \frac{g_b^3}{6} \int \int \int_{S^{d - 1}} \langle O O O \rangle.  
\end{equation}
These integrals over the sphere can be evaluated explicitly. They are divergent as $\epsilon \rightarrow 0$, and the divergences get cancelled when we plug in the bare coupling in terms of the renormalized coupling. All in all, the answer to leading order in $\epsilon$ turns out to be \cite{Fei:2015oha}
\begin{equation}
\delta \tilde{F} = \tilde{F}^{UV} - \tilde{F}^{IR} = \frac{\pi \Gamma \left(\frac{d - 1}{2} \right)^2 \mathcal{C}_2^3 }{3 \Gamma(d)  \mathcal{C}_3^2} \epsilon^3
\end{equation}
which is always positive since in an unitary theory $\mathcal{C}_2 > 0$ and $\mathcal{C}_3$ is real. Similar observations were made in \cite{Gaiotto:2014gha} about the hemisphere free energy. 
\subsection{Relation to trace anomaly coefficients in $d=3$} \label{SectionTraceAnomaly}
The boundary free energy is related to the conformal anomaly that appears in the trace of energy momentum tensor when we put the theory on a curved space with a boundary \cite{Graham:1999pm, Jensen:2015swa, Solodukhin:2015eca, Fursaev:2016inw, Herzog:2017kkj, Prochazka:2018bpb, Fursaev:2015wpa}. We confine to $d = 3$ for this discussion, where there is no conformal anomaly without the presence of a boundary. In the presence of a boundary, there is a conformal anomaly localized at the boundary, and the trace of the energy momentum tensor takes the form \cite{Graham:1999pm, Jensen:2015swa, Herzog:2017kkj}
\begin{equation} \label{TraceAnomaly3d}
\langle {T^{\mu}}_{\mu} \rangle^{d = 3} = \frac{\delta(x_{\perp})}{4 \pi} \left( a_{3 d} \hat{{\cal R}} + b \tr \hat{K}^2 \right)
\end{equation}
where $\hat{{\cal R}}$ is the boundary Ricci scalar and $\hat{K}_{ij}$ is the traceless part of the extrinsic curvature $K_{i j}$ associated to the boundary
\begin{equation}
\hat{K}_{ij} = K_{ij} - \frac{1}{2} \gamma_{i j} K \implies \tr \hat{K}^2 = \tr K^2 - \frac{1}{2} K^2
\end{equation}
with $\gamma_{ij}$ being the boundary metric.  The coefficient $b$ is related to the displacement operator two-point function, and we come back to it in appendix \ref{AppendixDisplacement}. The other coefficient $a_{3d}$ is proportional to the logarithmic divergence in the AdS free energy as we now discuss. The change in the free energy under a Weyl transformation, $g_{\mu \nu} \rightarrow e^{2 \sigma} g_{\mu \nu}$ is given by
\begin{equation}
\delta^W F = -\frac{1}{2} \int d^d x \sqrt{g} \delta g_{\mu \nu} \langle {T^{\mu}}_{\nu} \rangle = -\sigma \int d^d x \sqrt{g}  \langle {T^{\mu}}_{\mu} \rangle 
\end{equation}
In our case of Euclidean hyperbolic ball in three dimensions (we restore the AdS radius just for this discussion), the boundary is just a two-sphere of radius $R$, so the Ricci scalar is $2/R^2$ and the extrinsic curvature is $K_{ij} = \frac{1}{2} \gamma_{i j} K$ so that the traceless part vanishes. So we have 
the change in free energy under the Weyl transformation in terms of the anomly coefficient
\begin{equation}
\delta^W F = - 2 \sigma a_{3d}.
\end{equation}
Under the above Weyl transformation, $R \rightarrow e^{\sigma} R $. The regularized volume of AdS$_3$ can be computed by imposing a radial cutoff, and it is equal to $- 2 \pi \log (R/\epsilon)$ \cite{Casini:2010kt, Klebanov:2011uf} where $\epsilon$ is a UV cutoff and $R$ the radius of the boundary sphere. This gives the free energy and its change under the Weyl transformation using eq. \eqref{FreeEnergy3dFreeMassive} as 
\begin{equation}
F(\hat{\Delta}) = \frac{\log \left( R/\epsilon \right)}{6} (\hat{\Delta} - 1)^3, \hspace{1cm} \delta^W F (\hat{\Delta}) = \frac{\sigma}{6} (\hat{\Delta} - 1)^3
\end{equation}
which implies that 
\begin{equation} \label{a3dformula}
 a_{3d} = - \frac{1}{12} ( \hat{\Delta} - 1)^3\,.
\end{equation}
This tells us that in the free theory of a single scalar, $a^N_{3d} = 1/96$ and $a^D_{3d} = -1/96$ for Neumann and Dirichlet boundary conditions respectively, in agreement with the known results \cite{Jensen:2015swa, Herzog:2017kkj}. 
%We come back to computing free energy and $a_{3d}$ for fixed points in interacting theories in the next section. 

%\section{Boundary fixed points and RG flows in the critical $O(N)$ model} \
\section{Large $N$ $O(N)$ model in AdS: boundary critical points and free energy}
\label{SectionO(N)}
In this section, we study the critical $O(N)$ model in AdS, focusing on the large $N$ expansion. As explained in the introduction, this is related by a Weyl transformation to studying the $O(N)$ model on the flat half-space (or on a ball with spherical boundary). Mapping the action (\ref{ActionLargeN-flat}) to AdS by a Weyl transformation, we obtain the action for the critical $O(N)$ model in hyperbolic space as
\begin{equation} \label{ActionLargeN}
S = \int d^d x \sqrt{g} \left( \frac{1}{2} (\partial \phi^I)^2 - \frac{d (d - 2)}{8} \phi^I \phi^I + \frac{1}{2} \sigma \phi^I \phi^I\right)\,.
\end{equation}
The various boundary critical points of the model can be then recovered by solving the saddle point equations arising by integrating out the scalar fields. 

Before moving on to the interacting theory, let us first discuss in a bit more detail the case of free scalar BCFT, viewed from the AdS approach. The free scalar in half-space is related to a conformally coupled scalar in AdS. As expected from the Weyl transformation, it is easy to see explicitly that the two-point function of a massless scalar on the half-space with Neumann or Dirichlet boundary conditions is the same as the bulk-to-bulk propagator in AdS up to an overall conformal factor, namely 
\begin{equation}
\begin{split}
\langle \phi(x_1) \phi (x_2) \rangle^{\rm flat}_{N/D} &= \frac{\Gamma\left(\frac{d}{2} - 1 \right)}{4 \pi^{\frac{d}{2}} (4 y_1 y_2)^{\frac{d}{2} - 1}} \left( \frac{1}{\xi^{\frac{d}{2} - 1}} \pm \frac{1}{(\xi + 1)^{\frac{d}{2} - 1}} \right), \ \ \ \xi = \frac{\mathbf{x}_{12}^2 + y_{12}^2}{4 y_1 y_2} \\
& = \frac{1}{(y_1 y_2)^{\frac{d}{2} - 1}} G^{bb}_{\hat{\Delta}_{N/D}}, \ \ \hat{\Delta}_{N} = \frac{d}{2} - 1\,,\quad \hat{\Delta}_D=\frac{d}{2}
\end{split}
\end{equation}
where $G^{bb}_{\hat{\Delta}}$ is the well-known bulk-to-bulk propagator in AdS given by 
\begin{equation} \label{Bulk-BulkPropagator}
G^{bb}_{\hat{\Delta}} = \frac{\Gamma({\hat{\Delta}})}{2 \pi^{\frac{d - 1}{2}} \Gamma \left({\hat{\Delta}} + \frac{3 - d}{2} \right) (4 \xi)^{{\hat{\Delta}}}} {}_2F_1 \left( {\hat{\Delta}}, {\hat{\Delta}} - \frac{d}{2}  + 1, 2 {\hat{\Delta}} - d + 2, - \frac{1}{\xi}\right).
\end{equation}
and the values of the conformal dimensions $\hat{\Delta}_{N/D}$ corresponding to Neumann and Dirichlet boundary conditions can be obtained by the AdS/CFT mass/dimension relation in eq. \eqref{AdS/CFTMassDimension}.

In any BCFT the bulk two-point function can be expanded in conformal blocks in two different channels: 1) Bulk channel, which corresponds to taking the two operators close to each other in the bulk, i.e. $\xi \rightarrow 0$, or 2) Boundary channel, which corresponds to taking both the operators close to the boundary and then using their boundary operator expansion (BOE), i.e. $\xi \rightarrow \infty$ (see for instance \cite{Billo:2016cpy})
\begin{equation} \label{BlockExpansion}
\begin{split}
\langle O(x) O(x') \rangle &= \frac{A}{(4 y y')^{\Delta_O}} \xi^{- \Delta_O} G(\xi) \\
G &= 1 + \sum_{k} \lambda_k  f_{\textrm{bulk}} (\Delta_k ; \xi) = \xi^{\Delta_O} (a_O^2 + \sum_{l} \mu_l^2 f_{\textrm{bdry}} (\hat{\Delta}_l;\xi)). 
\end{split}
\end{equation}
The bulk and boundary blocks have following expressions \cite{McAvity:1995zd}
\begin{equation}
\begin{split}
f_{\textrm{bulk}} (\Delta_k ; \xi) &= \xi^{\frac{\Delta_k}{2}} \ \   _2F_1 \bigg(\frac{\Delta_k}{2},\frac{\Delta_k}{2}; \Delta_k + 1 - \frac{d}{2} ; -\xi \bigg) \\
f_{\textrm{bdry}} ({\hat{\Delta}}_l;\xi) &= \xi^{-{\hat{\Delta}}_l} \ \  _2 F_1 \bigg( {\hat{\Delta}}_l, {\hat{\Delta}}_l + 1 - \frac{d}{2}; 2 {\hat{\Delta}}_l + 2 - d; -\frac{1}{\xi}  \bigg) .
\end{split}
\end{equation}
To express the two-point function (\ref{BlockExpansion}) in AdS, we simply strip off the conformal factor of $(y y')^{\Delta_O}$, and everything else stays the same. Note that the boundary conformal block is proportional to the AdS bulk-to-bulk propagator, which is consistent with the fact that a free bulk field induces a single operator on the boundary of dimension $d/2 - 1$ or $d/2$.  

When we add interactions in the bulk and tune to criticality, we can have phases with more interesting boundary conditions. We can have phases that preserves the $O(N)$ symmetry, and also phases that spontaneously breaks it to $O(N-1)$. These phases correspond to different boundary critical behaviors of the model, and we will present their large $N$ analysis below.

\subsection{$O(N)$ invariant boundary fixed points} \label{SectionLargeNND}
Assuming that the $O(N)$ symmetry is preserved, we can start from the action in eq. \eqref{ActionLargeN} and integrate out the $N$ fundamental fields $\phi^I$ to get an effective action for $\sigma$
\begin{equation} \label{EffectiveActionSigmaN}
Z = \exp[- F] =  \int [d \sigma] \exp \left[-  \frac{N}{2} \textrm{tr} \log \left( - \nabla^2 + \sigma - \frac{d (d - 2)}{4}  \right)  \right].
\end{equation}
At large $N$, we can use a saddle point approximation to do the integral over $\sigma$ and look for a field configuration with a constant value of $\sigma = \sigma_*$.\footnote{In the flat half-space picture, one would instead find that at the saddle point $\sigma=\sigma_*/y^2$, with $\sigma_*$ the same constant found in the AdS calculation.}  Therefore, at leading order in large $N$, $\sigma_*$ just acts like a bulk ``mass" for the $\phi^I$ fields (note, however, that we are still describing a BCFT, i.e. the bulk theory remains critical: the non-zero expectation value for $\sigma$ is simply a reflection of the non-zero one-point function of the operator $\phi^2\sim \sigma$). As in section \ref{SectionGeneralRemarks}, the free energy can then be written as 
\begin{equation}
\begin{split}
    F (\sigma) &=\frac{N}{2} \textrm{tr} \log \bigg( -\nabla^2 +\sigma - \frac{d (d - 2)}{4} \bigg) \\
    &= \frac{ N \textrm{Vol}(H^d)}{ 2 (4 \pi)^{d/2} \Gamma(\frac{d}{2})} \int_{-\infty}^{\infty} d \nu \frac{\Gamma( i \nu + \frac{d - 1}{2}) \Gamma(- i \nu + \frac{d - 1}{2})}{\Gamma(i \nu) \Gamma(- i \nu)}  \log \bigg( \nu^2 + \sigma + \frac{1}{4}\bigg).
\end{split}
\end{equation}
The constant $\sigma_*$ can then be fixed by demanding that it extremizes the free energy, which happens when the following derivative of free energy with $\sigma$ vanishes
\begin{equation}
\begin{split}
 &\frac{\partial F (\sigma)}{\partial \sigma}\bigg|_{\sigma = \sigma_*} = \frac{ N \textrm{Vol}(H^d)}{ 2 (4 \pi)^{d/2} \Gamma(\frac{d}{2})} \int_{-\infty}^{\infty} d \nu \frac{\Gamma \left( i \nu + \frac{d - 1}{2} \right) \Gamma \left(- i \nu + \frac{d - 1}{2} \right)}{\Gamma(i \nu) \Gamma(- i \nu)}  \frac{1}{\nu^2 + \sigma_* + \frac{1}{4}} \\ 
 & = \frac{ N \textrm{Vol}(H^d)}{ 2 (4 \pi)^{d/2} \Gamma(\frac{d}{2})} \bigg[ -\sin \left(\pi  \sqrt{\sigma_* +\frac{1}{4}}\right) \Gamma \left(\frac{d - 1 - \sqrt{4\sigma_* + 1}}{2}\right) \Gamma \left(\frac{d - 1 + \sqrt{4\sigma_* + 1}}{2} \right) \\
 & - \sum_{n = 0}^{\infty} \frac{4  (-1)^n (d+2 n-1) \cos \left(\frac{\pi}{2}  (d+2 n)\right) \Gamma (d+n-1)}{ \Gamma (n+1) \left(4 d n + (d-2) d+4 n^2-4 (n + \sigma_* )\right)}  \bigg]
\end{split}
\end{equation}
To go from the first line to second line, we performed the $\nu$-integral by closing the contour in the complex $\nu$ plane and summing over residues. The arc at infinity can be dropped for $d<2$, but in dimensional regularization we may continue the final result to $d>2$. Note that one of the Gamma functions also introduces poles at $\nu = i (d-1 + 2 n)/2$, which lie on the upper half plane for $d>1$: these poles give the sum over $n$ above. The sum can be  performed by analytic continuation in $d$, and we get the final result
\begin{equation} 
\begin{aligned}
\frac{\partial F (\sigma)}{\partial \sigma}\bigg|_{\sigma = \sigma_*} &= \frac{ N \textrm{Vol}(H^d)}{ 2 (4 \pi)^{d/2} \Gamma(\frac{d}{2})} \frac{\Gamma \left(\frac{d - 1}{2} + \sqrt{\sigma_* + \frac{1}{4}} \right) \Gamma \left(\frac{d - 1}{2} - \sqrt{\sigma_* + \frac{1}{4}} \right) \sin  \left( \pi \left(\frac{d - 1}{2} - \sqrt{\sigma_* + \frac{1}{4}} \right) \right)}{\sin \left(\frac{\pi d}{2} \right)} \\
& = \frac{ N \textrm{Vol}(H^d)}{ 2 (4 \pi)^{d/2}} \frac{\Gamma(\hat{\Delta}) \Gamma \left( 1- \frac{d}{2} \right)}{ \Gamma (-d+\hat{\Delta} +2)}
\label{BoundaryFreeEnergyDerivative}
\end{aligned}
\end{equation}
where we used again the familiar AdS/CFT relation
\begin{equation}
\hat{\Delta} (\hat{\Delta}
 - (d - 1)) = \sigma_* - \frac{d (d - 2)}{4}  .
\end{equation}
We also had to use $\hat{\Delta} > (d - 1)/2$ to get to the last line in eq. \eqref{BoundaryFreeEnergyDerivative} which is where the above spectral representation is valid, but the final result can be analytically continued in $\hat{\Delta}$. 

Another way to arrive at the same result, that does not involve the spectral representation, is to note that $\partial F/ \partial \sigma$ is just the integral over AdS of the one point function $\langle \phi^I \phi^I (x) \rangle /2$, and the integral only produces the volume factor since the one-point functions are constant on AdS. At leading order in large $N$, $\sigma=\sigma_*$ acts as a constant mass term, so $\phi$ is a free massive field in AdS and its propagator must be the usual bulk-bulk propagator, eq. \eqref{Bulk-BulkPropagator}. The required one-point function $\langle \phi^I \phi^I (x) \rangle /2$ then is equal to the coincident point limit of the two-point function, and can be obtained from its $\xi \rightarrow 0$ limit
\begin{equation} \label{BulkBulkShortDistance}
\begin{split}
\langle \phi^I (x_1) \phi^J (x_2) \rangle &= \delta^{IJ} G_{b b}^{\hat{\Delta}} \\
& = \frac{\delta^{IJ}}{\xi^{\frac{d}{2} - 1}} \left( \frac{\Gamma \left(\frac{d}{2}-1\right) }{  (4 \pi)^{d/2} } + O(\xi)  \right) + \left( \frac{ \Gamma (\hat{\Delta} ) \Gamma \left(1-\frac{d}{2}\right)}{  (4 \pi)^{d/2} \Gamma (-d+\hat{\Delta} +2) } + O(\xi) \right).
\end{split} 
\end{equation} 
One can see that the constant piece of the above expression, which is the coincident limit of the two point function, is the same as the derivative of free energy in eq. \eqref{BoundaryFreeEnergyDerivative} up to a factor of $N \textrm{Vol}(H^d)/2$.  

The saddle point requirement of the vanishing of the free energy derivative in eq. \eqref{BoundaryFreeEnergyDerivative} or equivalently vanishing of the constant piece in eq. \eqref{BulkBulkShortDistance} can also be now motivated in another way: in a BCFT, the bulk OPE data should be unaffected by the boundary, and hence the operator spectrum encoded in the $\phi\phi$ bulk OPE should be the same as the one for the critical $O(N)$ model in flat space with no boundary. In particular, the bulk spectrum should be such that the operator $\phi^2$ of dimension $d-2$ in the free theory is replaced by the operator $\sigma$ of dimension 2. The $\xi\rightarrow 0$ expansion in eq. \eqref{BulkBulkShortDistance} is the same as doing the bulk OPE, and we recognize that the second term would correspond to the contribution of an operator of dimension $d-2$, which we must then set to zero. This yields the same condition as the above saddle point analysis, and fixes the dimension of the leading boundary operator $\hat{\Delta}$ and hence the value of $\sigma^*$ at leading order in large $N$. We will also use the same argument in subsection \ref{SectionBulkEOM} to find the $1/N$ corrections to the dimension of the leading boundary operator. 

Setting eq. \eqref{BoundaryFreeEnergyDerivative} equal to zero requires $\hat{\Delta} = d - n$, for integer $n$ with $ n \ge 2$. We will restrict to the case of unitary theories, and so we will only consider solutions satisfying the boundary unitarity bound. The existence of these saddles was also noted in \cite{Herzog:2020lel}.
\begin{comment}
For $d < 3$, the only allowed dimension of the boundary operator of this form is $d - 2$ which gives $\sigma_* = (d - 2)(d-4)/4$. We will see below that this large $N$ fixed point, near two dimensions, matches onto the non-linear sigma model in $d = 2 + \epsilon$ with Neumann boundary conditions on the unconstrained fields. We compute the anomalous dimension of the leading boundary operator in non-linear sigma model in eq. \eqref{AnomalousDimensionNLSM} and that is also consistent with this large $N$ result. 
\end{comment}

In dimensions $3 \le d \le 5$, there are two unitary solutions with dimension $\hat{\Delta}$ of the form $d - n$:
\begin{eqnarray}
\hat{\Delta} &=& d-2 \implies  \sigma_* = \frac{(d 
-2)(d-4)}{4}\label{delO} \\
\hat{\Delta} &=& d-3 \implies \sigma_* = \frac{(d-4)(d-6)}{4}  \label{delS}
\end{eqnarray}
The $\hat{\Delta} = d - 2$ solution describes the ordinary transition, while the $\hat{\Delta}= d - 3$ solution describes the special transition. In $d = 4 - \epsilon$, these match onto the possible boundary critical behaviors of the weakly coupled Wilson-Fisher fixed point in the $\phi^4$ theory \eqref{ActionPhi4} (see figure \ref{SurfaceFlowsInteracting}): the $\hat\Delta=d - 2$ and $\hat\Delta=d - 3$ solutions correspond to the Wilson-Fisher BCFTs obtained by perturbing the free theory respectively with Dirichlet or Neumann boundary conditions on the $N$ fundamental fields (note that the description in terms of Dirichlet or Neumann boundary conditions is really only appropriate in the vicinity of $d=4$, where we perturb a free scalar field theory). These results agree with what was found in \cite{McAvity:1995zd} in the flat space setup. As we review in section \ref{SectionBoundarySpectrum} below, computing the $\sigma$ two-point function around these saddle points, one finds that in the case of the special transition, $\sigma$ induces at the boundary an operator of dimension 2 and an operator of dimension $d$, while for the ordinary transition it induces only an operator of dimension $d$ (the latter is related to the displacement operator). Therefore, at the special transition ($\hat\Delta=d-3$) there is a single $O(N)$ invariant relevant operator at the boundary, which can be used to trigger a flow from the special to the ordinary transition ($\hat\Delta=d-2$). We may think of such operator as $\hat\sigma$, defined by the boundary limit of the bulk field $\sigma$. Since in the Hubbard-Stratonovich description $\sigma$ plays the role of $\phi^2$, the deformation by $\hat\sigma$ can be viewed as the large $N$ counterpart of adding a boundary mass term. Hence, we expect that the AdS free energy at the two saddle points (\ref{delO})-(\ref{delS}) should satisfy $\tilde F^{\hat\Delta=d-3} > \tilde F^{\hat\Delta=d-2}$. We will verify this explicitly below.

While the solution $\hat{\Delta}=d-3$ for the special transition does not extend to $d<3$,\footnote{The lower critical dimension for special transition is $d=3$ for $N>1$. In the $N=1$ case, however, the lower critical dimension is $d=2$ \cite{Die97}. In this paper we focus on the large $N$ theory.} the solution $\hat{\Delta}=d-2$ corresponding to the ordinary transition smoothly continues to $d<3$. As we will discuss below, this solution matches near $d=2$ with the critical point of the $O(N)$ non-linear sigma model in $d=2+\epsilon$, for the case of Neumann boundary conditions on the unconstrained $N-1$ fields \cite{PhysRevLett.56.2834}.  We will compute the anomalous dimension of the leading boundary operator in the non-linear sigma model in eq. \eqref{AnomalousDimensionNLSM} below, which is seen to be precisely consistent with the large $N$ result.    

As we go above $d = 5$, for $5 < d < 7$, we have now three possible solutions consistent with unitarity bounds. Besides (\ref{delO}) and (\ref{delS}), there is an additional one
\begin{equation}
\hat{\Delta} = d - 4 \implies  \sigma_* = \frac{(d 
-6)(d-8)}{4}.
\end{equation}
In $d = 6 - \epsilon$, all these solutions match onto possible phases of the cubic theory of eq. \eqref{Action6d}, as we verify below, see figure \ref{SurfaceFlowsInteracting} (since we are interested in unitary theories, we will restrict our attention to the case of $d\le 6$ in this paper). In particular, the new phase with $\hat\Delta=d-4+O(1/N)$ corresponds in $d=6-\epsilon$ to the fixed point of the cubic theory with Neumann boundary conditions on the $N$ fields $\phi^I$, while the $\hat\Delta=d-3+O(1/N)$ phase corresponds to Dirichlet boundary conditions. Therefore, at least in the perturbative description near $d=6$, we expect that we can flow from the $\hat\Delta=d-4$ to the $\hat\Delta=d-3$ phase by adding a boundary mass term of dimension $\sim 4$ (in the large $N$ description, this should correspond to an operator of dimension $4+O(1/N)$ contained in the bulk-boundary operator expansion of $\sigma$). Finally, the phase with $\hat\Delta=d-2$, which is the smooth continuation of the ordinary transition, corresponds in the cubic model description to a $O(N)$ invariant saddle point with non-zero expectation value for the $\sigma$ field. We expect that this phase can be reached by perturbing either the $\hat\Delta=d-4$ or $\hat\Delta=d-3$ phases by the dimension 2 operator $\sim h \hat\sigma$. To summarize, if the boundary $F$-theorem holds, we then expect that the free energies should satisfy $\tilde F^{\hat\Delta=d-4}> \tilde F^{\hat\Delta=d-3}>\tilde F^{\hat\Delta=d-2}$. We will verify this shortly.

Having identified the various boundary fixed points with $O(N)$ symmetry, we can go on and compute the corresponding values of the AdS free energy. In $d = 3$, recall from section \ref{SectionGeneralRemarks} that the free energy can be computed exactly for any value of $\hat{\Delta}$  and we can directly use eq. \eqref{FreeEnergy3dFreeMassive}, or, in terms of $\tilde F$:
\begin{equation}
\tilde F(\hat\Delta) = -\frac{\pi}{12}\left(\hat\Delta-1\right)^3\,,
\end{equation}
where we used that $-\sin(\pi(d-1)/2){\rm Vol}(H^d)|_{d\rightarrow 3}=\pi^2$. 
Special and ordinary transitions correspond to $\hat{\Delta} = d - 3$ and $\hat{\Delta} = d - 2$ respectively, and hence
%\begin{equation} \label{LargeNFreeEnergy3dSO}
%F^{S} =  \frac{N \textrm{Vol}(H^3)}{12 \pi}, \hspace{1cm} 
%F^{O} = 0 
%\end{equation}
%and
\begin{equation}
\tilde{F}^{S} =  \frac{N \pi}{12 },\hspace{1cm} \tilde{F}^{O} = 0.
\end{equation}
So clearly $\tilde{F}^{S} > \tilde{F}^{O}$. We can also immediately get the anomaly coefficient $a_{3d}$ by using eq. \eqref{a3dformula} 
\begin{equation} \label{FreeEnergy3dInt}
a_{3d}^{S} = \frac{N}{12}, \hspace{1cm} a_{3d}^{O} = 0.
\end{equation}
For $d = 5$ as well, we can use eq. \eqref{FreeEnergy5dFreeMassive} to calculate the free energy for all three symmetry preserving phases 
\begin{equation} \label{FreeEnergy5dInt}
\tilde{F} (\hat{\Delta} = 1) = \frac{\pi}{360}, \hspace{1cm } \tilde{F} (\hat{\Delta} = 2) = 0, \hspace{1cm }\tilde{F} (\hat{\Delta} = 3) = - \frac{\pi}{360}.
\end{equation}

\begin{figure}
\centering
\begin{subfigure}{0.5\textwidth}
\includegraphics[width = \textwidth]{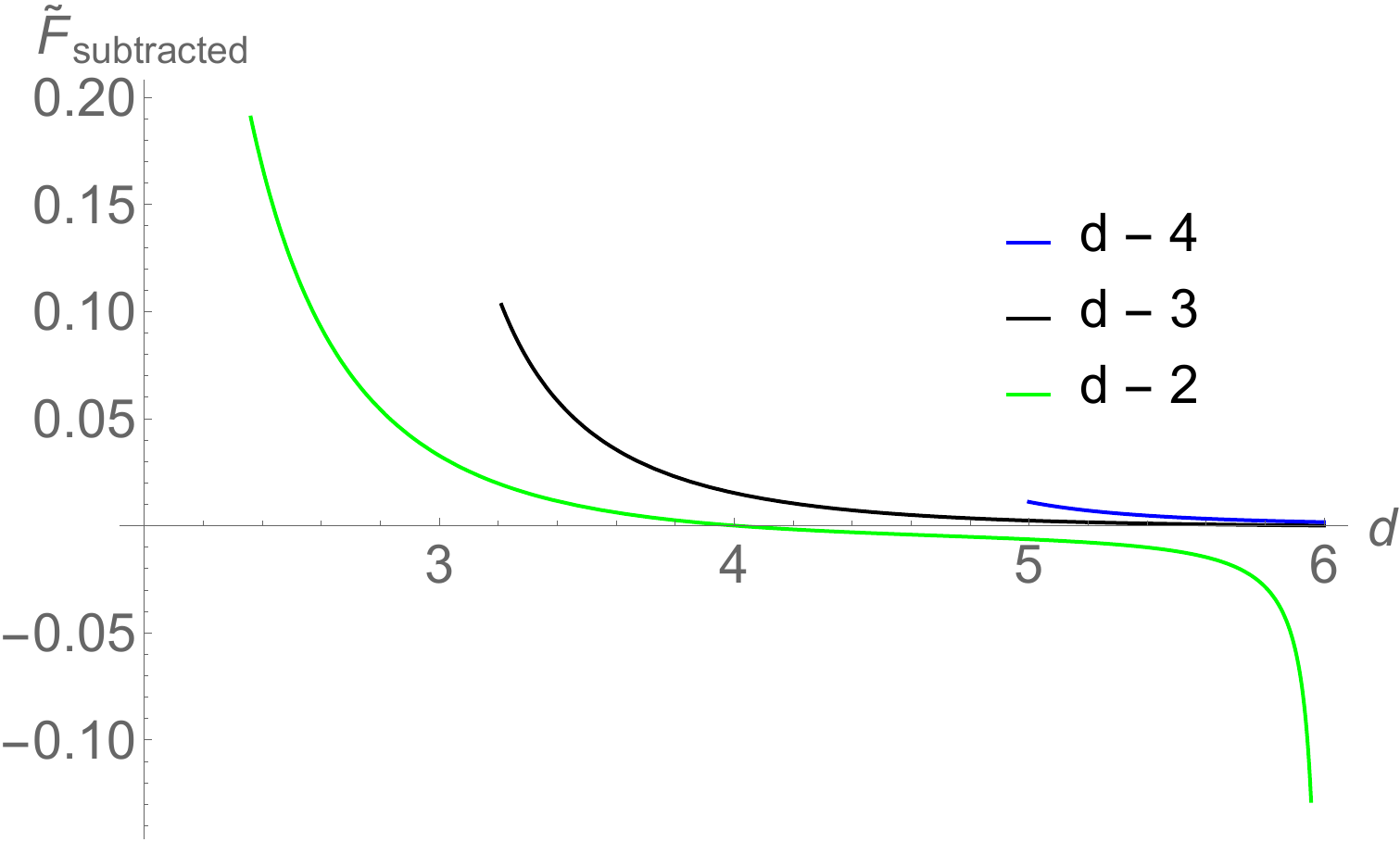}
\end{subfigure}
\begin{subfigure}{0.45\textwidth}
\includegraphics[width = \textwidth]{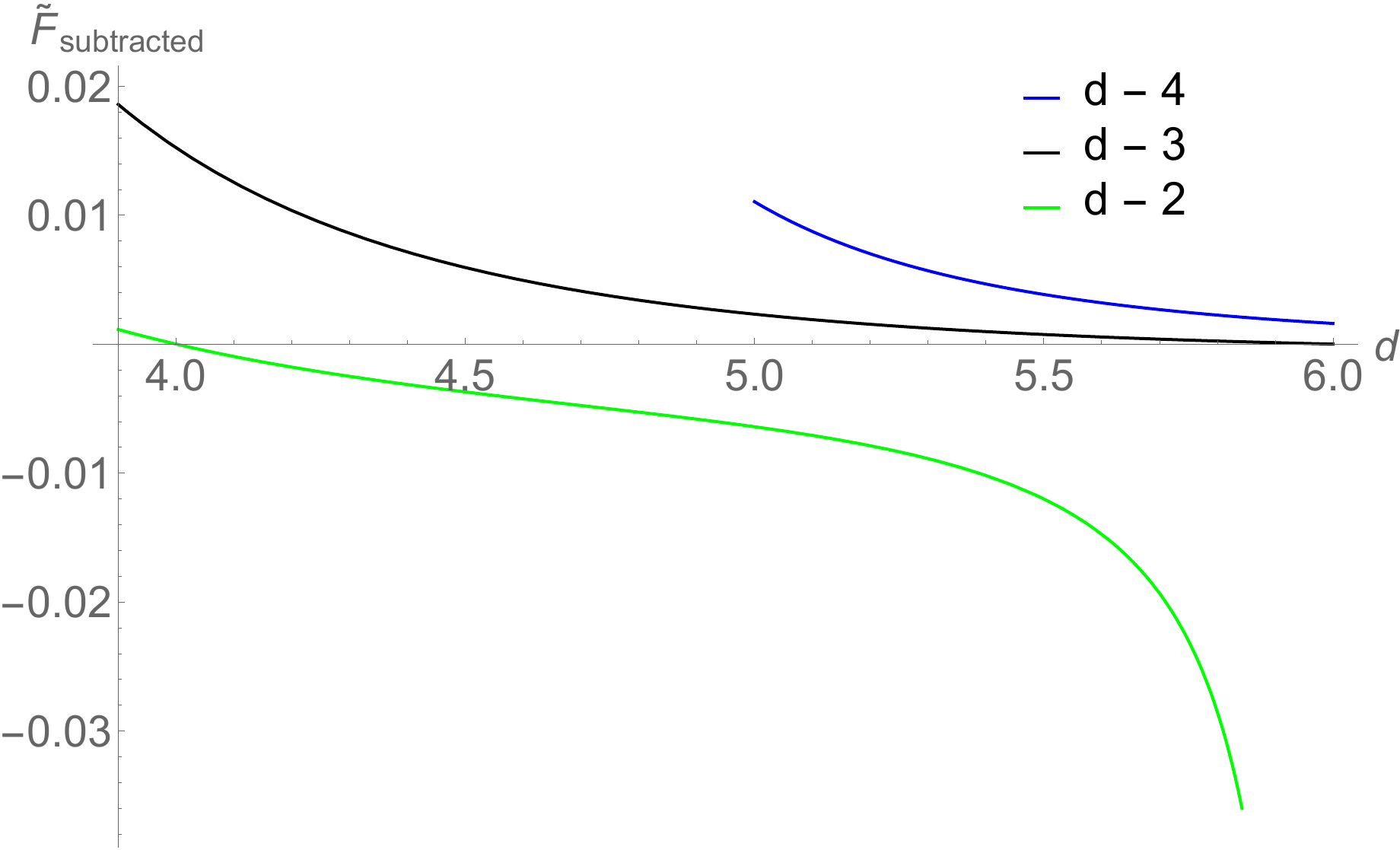}
\end{subfigure}
\caption{Large $N$ free energy between $2<d<6$ for the different boundary fixed points. We are plotting $\tilde{F}_{\textrm{subtracted}} = (\tilde{F} - N \tilde{F} (d/2))/N $ on $y-$ axis. On right, we zoom in to the region between $4$ and $6$ dimensions for clarity.}
\label{FreeEnergyPlotSymmetric}
\end{figure}

To make progress for other values of $d$, we can use the derivative of the free energy in eq. \eqref{BoundaryFreeEnergyDerivative} and express the free energy as a function of $\hat{\Delta}$ in terms of a reference value, say for a conformally coupled scalar with Dirichlet boundary, $F (\hat{\Delta} = d/2)$, and we find  
\begin{equation} \label{FreeEnergyDeltaIntegral}
F(\hat{\Delta}) = N \ F (d/2) +  \int_{\frac{d}{2}}^{\hat{\Delta}}  \frac{\partial F (\hat{\Delta})}{\partial \hat{\Delta}}   \ d \hat{\Delta}, \hspace{1cm} \frac{\partial F (\hat{\Delta})}{\partial \hat{\Delta}}  = (2 \hat{\Delta} - d + 1)\frac{\partial F (\sigma(\hat{\Delta}))}{\partial \sigma}  
\end{equation} 
The integral can be evaluated numerically, and we plot the result for all the large $N$ phases we discussed above in figure \ref{FreeEnergyPlotSymmetric}. The values of the free energy are indeed consistent for all $d$ in $2<d<6$ with the RG flows discussed above and the boundary $F$-theorem in terms of $\tilde F$ in AdS.  

In subsection \ref{SectionFreeEnergyEpsilon}, we will compute these free energies in an $\epsilon$ expansion near even dimensions and verify that they are consistent with the results from the large $N$ expansion. For comparison, we present here some of our large $N$ result near $d = 4$ and $6$ by performing the integral in eq. \eqref{FreeEnergyDeltaIntegral} near these even dimensions. In $d = 4 - \epsilon$, we have two possible phases of the large $N$ theory which preserve $O(N)$ symmetry
\begin{equation} \label{LargeNFreeEnergy4d}
\begin{split}
F(\hat{\Delta} = d - 3) &= N F \left(\frac{d}{2} - 1 \right) + N \textrm{Vol}(H^d) \bigg[  \frac{\epsilon}{128 \pi^2 } + \frac{\epsilon^2  (\gamma + 1+\log 4 \pi)}{256 \pi ^2} \bigg]\\
F(\hat{\Delta} = d - 2) &= N F \left(\frac{d}{2} \right) +  N \textrm{Vol}(H^d) \bigg[  \frac{\epsilon}{128 \pi^2 } + \frac{\epsilon^2  (\gamma -1+\log 4 \pi)}{256 \pi ^2} \bigg]\\
\end{split}
\end{equation}
where we express the $\hat{\Delta} = d - 3$ free energy in terms of $\hat{\Delta} = d/2 - 1$ because that is what we will directly obtain from $\epsilon$ expansion. This is because, as we said before, this phase is obtained by perturbing free theory in $4$ dimensions with Neumann boundary conditions. In $d = 6 - \epsilon$, we have three possible phases
\begin{equation}\label{LargeNFreeEnergy6d}
\begin{split}
F(\hat{\Delta} = d - 4) &= N F \left(\frac{d}{2} - 1 \right)  \ + \ \frac{N \textrm{Vol}(H^d) \epsilon}{512 \pi^3 }\\
F(\hat{\Delta} = d - 3) &= N F \left(\frac{d}{2} \right)  \ + \ \frac{ N \textrm{Vol}(H^d) \epsilon}{512 \pi^3 } \\
F(\hat{\Delta} = d - 2) &= N F \left(\frac{d}{2} \right) \ \ - \ \frac{N \textrm{Vol} (H^d)}{96 \pi^3 \epsilon} 
\end{split}
\end{equation}
Lastly, in $d = 2$, we have one phase with $\hat{\Delta} = d - 2$
\begin{equation}
F(\hat{\Delta} = d - 2) = N F \left(\frac{d}{2} - 1 \right)  \ + \ \frac{N \textrm{Vol}(H^d) (1 - 2 \log 2)}{8 \pi}.
\end{equation}
Note that there are higher order corrections to all of these formulas which go like higher powers in $\epsilon$. 

\subsection{$O(N)$ symmetry breaking phase: extraordinary transition} \label{SectionLargeNE}
There is also a phase of this model when the $O(N)$ symmetry is spontaneously broken to $O(N - 1)$ and the fundamental field $\phi^I$ also gets a one-point function, in addition to $\sigma$. This is what is referred to in the literature as the ``extraordinary transition". As explained in the introduction, it can be obtained by either perturbing the special transition by a boundary mass term with negative coefficient, or by adding a ``boundary magnetic field" term $\sim h \phi^N$ (picking a particular direction in the scalar field space). See e.g. \cite{Liendo:2012hy} for discussion of this phase in the $\epsilon$ expansion, and \cite{10.1143/PTP.72.736} for a large $N$ treatment in the flat space setup. In this section we discuss the large $N$ description of this phase, using the AdS setup. In section \ref{SectionFreeEnergyEpsilon} below we then discuss how this phase arises in the various $\epsilon$ expansions near the even dimensions $d = 2, 4$ and $6$, and match the results with large $N$ wherever applicable. 

We again start with eq. \eqref{ActionLargeN}, but unlike the previous section we do not assume $O(N)$ symmetry and only integrate out $N - 1$ fields 
\begin{equation}
\begin{split}
&Z = \exp[- F^{E}] =  \int d[\phi^N] d 
[\sigma] \exp \left[- S  \right] \\
&S =  \int d^d x \sqrt{g} \left( \frac{1}{2} (\partial \phi^N)^2 + \frac{1}{2}  (\phi^N)^2  \left(\sigma - \frac{d (d - 2)}{4}  \right) \right) + \frac{N - 1}{2} \textrm{tr} \log \left( - \nabla^2 + \sigma - \frac{d (d - 2)}{4}  \right)
\end{split}
\end{equation}
At large $N$, we look for a saddle point with constant one point functions $\sigma = \sigma_*$ and $\phi^N = \phi^N_*$ and require that the derivatives vanish 
\begin{equation}
\begin{split}
&\frac{\partial F^E}{\partial \sigma} \bigg|_{\sigma_*, \phi^N_* } = \frac{\textrm{Vol}(H^d) (\phi^N_*)^2}{2}  + \frac{ (N - 1)  \textrm{Vol}(H^d)}{ 2 (4 \pi)^{d/2} \Gamma(\frac{d}{2})} \\
& \times \frac{\Gamma \left(\frac{d - 1}{2} + \sqrt{\sigma_* + \frac{1}{4}} \right) \Gamma \left(\frac{d - 1}{2} - \sqrt{\sigma_* + \frac{1}{4}} \right) \sin  \left( \pi \left(\frac{d - 1}{2} - \sqrt{\sigma_* + \frac{1}{4}} \right) \right) }{\sin\left(\frac{\pi d}{2}\right)}  \\
&\frac{\partial F^E}{\partial \phi^N} \bigg|_{\sigma_*, \phi^N_* } = \textrm{Vol}(H^d) \phi^N_* \left(\sigma_* - \frac{d (d - 2)}{4} \right) 
\end{split}
\end{equation}
Assuming $\phi^N_*\neq 0$ (otherwise we fall back to the $O(N)$ invariant phases discussed earlier), the second equation gives us $\sigma_* = d (d - 2)/4$, which when plugged into the first equation yields
\begin{equation}
\label{phiN-largeN}
(\phi^N_*)^2 = - \frac{(N - 1) \Gamma(d - 1) \Gamma\left( 1 - \frac{d}{2} \right)}{(4 \pi)^\frac{d}{2}}.
\end{equation}
We can expand this result near even dimensions $d = 2, 4$ and $6$
\begin{equation} \label{ExtraordinaryOnePointLargeN}
(\phi^N_*)^2 \Big|_{d = 2 + \epsilon} = \frac{N -1}{2 \pi \epsilon}, \ \ \ (\phi^N_*)^2 \Big|_{d = 4 - \epsilon} = \frac{N -1}{4 \pi^2 \epsilon}, \ \ \ (\phi^N_*)^2 \Big|_{d = 6 - \epsilon} = -\frac{3 (N -1)}{8 \pi^3 \epsilon}
\end{equation} 
which, as we will see below, match the various $\epsilon$ expansions. Note that the result (\ref{phiN-largeN}) is negative for $4<d<6$, indicating that this phase is non-unitary in that range of dimensions. We will still discuss below, the $d=6-\eps$ description of this phase as a useful cross-check of our results. 

At large $N$, the $N - 1$ transverse fields are just free fields in AdS with their dimensions given by 
\begin{equation}
\hat\Delta^T (\hat\Delta^T - d + 1) = \sigma_* - \frac{d (d - 2)}{4} = 0 \implies \hat\Delta^T = d - 1 + O(1/N).
\end{equation} 
The fact that the fields are massless is related to the fact that these are Goldstone modes for the spontaneously broken $O(N)$ symmetry. Therefore, we expect that the relation $\hat\Delta^T=d-1$ may hold to all orders in $1/N$ (this was also observed in \cite{10.1143/PTP.72.736}). 

\begin{figure}
\centering
\includegraphics[scale = 0.4]{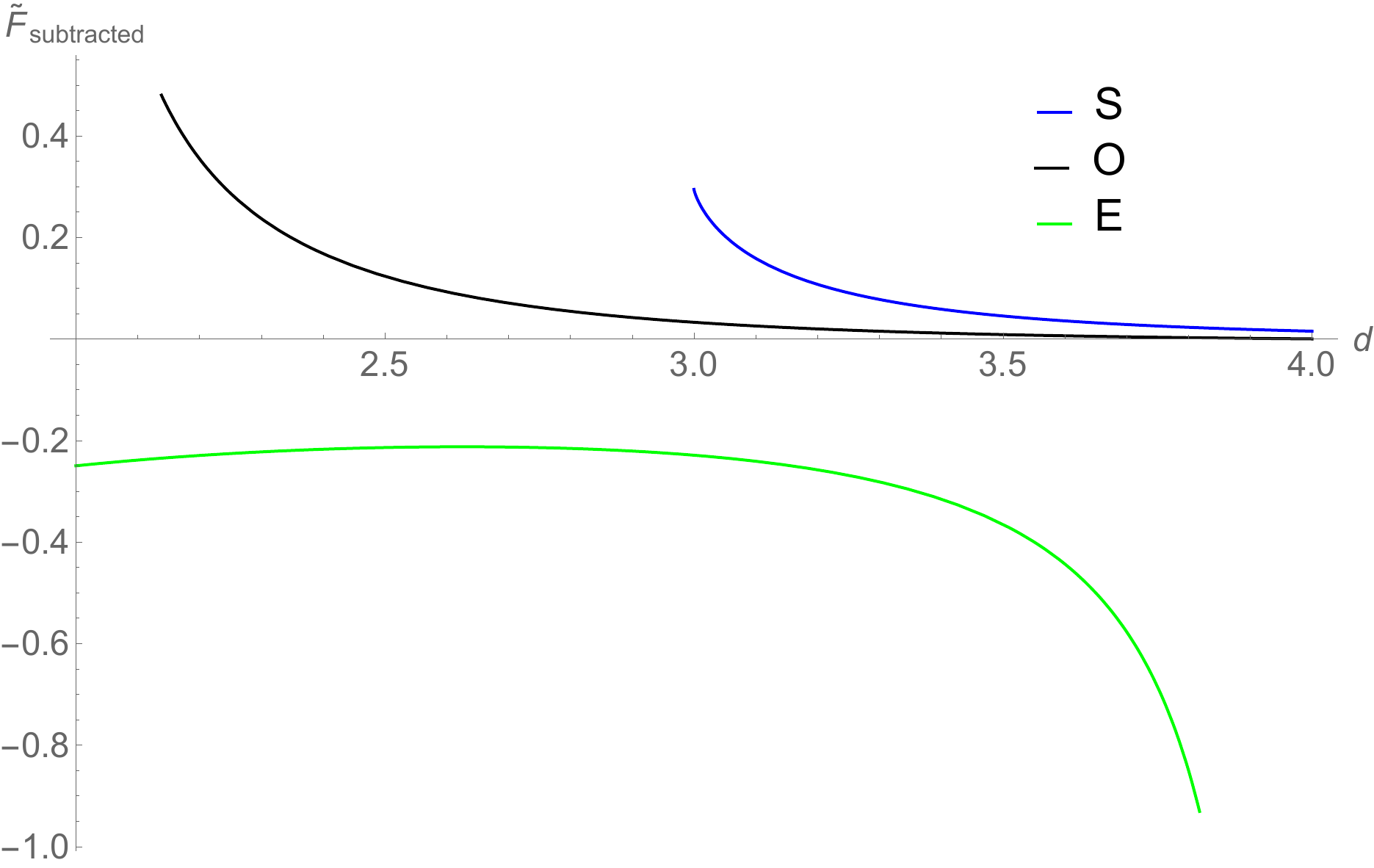}
\caption{Large $N$ free energy between $2<d<4$ for special(S), ordinary(O) and extraordinary(E) transitions. We are plotting $\tilde{F}_{\textrm{subtracted}} = (\tilde{F} - N \tilde{F} (d/2))/N $ on $y-$ axis. }
\label{FreeEnergyPlotSymmetricBreaking}
\end{figure}

At leading order at large $N$, the free energy only receives contribution from the determinants of these transverse fields, and knowing their dimension, we can compute its value using \eqref{FreeEnergy3dFreeMassive}. In $d = 3$ we get
\begin{equation} \label{LargeNFreeEnergy3dE}
F^{E} =  - \frac{N \textrm{Vol}(H^3)}{12 \pi}, \hspace{1cm} \tilde{F}^{E} = - \frac{N \pi }{12 }, \hspace{1cm} a_{3d}^{E} = - \frac{N}{12}.
\end{equation}
So clearly $\tilde{F}^{S} > \tilde{F}^{O} >  \tilde{F}^{E} $ consistent with the expected $F$-theorem for $\tilde F$. For other values of $d$, we can use eq. \eqref{FreeEnergyDeltaIntegral} to get
\begin{equation}
F(\hat{\Delta} = d - 1) = N \ F (d/2) +  \int_{\frac{d}{2}}^{d - 1}  \frac{\partial F (\hat{\Delta})}{\partial \hat{\Delta}}   \ d \hat{\Delta})\,.
\end{equation}
We plot the free energies for three phases (special, ordinary and extraordinary) in figure \ref{FreeEnergyPlotSymmetricBreaking}, again showing agreement with the boundary $F$-theorem in the continuous range $2<d<4$. We do not plot extraordinary above $d = 4$ because this phase becomes non-unitary for $d>4$, as mentioned above. Hence, we do not expect the conjectured $F$ theorem to hold for this phase in $4<d<6$. 
%We can see this for instance by computing $\tilde{F}$ for this phase in $d = 5$ using eq. \eqref{FreeEnergy5dFreeMassive}
%\begin{equation}
%\tilde{F} (\hat{\Delta} = 4) = \frac{7 \pi}{90}
%\end{equation}
%which is higher than all the other three symmetry preserving phases in $d = 5$ %in eq. \eqref{FreeEnergy5dInt}.

In preparation for the $\epsilon$ expansion calculation, we report the results for the large $N$ free energy for the extraordinary phase near dimensions $2, 4$ and $6$ to leading order in $\epsilon$
\begin{equation} \label{LargeNFreeEnergyEEpsilon}
\begin{split}
F(\hat{\Delta} = d - 1)|_{d = 2 + \epsilon} &= N F \left(\frac{d}{2} \right) \ - \frac{N  \textrm{Vol}(H^d)}{8 \pi} \\
F(\hat{\Delta} = d - 1)|_{d = 4 - \epsilon} &= N F \left(\frac{d}{2} \right) \ - \frac{N \textrm{Vol}(H^d)}{8 \pi^2 \epsilon} \\
F(\hat{\Delta} = d - 1)|_{d = 6 - \epsilon} &= N F \left(\frac{d}{2} \right) \ \ + \ \frac{9 N \textrm{Vol} (H^d)}{32 \pi^3 \epsilon}.
\end{split}
\end{equation}
Note that the free energy in extraordinary transition in $d  = 6 - \epsilon$ is actually higher than all the other phases in $6 - \epsilon$ dimensions computed in eq. \eqref{LargeNFreeEnergy6d}. This should be related to the non-unitarity of the extraordinary transition above $4$ dimensions. We can also check this explicitly in $d=5$, where using \eqref{FreeEnergy5dFreeMassive} we get
\begin{equation}
\tilde{F} (\hat{\Delta} = 4) = \frac{7 \pi}{90}
\end{equation}
which is again higher, rather than lower, than all the other three symmetry preserving phases in $d = 5$ in eq. \eqref{FreeEnergy5dInt}. We view this as an indication that the validity of the boundary $F$-theorem is tied with unitarity of the BCFT. 

\subsection{$\epsilon$ expansion near even dimensions} \label{SectionFreeEnergyEpsilon}
\begin{figure}
\centering
\includegraphics[scale = 1]{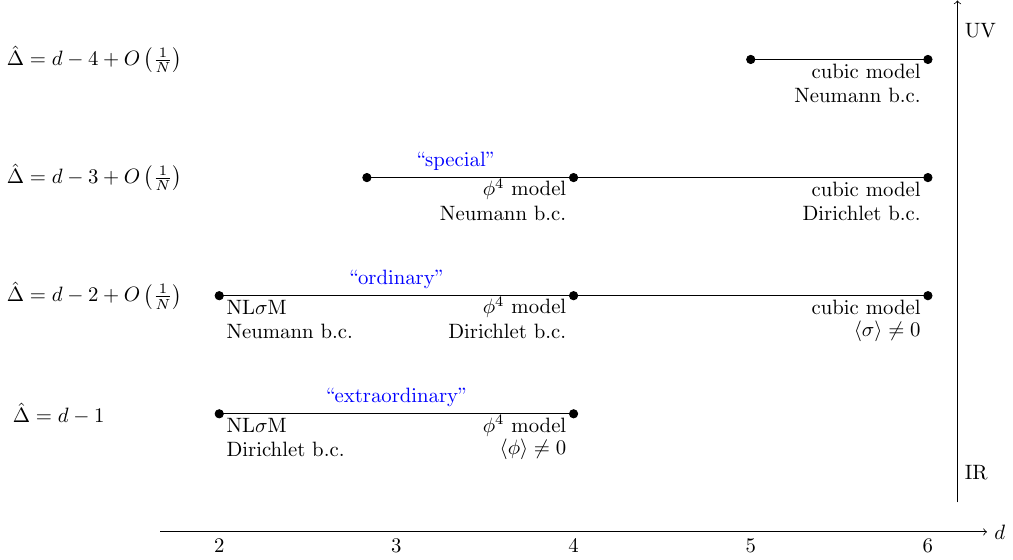}
\caption{All the large $N$ unitary boundary fixed points and their weakly coupled descriptions. The values of $\hat\Delta$ on the left side are the large $N$ scaling dimensions of the leading boundary operator induced by $\phi^I$.}
\label{SurfaceFlowsInteracting}
\end{figure}

The various phases discussed in the previous two subsections all have weakly coupled descriptions in the $\epsilon$ expansion near even dimensions for arbitrary $N$. We discuss those descriptions in this subsection, and in particular match the free energy results that we computed above by large $N$ methods. A summary of all the phases and their perturbative descriptions is given in figure \ref{SurfaceFlowsInteracting}.

\subsubsection{$d = 4 - \epsilon$ dimensions}
In $d = 4 - \epsilon$, the critical behavior of $O(N)$ model is described by the Wilson-Fisher fixed point of the quartic $O(N)$ model written in eq. \eqref{ActionPhi4-flat}. Mapping the action to AdS by including the appropriate conformal coupling term, we have
\begin{equation} \label{ActionPhi4}
S = \int d^d x \sqrt{g} \left( \frac{1}{2} (\partial_{\mu} \phi^I)^2 - \frac{d (d - 2)}{8} \phi^I \phi^I +  \frac{\lambda}{4} (\phi^I \phi^I)^2  \right)\,.
\end{equation}
Ordinary and special transition can be described by the fixed points obtained by perturbing the free theory with Neumann or Dirichlet boundary condition, respectively. We can compute their free energy in perturbation theory as 
\begin{equation}
\begin{split}
    F  &= N F_{\textrm{free}} \ +  \frac{\lambda}{4} \int d^d x \sqrt{g_x} \langle (\phi^I \phi^I(x) )^2 \rangle_0 + \frac{\delta_{\lambda}}{4} \int d^d x \sqrt{g_x} \langle (\phi^I \phi^I(x) )^2 \rangle_0 \\
    &- \frac{\lambda^2}{32} \int d^d x \int d^d x' \sqrt{g_x} \sqrt{g_x'}  \langle (\phi^I \phi^I(x) )^2 (\phi^J \phi^J(x') )^2  \rangle_0
\end{split}
\end{equation}
where the expectation values are taken in the free theory and we have introduced counterterms to deal with the divergences that arise (note that the counterterm $\delta_{\lambda}$ is fixed by the flat space divergences and is unaffected by the presence of the boundary). To do the integrals on the second line, it will be convenient to work with the ball coordinates in AdS introduced in eq. \eqref{BallCoordinates}. %where the determinant of metric is $2^d \ u^{d - 1} \ (1 - u^2)^{-d}$.  
In terms of these coordinates, we can put one of the points at the center of the ball ($u' = 0$) and then it can be checked that the cross ratio becomes
 \begin{equation}
    \xi = \frac{(\textbf{x} - \textbf{x}')^2 + (y - y')^2}{4 y y'} \rightarrow \frac{ u^2}{1 - u^2}\,.
 \end{equation}
The two-point function becomes (here and elsewhere in this section, the top sign refers to Neumann and the bottom sign to Dirichlet boundary condition)
\begin{equation}
   \langle \phi^I (x) \phi^J (x') \rangle_{N/D}  = \delta^{IJ} G(x,x') =  \frac{ \delta^{IJ} \Gamma \left( \frac{d}{2} - 1 \right) (1 - u^2)^{\frac{d}{2} - 1} }{ (4 \pi)^{\frac{d}{2}} } \left( u^{2 -d} \pm 1 \right). 
\end{equation}
Plugging these in, we get
\begin{equation}
\begin{split}
	& F = N F_{\textrm{free}} \ +  \frac{(\lambda + \delta_{\lambda}) \textrm{Vol} (H^d)}{4}  N(N+2) \frac{\Gamma\left(\frac{d}{2} - 1 \right)^2}{ (4\pi)^d }  - \frac{\lambda^2 \textrm{Vol} (H^d)}{4} \frac{\Gamma(\frac{d}{2} - 1)^4}{ (4 \pi)^{2 d} } \frac{2 \pi^{\frac{d}{2}} 2^d}{\Gamma \left( \frac{d}{2} \right)}  \times \\
	& \int_0^1  \frac{u^{d-1} du}{(1-u^2)^d} \left[N(N^2 + 4(N + 1)) (1-u^2)^{d-2} (u^{2-d} \pm 1)^2 + N(N + 2) (1-u^2)^{2d-4} (u^{2-d} \pm 1)^4 \right]
\end{split}
\end{equation}
where the integral over one of the points gave the volume of AdS and the integral over spherical coordinates of the second point gave the volume of the sphere, leaving behind a single integral over $u$ in the last term. The integrals above diverge in the case of Neumann boundary condition. The divergence occurs as the insertion approaches the boundary, $u \rightarrow 1 $. However we can regularize it by introducing a regulator $\delta$ and doing instead the following integral
\begin{equation}
\begin{split}
&F  = N F_{\textrm{free}} \ +  \frac{(\lambda + \delta_{\lambda}) \textrm{Vol} (H^d)}{4}  N(N+2) \frac{\Gamma\left(\frac{d}{2} - 1 \right)^2}{ (4\pi)^d }  - \frac{\lambda^2 \textrm{Vol} (H^d) \Gamma(\frac{d}{2} - 1)^3}{\pi^{\frac{3 d}{2}} (d-2) 2^{3 d}} \int_0^1 \frac{u^{d-1} d u}{(1-u^2)^d}  \\
& \times  \left[N (N^2 + 4 (N + 1)) (1-u^2)^{d-2 + \delta} (u^{2-d - \delta} \pm 1)^2 + N(N + 2) (1-u^2)^{2d-4 + 2 \delta} (u^{2-d - \delta} \pm 1)^4 \right]
\end{split}
\end{equation}
and then taking limit $\delta \rightarrow 0$. We can then plug in $d = 4 - \epsilon$ and $\delta_{\lambda} = \frac{N+8}{8 \pi^2} \frac{\lambda^2}{\epsilon}$ (see for instance \cite{Kleinert:2001ax}), and with Neumann boundary conditions, we get
\begin{equation}
\begin{split}
F^N  &= N F \left(\frac{d}{2} - 1 \right) \ + \textrm{Vol}(H^d) \bigg[  \frac{N(N+2)}{4}  \bigg(\lambda + \frac{N+8}{8 \pi^2} \frac{\lambda^2}{\epsilon}  \bigg) \bigg( \frac{1}{256 \pi^4} + \frac{\gamma + \log 4 \pi}{256 \pi^4} \epsilon \bigg) -  \\
& \lambda^2 \bigg(\frac{N (N+2) (N+8)}{8192 \pi ^6 \epsilon }+\frac{N (N+2) (9 (N+8) (\gamma + \log 4 \pi ) -3 N +25)}{49152 \pi ^6} \bigg) \bigg] \\
&= N F \left(\frac{d}{2} - 1 \right) \ + \textrm{Vol}(H^d) \bigg[  \lambda \frac{N(N+2)}{1024 \pi^4} (1 + \epsilon (\gamma + \log 4 \pi)) \\
& - \frac{\lambda^2 N (N + 2)}{49152 \pi^6}   ( 3 (N + 8) (\gamma + \log 4 \pi ) -3 N + 25  )\bigg]
\end{split}
\end{equation}
where we used the fact that in the free theory with Neumann boundary condition the dimension of the $\hat\phi$ boundary operator is $\hat\Delta=d/2 - 1$. Plugging in the critical point value of the coupling $\lambda = \lambda_* = \frac{8 \pi^2}{N+8} \epsilon   + \frac{24 (3N + 14) \pi^2}{(N + 8)^3} \epsilon^2$, we find
\begin{equation}
\begin{split}
   F^N  = N F \left(\frac{d}{2} - 1 \right) \ +\textrm{Vol}(H^d) \bigg[& \frac{\epsilon }{128 \pi^2} \frac{N(N + 2)}{(N+8)} + \frac{3 \epsilon^2}{128 \pi^2} \frac{N(N + 2) (3 N + 14)}{(N + 8)^3}  \\ 
    & + \frac{\epsilon^2 N (N+2)}{(N + 8)^2 768 \pi^2} ( 3 (N + 8) (\gamma + \log 4 \pi) - 25 + 3 N )  \bigg]
\end{split}    
\end{equation}
which to leading order in $N$ becomes 
\begin{equation}
F^N  = N F \left(\frac{d}{2} - 1 \right) \ + N \textrm{Vol}(H^d) \bigg[ \frac{\epsilon }{128 \pi^2}  + \frac{\epsilon^2} { 256 \pi^2} ( \gamma + 1 + \log 4 \pi)  \bigg].
\end{equation}
This agrees with $d = 4-\epsilon$ value of the large $N$ expansion found above in eq. \eqref{LargeNFreeEnergy4d}. With Dirichlet boundary conditions, the integral gives 
\begin{equation}
\begin{split}
F^D  &=  N F \left(\frac{d}{2} \right) \ + \textrm{Vol}(H^d) \bigg[  \lambda \frac{N(N+2)}{1024 \pi^4} (1 + \epsilon (\gamma + \log 4 \pi)) \\
& - \frac{\lambda^2 N (N + 2)}{49152 \pi^6}   ( 3 N+3(N+8) (\gamma + \log 4 \pi)+13))\bigg]
\end{split}
\end{equation}
Again, plugging in the critical point value, we find
\begin{equation}
\begin{split}
F^D  = N F \left(\frac{d}{2} \right) \ + \textrm{Vol}(H^d) &\bigg[ \frac{\epsilon }{128 \pi^2} \frac{N(N + 2)}{(N+8)} + \frac{3 \epsilon^2}{128 \pi^2} \frac{N(N + 2) (3 N + 14)}{(N + 8)^3}  \\
& + \frac{\epsilon^2 N (N+2)}{(N + 8)^2 768 \pi^2} (  -3 N+3(N+8) (\gamma + \log 4 \pi)-13) )  \bigg]
\end{split}
\end{equation}
which at large $N$ becomes
\begin{equation}
F^D  = N F \left(\frac{d}{2} \right) \ + N\textrm{Vol}(H^d) \bigg[ \frac{\epsilon }{128 \pi^2} + \frac{\epsilon^2 }{ 256 \pi^2} (  \gamma + \log 4 \pi -1 ) \bigg]\,.
\end{equation}
This again  agrees with the $4-\epsilon$ value of the large $N$ expansion result, see eq. \eqref{LargeNFreeEnergy4d}. 

Let us now discuss the phase of this model where $O(N)$ symmetry is broken and which describes the extraordinary transition in $d = 4- \epsilon$ dimensions. In the AdS approach, this is simply obtained by minimizing the potential on the hyperbolic space for the action in eq. (\ref{ActionPhi4}), $V(\phi)=-\frac{d(d-2)}{8}\phi^I\phi^I+\frac{\lambda}{4}(\phi^I\phi^I)^2$. Extremizing the potential one finds\footnote{In the flat space approach, this phase corresponds to the classical solution of the equation of motion $\nabla^2\phi^N = \lambda (\phi^N)^3$ given in $d=4$ by $\phi^N = \sqrt{2/\lambda}\frac{1}{y}$, see e.g. \cite{Liendo:2012hy}.}
\begin{equation} \label{ExtraordinaryOnePoint4d}
\phi^K \phi^K = \frac{d (d - 2)}{4 \lambda} = \frac{N + 8}{4 \pi^2 \epsilon} \implies  \phi^a = 0, \ \ a = 1,...N - 1; \ \phi^N = \sqrt{\frac{d (d - 2)}{4 \lambda}} = \sqrt{\frac{N + 8}{4 \pi^2 \epsilon}}
\end{equation}
where we used the value of the coupling at the fixed point $\lambda = \lambda_* = (8 \pi^2 \epsilon)/(N + 8)$. The value of the one-point function above is precisely consistent with the result of the large $N$ expansion in eq. (\ref{ExtraordinaryOnePointLargeN}). We can expand around this minimum in terms of $\phi^a$ and $\phi^N = \sqrt{\frac{d (d - 2)}{4 \lambda}} + \chi $
\begin{equation} \label{ActionExtra4d}
\begin{split}
S &= \int d^d x \sqrt{g} \left( \frac{1}{2} (\partial_{\mu} \phi^I)^2 - \frac{d (d - 2)}{8} \phi^I \phi^I +  \frac{\lambda_0}{4} (\phi^I \phi^I)^2  \right) \\
&= - \frac{d^2(d-2)^2}{64\lambda_0} \int d^d x \sqrt{g} \ + \  \int d^d x \sqrt{g} \bigg( \frac{1}{2} (\partial_{\mu} \phi^a)^2  + \frac{1}{2} (\partial_{\mu} \chi)^2  + \frac{d(d - 2)}{4} \chi^2 \\
& + \frac{\sqrt{d(d - 2) \lambda_0}}{2} (\chi^3 + (\phi^a \phi^a) \chi) +  \frac{\lambda_0}{4} \left( (\phi^a \phi^a)^2 + \chi^4 + 2 \chi^2 (\phi^a \phi^a) \right)  \bigg)
\end{split}
\end{equation}
where we use $\lambda_0$ to emphasize that we are using bare coupling here. So we are left with $N - 1$ massless fields with boundary dimension $3$ and a single massive field $\chi$. The mass terms of $\chi$ also tell us the dimensions of the boundary operator corresponding to $\chi$
\begin{equation}
\hat{\Delta} (\hat{\Delta} - d + 1) = \frac{d (d - 2)}{4}, \implies \hat{\Delta}_{\pm} = \frac{d - 1 \pm \sqrt{3 d^2 - 6 d + 1}}{2} \ \xrightarrow{d \rightarrow 4} \  4, - 1
\end{equation}
We obviously choose the $+$ boundary condition for $\chi$ for boundary unitarity, which gives a dimension $4$ operator at the boundary. To leading order in $\epsilon$, the free energy can then be written as
\begin{equation}
F^E = (N - 1) F_{\phi^a} (\hat{\Delta} = 3) + F_{\chi}(\hat{\Delta} = 4) + S_{\textrm{clas.}} = (N - 1) F_{\phi^a} (\hat{\Delta} = 3) + F_{\chi}(\hat{\Delta} = 4) - \frac{\textrm{Vol}(H^d)}{\lambda_0}.
\end{equation}
Using eq. \eqref{FreeEnergyDeltaIntegral} in $4 - \epsilon$ dimensions, we get
\begin{equation}
F_{\phi^a} (\hat{\Delta} = 3) =  F \left(\frac{d}{2} \right) \ - \frac{ \textrm{Vol}(H^d)}{8 \pi^2 \epsilon}; \hspace{1cm} F_{\chi}(\hat{\Delta} = 4) = F \left(\frac{d}{2} \right) - \frac{9  \textrm{Vol}(H^d) }{8 \pi^2 \epsilon}; 
\end{equation}
where both of these results are true up to terms that are finite as $\epsilon \rightarrow 0$. These $1/\epsilon$ poles in the free energy must be cancelled by the ones present in the bare coupling coming from the classical action. Recall that in this model, the coupling gets renormalized as  (see for instance, \cite{Fei:2015oha})
\begin{equation}
\frac{1}{\lambda_0} = \mu^{- \epsilon} \left( \frac{1}{\lambda} - \frac{N + 8}{8 \pi^2 \epsilon} + O(\lambda) \right).
\end{equation}
The pole in $\epsilon$ here clearly cancels the ones coming from one-loop determinants for $\chi$ and $\phi^a$, so that the free energy becomes 
\begin{equation}
F^E = N  F \left(\frac{d}{2} \right) - \frac{\textrm{Vol}(H^d)}{ \lambda_*} = N  F \left(\frac{d}{2} \right) \ - \ \frac{\textrm{Vol}(H^d) (N + 8)}{8 \pi^2 \epsilon}
\end{equation}
where we finally plugged in the value of the renormalized coupling at fixed point. At large $N$, this precisely matches the result of large $N$ computation in eq. \eqref{LargeNFreeEnergyEEpsilon} near $4$ dimensions.

\subsubsection{$d = 6 - \epsilon$ dimensions}
Near $6$ dimensions, the large $N$ $O(N)$ model can be described by the fixed point of the cubic $O(N)$ scalar theory written in \eqref{Action6d-flat} in $d = 6 -\epsilon$ dimensions. In AdS, we work with the action 
\begin{equation} \label{Action6d}
    S = \int d^d x\sqrt{g} \bigg[ \frac{1}{2} (\partial_{\mu} \phi^I)^2 - \frac{d (d - 2)}{8} (\phi^I \phi^I + \sigma^2) + \frac{1}{2} (\partial_{\mu} \sigma)^2 + \frac{g_1}{2} \sigma \phi^I \phi^I + \frac{g_2}{6} \sigma^3 \bigg].
\end{equation}
At large enough $N$, one finds a fixed point at \cite{Fei:2014yja} 
\begin{equation} \label{FixedPoint6d}
(g_1)_* = \sqrt{\frac{6 \epsilon (4 \pi)^3}{N}}, \ \ (g_2)_* = 6 (g_1)_*.
\end{equation}
The $\hat{\Delta} = d - 4$ and $d - 3$ phases can be described by imposing Neumann or Dirichlet boundary conditions on N $\phi^I$ fields, respectively. We will only do the leading order in $N$ calculation here, and this is sufficient for our purposes of comparing with the large $N$. To leading order, the free energy is given by
\begin{equation}
F = N F_{\textrm{free}} \ -  \frac{g_1^2 N ^2}{8} \int d^d x d^d x' \sqrt{g_x} \sqrt{g_{x'}} G(x,x) G(x', x') G(x,x').  
\end{equation}
Plugging in the correlators gives
\begin{equation}
 F  = N F_{\textrm{free}} \ -  \ \frac{( g_1^2 N^2) \textrm{Vol} (H^d) \Gamma(\frac{d}{2} - 1)^2 }{2^{2 d +1}  \pi^d (d - 2)} \int_0^1 d u \ \frac{u^{d - 1}}{(1 - u^2) ^d} (1 - u^2) ^{\frac{d + \delta}{2} -1} (u ^{2 - d - \delta} \pm 1) 
\end{equation}
where we already introduced the regulator $\delta$ as before, since this integral also diverges in Neumann case. By $\pm$ above, we really mean Neumann or Dirichlet boundary condition on the $\sigma$ correlator, since $\phi^I$ only contributes through one-point function at this order. We should not be able to tell which boundary condition we are using for $\sigma$ just from the leading large $N$ calculation, and indeed both the signs give the same answer. In the case of Neumann condition on $\phi$, we get
\begin{equation}
F^N  = N F\left(\frac{d}{2} - 1 \right) \ + \ \frac{ g_1^2 N^2 \textrm{Vol} (H^d) }{6 (2)^{15} \pi^6} = N F\left(\frac{d}{2} - 1 \right) \ + \ \frac{\epsilon N \textrm{Vol} (H^6)}{ 512 \pi^3}
\end{equation}
while in the Dirichlet case
\begin{equation}
   F^D  = N F\left(\frac{d}{2} \right) \ + \ \frac{ g_1^2 N^2 \textrm{Vol} (H^d) }{6 (2)^{15} \pi^6} = N F\left(\frac{d}{2} \right) \ + \ \frac{\epsilon N \textrm{Vol} (H^6)}{ 512 \pi^3}
\end{equation}
This agrees with what we found in eq. \eqref{LargeNFreeEnergy6d}.

There is another phase of this model that preserves $O(N)$ symmetry, and which turns out to be counterpart of the large $N$ phase with $\hat{\Delta} = d - 2$ (i.e., this is the smooth continuation of the ``ordinary" transition above $d=4$). It corresponds to the following extremum of the potential in AdS where the field $\sigma$ gets a one-point function\footnote{In $d=6$ flat space with flat boundary, this corresponds to the solution of $\nabla^2\sigma = \frac{g_2}{2}\sigma^2$ given by $\sigma=\frac{12}{g_2 y^2}$.}
\begin{equation}
\sigma = \frac{d (d - 2)}{2 g_2}; \hspace{1cm} \phi^I = 0. 
\end{equation}
We can expand the action around this solution as 
\begin{equation}
\begin{split}
S &= \int d^d x \sqrt{g} \left( \frac{1}{2} (\partial_{\mu} \phi^I)^2 + \frac{1}{2} (\partial_{\mu} \sigma)^2 - \frac{d (d -2)}{8} (\sigma^2 + \phi^I \phi^I) + \frac{g_1}{2} \sigma \phi^I \phi^I + \frac{g_2}{6} \sigma^3 \right) \\
S &= - \frac{d^3 (d - 2)^3}{96 \ (g_{2,0})^2} \int d^d x \sqrt{g} +  \int d^d x \sqrt{g} \bigg( \frac{1}{2} (\partial_{\mu} \phi^I)^2 + \frac{1}{2} (\partial_{\mu} \delta \sigma)^2 + \frac{d (d - 2)}{8} (\delta \sigma)^2  \\ & -  \left( 1 - \frac{ 2 g_{1,0}}{g_{2,0}} \right) \frac{d (d -2)}{8} \phi^I \phi^I  +  \frac{g_{1,0}}{2} \delta \sigma  \phi^I \phi^I + \frac{g_{2,0}}{6}  (\delta\sigma)^3 \bigg).
\end{split}
\end{equation}
where $\delta \sigma$ is the fluctuation of $\sigma$, and we emphasize that the coupling constants are bare. Given the mass of the fluctuations, we can read off the boundary dimensions
\begin{equation}
\begin{split}
\hat{\Delta}^{\sigma}(\hat{\Delta}^{\sigma} - d + 1) &= \frac{d (d - 2)}{4}, \implies \hat{\Delta}_{\pm}^{\sigma} = 6, - 1; \\
\hat{\Delta}^{\phi^I} (\hat{\Delta}^{\phi^I} - d + 1) &= -\frac{d (d - 2)}{4}\left( 1 - \frac{ 2 g_{1,0}}{g_{2,0}} \right), \implies \hat{\Delta}_{\pm}^{\phi^I} = 4, 1
\end{split}
\end{equation}
We obviously choose the $\hat{\Delta}_+$ boundary condition for both $\sigma$ and $\phi^I$. The dimension for $\phi^I$ is indeed consistent with the phase of the large $N$ theory with leading boundary dimension being $d - 2$. The $\sigma$ dimension is consistent with the fact that for this phase, as we will review in section \ref{SectionBoundarySpectrum} below, the leading boundary operator induced by $\sigma$ has dimension $\hat\Delta = d$. 

The free energy at leading order in $\epsilon$ then becomes 
\begin{equation}
F^O = N  F_{\phi^I} (\hat{\Delta} = 4) \ + \ F_{\delta \sigma}(\hat{\Delta} = 6) \ + \ S_{\textrm{clas.}} = N F_{\phi^I} (\hat{\Delta} = 4) \ + \ F_{\delta \sigma}(\hat{\Delta} = 6) -  \frac{144 \ \textrm{Vol}(H^d) }{(g_{2,0})^2}.
\end{equation}
At large $N$, we only need to take care of the $N$ $\phi^I$ fields, and using eq. \eqref{FreeEnergyDeltaIntegral} 
\begin{equation}
F_{\phi^I} (\hat{\Delta} = 4) =  F \left(\frac{d}{2} \right) \ - \frac{ \textrm{Vol}(H^d)}{96 \pi^3 \epsilon}
\end{equation}
This $1/\epsilon$ pole gets cancelled when we plug in the bare coupling in terms of the renormalized coupling \cite{Fei:2014xta}
\begin{equation} \label{BareCoupling6d}
\begin{split}
g_{1,0} &= \mu^{\frac{\epsilon}{2}} \left(g_1 + \frac{N g_1^3 + g_1 g_2^2 - 8 g_1^3 - 12 g_1^2 g_2}{12 (4 \pi)^3 \epsilon} + ...  \right) \\
g_{2,0} &= \mu^{\frac{\epsilon}{2}} \left(g_2 + \frac{N g_1^2 g_2 - 3 g_2^3 - 4 N g_1^3 }{4 (4 \pi)^3 \epsilon} + ...  \right).
\end{split}
\end{equation}
The free energy then becomes a finite function of the renormalized coupling, and plugging in the fixed point value we find
\begin{equation}
F^O = N  F \left(\frac{d}{2} \right) - \frac{ 144 \ \textrm{Vol}(H^d)}{ (g_2)^2_*} = N  F\left(\frac{d}{2} \right) \ - \ \frac{N \textrm{Vol}(H^d)  }{96 \pi^3 \epsilon}
\end{equation}  
This agrees with the large $N$ result in eq. \eqref{LargeNFreeEnergy6d}.

As discussed above, there is also a (non-unitary) $O(N)$ symmetry breaking vacuum of the cubic theory in eq. (\ref{Action6d}) which describes the extraordinary transition in $d = 6 - \epsilon$. It corresponds to an extremum of the potential in AdS at the following complex values
\begin{equation} \label{ExtraordinaryOnePoint6d}
\sigma = \frac{d (d - 2)}{4 g_1}; \ \ \ \phi^N = \pm \frac{d (d - 2) \sqrt{2 g_1 - g_2}}{4 g_1^{3/2}} = \pm 12 i \sqrt{\frac{N }{ 6 \epsilon (4 \pi)^3}} ; \ \ \ \phi^a = 0, \ \ a = 1, .. , N -1.
\end{equation}
The one-point function of $\phi^N$ agrees with the large $N$ result above \eqref{ExtraordinaryOnePointLargeN}, and its being complex indicates that the theory is non-unitary. We can expand around this classical solution in terms of fluctuations
\begin{equation}
\begin{split}
S &= \frac{36 (g_{2,0} - 3 g_{1,0})}{g_{1,0}^3} \int d^d x \sqrt{g} +  \int d^d x \sqrt{g} \bigg( \frac{1}{2} (\partial_{\mu} \phi^a)^2 + \frac{1}{2} (\partial_{\mu} \delta \sigma)^2 + \frac{1}{2} (\partial_{\mu} \chi)^2 - \\ &  \left( 1 - \frac{g_{2,0}}{g_{1,0}} \right) \frac{d (d -2)}{8} (\delta\sigma)^2  + \frac{d (d - 2) \sqrt{2 g_{1,0} - g_{2,0}}}{4 \sqrt{g_{1,0}}} \delta \sigma \chi +  \frac{g_{1,0}}{2} \delta \sigma ( \phi^a \phi^a + \chi^2) + \frac{g_{2,0}}{6}  (\delta\sigma)^3 \bigg).
\end{split}
\end{equation}
where $\delta \sigma, \chi$ and $\phi^a$ are fluctuations of $\sigma, \phi^N$ and the $N-1$ transverse fields respectively. The transverse fields are massless Goldstone bosons and the leading boundary operator in their boundary operator expansion has dimensions $5$. So the free energy is
\begin{equation}
F^E = (N - 1) F_{\phi^a} (\hat{\Delta} = 5) \ + \ F_{ \delta \sigma} \ + \ F_{\chi}   \ + \frac{36 (g_{2,0} - 3 g_{1,0}) \textrm{Vol}(H^d) }{g_{1,0}^3}.
\end{equation}
At large $N$, we only really need to take into account the contribution of $N - 1$ massless fields, and using eq. \eqref{FreeEnergyDeltaIntegral} in $d = 6 - \epsilon$, we get 
\begin{equation}
F_{\phi^a} (\hat{\Delta} = 5) =  F \left(\frac{d}{2} \right) \ + \ \frac{ 9  \textrm{Vol}(H^d)}{32 \pi^3 \epsilon} + .. 
\end{equation}
up to terms that are finite as $\epsilon \rightarrow 0$. This $1/\epsilon$ pole is cancelled by the one coming from classical action when we plug in the bare coupling in terms of renormalized coupling, thus rendering the free energy finite in terms of renormalized coupling. We can then plug in the fixed point value of the coupling to get at large $N$
\begin{equation}
F^E = N F \left(\frac{d}{2} \right) \ + \frac{36 ((g_{2})_* - 3 (g_{1})_*) \textrm{Vol}(H^d) }{g_{1,0}^3} = N F \left(\frac{d}{2} \right) \ + \ \frac{ 9 N \textrm{Vol}(H^d)}{32 \pi^3 \epsilon}.
\end{equation}
This again agrees with the result we found using a large $N$ expansion \eqref{LargeNFreeEnergyEEpsilon}. 

\subsubsection{$d = 2 + \epsilon$ dimensions}
A similar analysis can be done for the non-linear sigma model in $2 + \epsilon$ dimensions. Mapping the flat space action (\ref{Action2d-flat}) to AdS, we have
\begin{equation} \label{Action2d}
S = \int d^d x \sqrt{g} \left( \frac{1}{2} \partial_{\mu} \phi^I \partial^{\mu} \phi^I - \frac{d (d - 2)}{8} \phi^I \phi^I  + \sigma \left( \phi^I \phi^I  - \frac{1}{t^2} \right)\right).
\end{equation}
For the free energy calculation in this section, we will restrict to the simpler case of Dirichlet boundary conditions, which we expect to be related to the extraordinary transition. We can solve the constraint explicitly in terms of the following unconstrained variables
\begin{equation} \label{ExtraordinaryOnePoint2d}
\phi^a = \psi^a, \ \ a = 1,...,N -1; \ \ \ \phi^N = \frac{1}{t} \sqrt{1 - t^2 \psi^a \psi^a} \,.
%= \sqrt{\frac{N - 2}{2 \pi \epsilon}} .
\end{equation}
and the $\psi^a$ are quantized with Dirichlet boundary conditions. 
Recall that in $d = 2 + \epsilon$, this model has a fixed point at $t^2 = (t_*)^2 = (2 \pi \epsilon)/(N - 2)$ \cite{Brezin:1976ap}. Using this fixed point value, we find the one-point function of $\phi^N$ to leading order in $\epsilon$
\begin{equation}
\phi^N = \frac{1}{t_*} = \sqrt{\frac{N - 2}{2 \pi \epsilon}}\,,
\end{equation}
which is seen to match the large $N$ expansion result for the extraordinary transition, given in \eqref{ExtraordinaryOnePointLargeN}. Using the parametrization (\ref{ExtraordinaryOnePoint2d}) we can write down the action in terms of the unconstrained variables
 \begin{equation}
S = \int d^d x \sqrt{g} \left( \frac{1}{2} \partial_{\mu} \psi^a \partial^{\mu} \psi^a  - \frac{d (d - 2)}{8 t_0^2} +  \frac{t_0^2}{2} (\psi^a \partial_{\mu} \psi^a)^2 +\ldots \right)
\end{equation}
where we omitted  additional terms of higher order in $t$. So we are left with $N - 1$ massless fields $\psi^a$, and the extraordinary transition corresponds to choosing $\hat{\Delta}_+ = d - 1$ boundary condition for these fields in AdS. We will check this further by computing the anomalous dimensions of these transverse fields in eq. \eqref{AnomalousDimensionNLSM} and comparing it to the large $N$ result for the extraordinary transition.  To leading order, the free energy will be just the classical action plus the fluctuations of $N - 1$ massless scalars
\begin{equation}
F^E = (N - 1) F (\hat{\Delta} = d - 1)  \ - \ \frac{\epsilon \textrm{Vol}(H^d)}{4 t_0^2}
\end{equation}
The bare coupling in this case is (see for instance \cite{Giombi:2019enr})
\begin{equation}
t_0 = \mu^{-\frac{\epsilon}{2}} \left( t + \frac{t^3 (N - 1)}{4 \pi \epsilon} + ... \right)
\end{equation}
and using eq. \eqref{FreeEnergyDeltaIntegral} in $d = 2 + \epsilon$ gives 
\begin{equation}
F(\hat{\Delta} = d - 1) =  F \left(\frac{d}{2} \right) \ - \ \frac{ \textrm{Vol}(H^d)}{8 \pi} + .. 
\end{equation}
up to terms that vanish as $\epsilon \rightarrow 0$. Combining these results and using the fixed point value $t_*^2 = \frac{2 \pi \epsilon}{N - 2}$, we get a precise match with the large $N$ result in eq. \eqref{LargeNFreeEnergyEEpsilon}
\begin{equation}
F^E = (N - 1) F \left(\frac{d}{2} \right)  \ - \ \frac{(N - 2) \textrm{Vol}(H^d)}{8 \pi}.
\end{equation}

\section{Bulk correlators and extracting BCFT data} \label{SectionBoundarySpectrum}
In this section, we study in more detail the BCFT data of the models discussed in this paper. We start with the discussion of the bulk two-point functions of both $\phi$ and $\sigma$ at large $N$.  We then move on to the Wilson-Fisher model in $d = 4 - \epsilon$ dimensions, where we compute the bulk two-point function of the fundamental field $\phi$, to second order in $\epsilon$, for both Neumann and Dirichlet boundary conditions on the fundamental field. This can be used to calculate the anomalous dimensions and OPE coefficients of various boundary operators appearing in boundary operator expansion of $\phi$. Instead of directly computing loop diagrams, we make essential use of the fact that $\phi$ satisfies an equation of motion in the bulk. We use similar ideas and the BCFT crossing equation to compute $1/N$ corrections to some of the BCFT data in the large $N$ description. We find that at order $1/N$, for the ordinary transition, the boundary operator expansion of $\phi$ contains a tower of operators with dimensions $2 d - 2 + 2 k$. From the boundary point of view, these can be schematically written as $\hat{\phi} (\vec{\partial}^2)^{k} \hat{\sigma}$ with $\hat{\phi}$ and $\hat{\sigma}$ being the leading boundary operators induced by $\phi$ and $\sigma$. We find the boundary operator expansion coefficients for this tower, and use this result to ``bootstrap" the $1/N$ correction to the scaling dimension of $\hat\phi$. For the special transition, we find that two such towers appear, with dimensions $d - 1 + 2 p$ and $2 d - 3 + 2 q$. This is in accordance with the fact that, as we will see, for special transtion, $\sigma$ induces two boundary operators with dimensions $2$ and $d$. 

We also give a formula for the anomalous dimensions of higher-spin displacement operators which are induced on the boundary by higher-spin currents in the bulk. We will find the anomalous dimensions by using the fact that the current conservation is weakly broken in the bulk, similarly to the analysis in \cite{Skvortsov:2015pea,Giombi:2016hkj}. Lastly, we calculate the boundary four-point function in the Wilson-Fisher model and in the non-linear sigma model, which in AdS language is given by a relatively simple contact Witten diagram. We extract the anomalous dimensions of boundary operators from these four-point functions using techniques familiar from AdS/CFT literature. 

\subsection{Bulk two-point functions}
Let us start by analyzing the ordinary ($ \hat{\Delta} = d - 2+O(1/N)$) and special ($\hat{\Delta} = d - 3+O(1/N)$) transition in a little more detail. Knowing the dimension of the boundary operator, we can immediately get the two point function of the bulk fundamental field $\phi$ to leading order in large $N$. For ordinary transition (O), it is  
\begin{equation}
\langle \phi^I \phi^J \rangle = \delta^{I J} G_{\phi} (\xi) =  \delta^{IJ} G_{b b}^{\hat{\Delta} = d - 2} = \frac{\Gamma\left( \frac{d}{2} - 1 \right)}{ (4 \pi)^\frac{d}{2}}\frac{1}{(\xi(1 + \xi))^{\frac{d}{2} -1}} 
\end{equation}
while for special transition(S), we have 
\begin{equation}
\langle \phi^I \phi^J \rangle = \delta^{I J} G_{\phi} (\xi) = \delta^{IJ} G_{b b}^{\hat{\Delta} = d - 3}(\xi) = \frac{\Gamma\left( \frac{d}{2} - 1 \right)}{ (4 \pi)^\frac{d}{2}}\frac{1 + 2 \xi}{(\xi(1 + \xi))^{\frac{d}{2} -1}} 
\end{equation}
In the boundary channel, clearly there is only a single operator in both cases (this is because, as mentioned above, the bulk-to-bulk propagator in AdS is proportional to a single boundary conformal block). In the bulk channel, there is a tower of operators with dimensions $2 n + 2$ with the following OPE coefficients
\begin{equation}
\begin{split}
(\lambda )_n^O &= -\frac{ (-1)^n \Gamma \left(\frac{d}{2}\right) \, _2F_1\left(-n-1,-n;\frac{1}{2} (d-4 n-2);1\right)}{\Gamma (n+2) \Gamma \left(\frac{d}{2}-n-1\right)} \\
(\lambda )^S_n &= \frac{(d^2 - 4 d (n + 2) + 8 (1 + n)^2 + 4) \Gamma(1 - \frac{d}{2} + n) \Gamma(2 - \frac{d}{2} + n)^2}{4 \Gamma(2 - \frac{d}{2})^2 \Gamma(n + 2) \Gamma(2 - \frac{d}{2} + 2 n)}. 
\end{split}
\end{equation} 
We show how to derive these formulae in the appendix \ref{BulkOPEAppendix}.

To do a large $N$ perturbation theory, we still need the $\sigma$ propagator. To calculate that, we decompose $\sigma = \sigma_* + \delta \sigma(x)$ into a constant background and fluctuations around 
it. We can plug this into eq. \eqref{EffectiveActionSigmaN} and read off the quadratic piece of the effective action for $\sigma$
\begin{equation}
S_2 = - \frac{N }{4} \int d^d x d^d y \sqrt{g_x} \sqrt{g_y} \delta \sigma(x) B(x,y) \delta \sigma(y), \ \ \ B(x,y) = \left[ \frac{1}{- \nabla^2 + \sigma_* - \frac{d (d - 2)}{4}} \right]^2.
\end{equation}
This tells us that the connected propagator of $\sigma$ must satisfy
\begin{equation} \label{SigmaPropEqN}
\int d^d x \sqrt{g_x} \ \left(G_{\phi}(\xi_{x_1, x}) \right)^2 G^0_{\sigma}(\xi_{x, x_2})  = -\frac{2}{N} \frac{\delta^{d}(x_1 - x_2)}{\sqrt{g_{x_1}}}.
\end{equation}
So we need to invert the square of the $\phi$ propagator in order to get the $\sigma$ propagator. The general method to invert functions on half-space was described in \cite{McAvity:1995zd}. The procedure can be straightforwardly adapted to AdS and we show how to do that in Appendix \ref{SigmaCorrelatorAppendix}. Here we just report the results. The full two point function of $\sigma$ takes the form 
\begin{equation}
\langle \sigma(x) \sigma(x') \rangle = (\sigma_*)^2 +  G_{\sigma} (\xi).
\end{equation}
For ordinary transition with $\sigma_* = \frac{(d - 2)(d - 4)}{4}$, we get 
\begin{equation}
G_{\sigma} (\xi) = B \frac{\Gamma(d) \Gamma(d - 2)}{\Gamma(2 d - 4)} \xi^{-d} \ \  _2F_1 (d - 2, d; 2 d -4; - \frac{1}{\xi})
\end{equation}
where 
\begin{equation}
B =  \frac{(4 - d)\Gamma(d - 2) }{ N \Gamma(2 - \frac{d}{2}) \Gamma(\frac{d}{2} - 1)^3}.
\end{equation}
In the limit when we push one of the points to the boundary, $y ' \rightarrow 0$, the leading term is 
\begin{equation} \label{SigmaBulkBoundaryDirichlet}
G_{\sigma} (\xi) = B \frac{\Gamma(d) \Gamma(d - 2)}{\Gamma(2 d - 4)} \bigg( \frac{4 y y'}{(\vec{x} - \vec{x} ')^2 + y ^2} \bigg)^d.
\end{equation}
This tells us that the dimension of leading boundary operator induced by $\sigma$ is $\hat\Delta=d$, and we identify it as the being proportional to the displacement operator (more comments on the displacement operators are collected in Appendix \ref{AppendixDisplacement}). For the special transition with $\sigma_* = \frac{(d - 4)(d - 6)}{4}$ , we get the more complicated expression
\begin{equation}
\begin{split}
G_{\sigma} (\xi) &= B \bigg[ \frac{1}{3} \frac{(6 - d) \Gamma(d) \Gamma(d - 2)}{(d -2) \Gamma( 2d - 5)} \frac{\xi + \frac{1}{2}}{(\xi (\xi + 1))^{\frac{d + 1}{2}}} \ \ _3 F_2 (\frac{d}{2} + \frac{1}{2},\frac{d}{2} - \frac{3}{2} , \frac{3}{2}; d - \frac{5}{2}, \frac{5}{2}; - \frac{1}{4 \xi (1 + \xi)}  )  \\ 
&+ \frac{\pi \Gamma (\frac{d}{2} -1)^2}{\Gamma(d - 3) \Gamma(\frac{d}{2} - \frac{3}{2}) \Gamma(\frac{7}{2} - \frac{d}{2})} \frac{8}{(1 + 2 \xi)^2} \ \ _2F_1 (\frac{3}{2}, 1 ;\frac{7}{2} - \frac{d}{2} ; \frac{1}{(1 + 2 \xi)^2})   \bigg]. 
\end{split}
\end{equation}
Again, if we push one of the points to the boundary, $y ' \rightarrow 0$, the leading terms are
\begin{equation} \label{SigmaBulkBoundaryNeumann}
\begin{split}
G_{\sigma} (\xi) &= B \bigg[ \frac{1}{3} \frac{(6 - d) \Gamma(d) \Gamma(d - 2)}{(d -2) \Gamma( 2d - 5)} \bigg( \frac{4 y y'}{(\vec{x} - \vec{x} ')^2 + y ^2} \bigg)^d \\ 
&+ \frac{8 \pi \Gamma (\frac{d}{2} -1)^2}{\Gamma(d - 3) \Gamma(\frac{d}{2} - \frac{3}{2}) \Gamma(\frac{7}{2} - \frac{d}{2})} \bigg( \frac{4 y y'}{(\vec{x} - \vec{x} ')^2 + y ^2} \bigg)^2  \bigg]. 
\end{split}
\end{equation}
So there are two operators of dimensions $d$ and $2$ induced by $\sigma$ on the boundary. The dimension $d$ operator is proportional to the displacement. The dimension $2$ operator is the boundary mass operator which drives the transition from special to ordinary transition.

Let us now turn to the extraordinary transition, where the $O(N)$ symmetry is broken to $O(N-1)$ and both $\sigma$ and $\phi^N$ acquire one-point functions. The correlator of the transverse $N-1$ fields to leading order at large $N$ is easily obtained by simply plugging into the bulk-to-bulk propagator the boundary dimension $\hat\Delta=d - 1$ 
\begin{equation}
\langle \phi^a (x_1) \phi^b (x_2) \rangle = \delta^{ab} G^T (\xi); \ \  G^T (\xi) = \frac{\Gamma\left( \frac{d}{2} \right)}{(4 \pi)^{\frac{d}{2}} (d - 1)} \xi^{-d + 1} {}_2F_1 \left( d - 1, \frac{d}{2}; d ; -\frac{1}{\xi} \right)
\end{equation}
To compute the correlators for the fields $\phi^N$ and $\sigma$, we can expand them around their background values, $\phi^N = \phi^N_* + \chi$ and $\sigma = \sigma_* + \delta \sigma$ and then the action up to the quadratic terms for the fluctuations becomes 
\begin{equation}
\begin{split}
S_2 &= \int d^d x \sqrt{g} \bigg( \frac{1}{2} (\partial \chi)^2 + \phi^N_* \chi \delta \sigma   \bigg) - \frac{N -1}{4} \int d^d x d^d y \sqrt{g_x} \sqrt{g_y} \delta \sigma(x) B(x,y) \delta \sigma(y) \\
B(x,y) &= \left[ \frac{1}{- \nabla^2 + \sigma_* - \frac{d (d - 2)}{4}} \right]^2 = \frac{1}{(\nabla^2)^2}
\end{split}
\end{equation}
where we already used the large $N$ values of $\sigma_*$ and $\phi^N_*$. To get the $\sigma$ correlator $G_{\sigma}$, we integrate out $\chi$ exactly, which gives the effective quadratic term for $\sigma$, and tells us that the correlator must satisfy
\begin{equation} \label{SigmaPropEqE}
\begin{split}
&\int d^d x \sqrt{g_x} \ \Pi(\xi_{x_1, x}) G_{\sigma}(\xi_{x x_2})  = \frac{\delta^{d}(x_1 - x_2)}{\sqrt{g_{x_1}}}\\
&\Pi(\xi_{x_1, x}) = - \frac{N -1}{2} (G^T (\xi))^2 + (\phi^N_*)^2 G^T (\xi)
\end{split}
\end{equation}
Then, following \cite{10.1143/PTP.72.736}, to invert $\Pi (\xi)$, it is convenient to first apply the differential operator corresponding to the massless equation of motion to it, which annihilates one of the terms and simplifies the other term in $\Pi(\xi)$
\begin{equation} \label{PiTilde}
\begin{split}
\frac{1}{\sqrt{g}} \partial_{\mu} (\sqrt{g} g^{\mu \nu} \partial_{\nu}) \Pi(\xi_{x_1, x}) &= \left( \xi (\xi + 1) \partial_{\xi}^2 + d  (\xi + \frac{1}{2} ) \partial_{\xi} \right)  \Pi(\xi_{x_1, x}) \\
&= -  \frac{(N - 1)\Gamma\left( \frac{d}{2} \right)^2}{ (4 \pi)^d \left(\xi ( \xi + 1) \right)^{d - 1}} = \tilde{\Pi} (\xi_{x_1, x}).
\end{split}
\end{equation}
When we apply this to the equation of motion of $\sigma$ in eq. \eqref{SigmaPropEqE}, we find
\begin{equation} \label{SigmaPropEqE2}
\int d^d x \ y^{- d} \ \tilde{\Pi}(\xi_{x_1, x}) G_{\sigma}(\xi_{x x_2})  = y_1^{d + 2} \left(\vec{\nabla}^2_{x_1} + \frac{\partial^2}{\partial y_1^2} + \frac{d + 2}{y_1} \frac{\partial}{\partial y_1} + \frac{d}{y_1^2} \right) \delta^{d}(x_1 - x_2)
\end{equation}
Now that we have a sufficiently simpler function to invert, we turn to the method reviewed in Appendix \ref{SigmaCorrelatorAppendix}, being careful about the modifications wherever necessary because of the differential operator present on the right hand side. Integrating the above equation over the boundary coordinates $\textbf{x}_1$, we get
\begin{equation}
\begin{split}
\int_0^{\infty} \frac{d y}{y} \ \tilde{\pi}(\rho_{y_1, y}) g_{\sigma}(\rho_{y y_2})  &= \frac{y_1^{\frac{d + 5}{2}} y_2^{\frac{1 - d}{2}}}{4^{d -1}} \left(\frac{\partial^2}{\partial y_1^2} + \frac{d + 2}{y_1} \frac{\partial}{\partial y_1} + \frac{d}{y_1^2} \right) \delta(y_1 - y_2)\\
&= \frac{y_1^{3} }{4^{d -1}} \left(\frac{\partial^2}{\partial y_1^2} + \frac{3}{y_1} \frac{\partial}{\partial y_1} + \frac{(-d^2 + 2 d + 3)}{4 y_1^2} \right) \delta(y_1 - y_2)
\end{split}
\end{equation}
Then making a change of variables to $y = e^{2 \theta}$ and doing a Fourier transform gives
\begin{equation}
\begin{split}
\int_{-\infty}^{\infty} d \theta \ \tilde{\pi}(\sinh^2 (\theta_1 - \theta)) g_{\sigma}(\sinh^2 (\theta - \theta_2))  &= \frac{1}{4^{d + 1}} \left( \frac{\partial^2}{\partial \theta_1^2} - (d - 1)^2 \right) \delta(\theta_1 - \theta_2) \\
\implies \hat{\tilde{\pi}}(k) \hat{g}_{\sigma} (k) &= \frac{- k^2 - (d-1)^2 }{4^{d + 1}} 
\end{split}
\end{equation}
Then, following the Appendix \ref{SigmaCorrelatorAppendix}, the transform $\hat{\tilde{\pi}}(k)$ corresponding to $\tilde{\Pi} (\xi)$ in eq. \eqref{PiTilde} can be worked out, and it gives 
\begin{equation}
\begin{split}
\hat{\tilde{\pi}}(k) &= -\frac{ (N - 1) \pi^{\frac{3 - d}{2}} \csc \left(\frac{\pi d}{2} \right) }{ 4^d \ \Gamma\left( \frac{d - 1}{2} \right)} \frac{\Gamma \left(\frac{ 3 d - 3 - i k}{4}\right) \Gamma \left(\frac{ 3 d - 3 + i k}{4}\right) }{\Gamma \left(\frac{d + 1 - i k}{4}\right) \Gamma \left(\frac{d + 1 + i k}{4}\right)}\\
\hat{g}_{\sigma} (k) &= \frac{ 4 \pi ^{\frac{d-3}{2}} \sin \left(\frac{\pi  d}{2}  \right) \Gamma \left(\frac{d-1}{2}\right)}{ (N-1)}  \left[ \frac{\Gamma \left(\frac{d + 5 - i k}{4}\right) \Gamma \left(\frac{d + 5 + i k}{4}\right)}{\Gamma \left(\frac{ 3 d - 3 - i k}{4}\right) \Gamma \left(\frac{ 3 d - 3 + i k}{4}\right)} - \frac{d}{4}  \frac{\Gamma \left(\frac{d + 1 - i k}{4}\right) \Gamma \left(\frac{d + 1 + i k}{4}\right)}{\Gamma \left(\frac{ 3 d - 3 - i k}{4}\right) \Gamma \left(\frac{ 3 d - 3 + i k}{4}\right)} \right]\,.
\end{split}
\end{equation}
This yields 
\begin{equation} \label{SigmaBulkBoundaryEx}
\begin{split}
G_{\sigma}(\xi) = \frac{2^{-d-1} \sin \left(\frac{\pi  d}{2}\right) \Gamma \left(\frac{d-1}{2}\right) \Gamma (d+2)}{\pi  (N-1) \Gamma \left(\frac{d}{2}-2\right) \Gamma \left(d+\frac{1}{2}\right)} \bigg[ \frac{1}{\xi^{d + 2}} \ {}_2 F_1 \left( d + 2, d ; 2d; -\frac{1}{\xi}\right)& \\
- \frac{4 (2 d - 1)}{(d + 1) (d - 4)} \frac{1}{\xi^d} \ {}_2 F_1 \left( d, d-1 ; 2d - 2; -\frac{1}{\xi}\right)  &\bigg].
\end{split} 
\end{equation}
This is the result for $\sigma$ correlator. We will not try to compute the correlator of $\chi$ here. This $\sigma$ correlator immediately shows that the leading operator induced by $\sigma$ on the boundary has dimensions $d$, which we identify to be proportional to the displacement operator. There are no relevant operators at the boundary, consistently with the fact that this phase has the lowest free energy in $2<d<4$. 

Let us also report the results for extraordinary transition in $d = 4 - \epsilon$ for comparison. Near $d = 4$, the dimension $d$ displacement operator can be seen in the boundary operator expansion of the fluctuation $\chi$. From our discussion below eq. \eqref{ActionExtra4d}, it is clear that the two-point functions of the fields $\phi^a$ and $\chi$ are just the bulk-to-bulk propagators with dimensions 3 and 4 respectively
\begin{equation}
\begin{split}
\langle \phi^a (x_1) \phi^b (x_2) \rangle &= \frac{\delta^{ab}}{16 \pi^2} \left( \frac{1}{\xi} + \frac{1}{\xi + 1} + 2  \log \frac{\xi}{ 1 + \xi} \right) \\
\langle \chi(x_1) \chi(x_2) \rangle &= \frac{1}{16 \pi^2} \left( \frac{1}{\xi} - \frac{1}{\xi + 1} + 12 + 6 (1 + 2 \xi) \log \frac{\xi}{ 1 + \xi} \right).
\end{split}
\end{equation}
To identify the contribution of the boundary operators, we can take one of the points to the boundary which corresponds to taking $\xi \rightarrow \infty$ 
\begin{equation} \label{BulkBoundaryChi}
\langle \chi(x_1) \chi(x_2) \rangle = \frac{1}{160 \pi^2} \frac{1}{\xi^4} + O\left(\frac{1}{\xi^5} \right) ; \ \ \ \langle \phi^a (x_1) \phi^b (x_2) \rangle = \frac{\delta^{ab}}{48 \pi^2} \frac{1}{\xi^3} + O\left(\frac{1}{\xi^4} \right).
\end{equation}
In this setting, the order $\epsilon$ corrections to these two-point functions require computing loop diagrams in AdS. We do not do that here. But in a flat space setting, these corrections were recently reported in \cite{Shpot:2019iwk, Dey:2020lwp}. 

\subsection{Using bulk equations of motion} \label{SectionBulkEOM}
To warm-up, we will first apply the equation of motion on the bulk-boundary two-point function, and then on the more complicated case of bulk two-point functions. 

Let us start with the case of the Wilson-Fisher fixed points in $d=4-\epsilon$, focusing on the $O(N)$ invariant phases corresponding to perturbing Neumann and Dirichlet boundary conditions. Consider the bulk-boundary two point function of the operators $\phi^I$ and $\hat\phi^I$, which is fixed by the conformal symmetry to take the following form in flat half-space
\begin{equation}
\langle \phi^I(x_1) \hat{\phi}^J (\textbf{x}_2) \rangle = \frac{B_{\phi \hat{\phi}} \delta^{IJ}}{(2 y_1)^{\Delta_{\phi} - \hat{\Delta}_{\phi}} (\textbf{x}_{12}^2 + y_1^2)^{\hat{\Delta}_{\phi}}}.
\end{equation}
Applying the Laplacian operator on the bulk point $x_1$ gives 
\begin{equation} \label{LaplacianBulkBoundary}
\begin{split}
\nabla^2 \langle \phi^I(x_1) \hat{\phi}^J (\textbf{x}_2) \rangle = \frac{B_{\phi \hat{\phi}} \delta^{IJ}}{(2 y_1)^{\Delta_{\phi} - \hat{\Delta}_{\phi}} (\textbf{x}_{12}^2 + y_1^2)^{\hat{\Delta}_{\phi} + 1}} \bigg(& \hat{\Delta}_{\phi} (3 - 2 d + \hat{\Delta}_{\phi}) + \Delta_{\phi} (1 + 2 \hat{\Delta}_{\phi} + \Delta_{\phi}) \\
&+ (\hat{\Delta}_{\phi} - \Delta_{\phi})(\hat{\Delta}_{\phi} - \Delta_{\phi} - 1) \frac{\textbf{x}_{12}^2} {y_1^2} \bigg).
\end{split}
\end{equation}
In the free theory, we have $\nabla^2 \phi^I(x_1)= 0$ and setting the right hand side above to $0$ gives two possibilities: 1) $\hat{\Delta}_{\phi} = \Delta_{\phi} = d/2 -1$ corresponding to Neumann boundary condition and, 2) $\hat{\Delta}_{\phi} = \Delta_{\phi} = d/2$ corresponding to Dirichlet boundary condition. In the interacting theory with $\lambda (\phi^I\phi^I)^2/4$ interaction, $\nabla^2 \phi^I(x_1)= \lambda \phi^I \phi^K \phi^K (x_1)$. Plugging this into the above equation gives 
\begin{equation} \label{EOMPhi4BulkBound}
\begin{split}
\frac{\lambda \langle \phi^I \phi^K \phi^K (x_1) \hat{\phi}^J (\textbf{x}_2) \rangle}{\langle \phi^I(x_1) \hat{\phi}^J (\textbf{x}_2) \rangle} &= 
\frac{\hat{\Delta}_{\phi} (3 - 2 d + \hat{\Delta}_{\phi}) + \Delta_{\phi} (1 + 2 \hat{\Delta}_{\phi} + \Delta_{\phi}) }{\textbf{x}_{12}^2 + y_1^2} \\
&+ \frac{(\hat{\Delta}_{\phi} - \Delta_{\phi})(\hat{\Delta}_{\phi} - \Delta_{\phi} - 1) \  \textbf{x}_{12}^2} {y_1^2 (\textbf{x}_{12}^2 + y_1^2)} 
\end{split}
\end{equation}
This equation is exact at the fixed point where the coupling becomes $\lambda_* = \frac{8 \pi^2}{N + 8} \epsilon$. So we can compute the correlators on the left hand side in the free theory in four dimension, and the factor of $\lambda$ in front will ensure that the BCFT data on the right is correct to order $\epsilon$. Recall the free theory correlators
\begin{equation}
\langle \phi^I \phi^K \phi^K (x_1) \hat{\phi}^J (\textbf{x}_2) \rangle^{N/D} = (N + 2) \frac{A_{\phi^2}^{N/D}}{(2 y_1)^{\Delta_{\phi^2}}} \langle \phi^I(x_1) \hat{\phi}^J (\textbf{x}_2) \rangle , \ \ \ \Delta_{\phi^2} = 2, \ \ \  A_{\phi^2}^{N/D} = \pm \frac{1}{4 \pi^2}. 
\end{equation}
Using the fact that at order $\epsilon$, the bulk dimension $\Delta_{\phi} = d/2 - 1$ does not get corrected (it is well-known that the anomalous dimension of $\phi$ at the Wilson-Fisher fixed point starts at order $\epsilon^2$), the above equation gives us 
\begin{equation} \label{PhiHatAnomPhi4}
\hat{\Delta}^N_{\phi}  = \frac{d}{2} - 1 + \hat{\gamma}_{\phi}^N, \ \ \hat{\Delta}^D_{\phi}  = \frac{d}{2} + \hat{\gamma}_{\phi}^D, \ \ \  \hat{\gamma}_{\phi}^N = \hat{\gamma}_{\phi}^D = - \frac{\epsilon}{2} \frac{N + 2}{N + 8}.
\end{equation}
This agrees with the results in \cite{McAvity:1995zd}. Notice that our approach using the equation of motion did not require us to do any loop calculations or regularization.  

We can also do a similar analysis in the non-linear sigma model in $d = 2 + \epsilon$. The equation of motion for $N - 1$ unconstrained fields in that case is $\nabla^2 \varphi^a(x_1)= - t^2 \varphi^a \partial_{\mu} \varphi^b \partial^{\mu} \varphi^b (x_1) + O(g^4)$. This then gives us
\begin{equation} \label{EOMNonLinearBulkBound}
\begin{split}
-\frac{t^2 \langle \varphi^a( \partial_{\mu} \varphi^b)^2 (x_1)\hat{\varphi}^c (\textbf{x}_2) \rangle}{\langle \varphi^a(x_1) \hat{\varphi}^c (\textbf{x}_2) \rangle} &= \frac{\hat{\Delta}_{\phi} (3 - 2 d + \hat{\Delta}_{\phi}) + \Delta_{\phi} (1 + 2 \hat{\Delta}_{\phi} + \Delta_{\phi})}{\textbf{x}_{12}^2 + y_1^2}  \\
&+ \frac{ (\hat{\Delta}_{\phi} - \Delta_{\phi})(\hat{\Delta}_{\phi} - \Delta_{\phi} - 1) \ \textbf{x}_{12}^2 }{y_1^2 (\textbf{x}_{12}^2 + y_1^2)}.
\end{split}
\end{equation}
The relevant correlation function can again be computed in the free theory to be 
\begin{equation}
\langle \varphi^a( \partial_{\mu} \varphi^b)^2 (x_1)\hat{\varphi}^c (\textbf{x}_2) \rangle = \pm \frac{ \langle \varphi^a(x_1) \hat{\varphi}^c (\textbf{x}_2) \rangle }{S_d (2 y_1)^d} \bigg(2 (N - 1) (d - 1) + 4 \Delta_{\phi} + 4 \hat{\Delta}_{\phi} \frac{y_1^2 - \textbf{x}_{12}^2 }{y_1^2 + \textbf{x}_{12}^2}   \bigg). 
\end{equation}
Using the bulk results to order $\epsilon$, $\Delta_{\phi} = \frac{d}{2} - 1 + \frac{\epsilon}{2 (N - 2)}$ and $t_*^2 = \frac{2 \pi \epsilon}{N - 2}$, we get 
\begin{equation}
\begin{aligned} 
\label{AnomalousDimensionNLSM}
&\hat{\Delta}^N_{\phi}  = \frac{d}{2} - 1 + \frac{N}{2 (N - 2)} \epsilon +O(\epsilon^2)
=\epsilon \frac{N-1}{N-2}+O(\epsilon^2), \\  
& \hat{\Delta}^D_{\phi}  = \frac{d}{2}  + \frac{\epsilon}{2}+O(\epsilon^2)=
1+\epsilon+O(\epsilon^2)
\end{aligned}
\end{equation} 
The Neumann case agrees with what was found in \cite{PhysRevLett.56.2834} and is consistent with the large $N$ result for the ordinary transition found in section \ref{SectionLargeNND}, i.e. $\hat\Delta=d-2+O(1/N)$. The Dirichlet case is consistent with the large $N$ result for extraordinary transition found in section \ref{SectionLargeNE}, i.e. $\hat\Delta=d-1$. 

It is possible to extend this idea of applying the equation of motion to the bulk two-point function. This method can be applied to either the AdS or flat half-space approach, but it is slightly more convenient to work in AdS, since then the two-point function is simply a function of the single cross-ratio $\xi$
\begin{equation}
\langle \phi^I (x_1) \phi^J (x_2) \rangle = \delta^{IJ} G_{\phi} (\xi).
\end{equation}   
Hence, the free equation of motion $(\nabla^2 - m^2) \phi^I = 0$ when applied to the two point function just gives 
\begin{equation}
\left( \nabla^2 + \frac{d (d - 2)}{4} \right) G_{\phi} (\xi) =  \left( \xi (\xi + 1) \partial_{\xi}^2 + d  (\xi + \frac{1}{2} ) \partial_{\xi} +  \frac{d (d - 2)}{4} \right)G_{\phi} (\xi) = D^{(2)} G_{\phi} (\xi) =  0 
\end{equation}
This differential equation has two solutions, which of course correspond to the Neumann or Dirichlet boundary conditions in the free theory \footnote{To be precise, the differential equation for the two-point function must have a delta function on the right, which is indeed reproduced by the solution we mention.}
\begin{equation} \label{EOMSolutionFree}
G_{\phi} (\xi) = b_1 \left( \frac{1}{(\xi)^{\frac{d}{2} - 1}} + \frac{1}{(\xi + 1)^{\frac{d}{2} - 1}}  \right) + b_2 \left( \frac{1}{(\xi)^{\frac{d}{2} - 1}} - \frac{1}{(\xi + 1)^{\frac{d}{2} - 1}}   \right).
\end{equation}
One of the constants above can be fixed by the normalization of the field $\phi$ and the other one can be fixed by demanding that the boundary spectrum contains a single operator of dimension $d/2 - 1$ or $d/2$ for Neumann or Dirichlet boundary conditions respectively. The canonical normalization corresponds to $b_1 = \Gamma \left(\frac{d}{2} \right)/ \left(2^{ d - 1} (d - 2) \pi^{\frac{d}{2}}\right), b_2  = 0$ for Neumann boundary condition and vice versa for Dirichlet boundary condition. However, in the rest of this subsection, we will find it convenient to work in a normalization in which $b_1 = 1$ for Neumann and $b_2 = 1$ for Dirichlet boundary conditions.

In the interacting theory in $d = 4 - \epsilon$ dimensions, the equation of motion gets modified to $(\nabla^2 - m^2) \phi^I (x) = \lambda \phi^I \phi^K \phi^K (x)$, which tells us that the two-point function must satisfy 
\begin{equation} \label{DiffEqSecondOrder}
D^{(2)} G_{\phi}^{N/D} (\xi) = \lambda_* (N + 2) A_{\phi^2}^{N/D} G_{\phi}^{N/D}(\xi)  + O(\lambda_*^2)
\end{equation}
and in this normalization, $\lambda_* = \epsilon/(2 (N + 8))$, and $A_{\phi^2}^{N/D} = \pm 1$. We can also apply the equation of motion operator on 
the other $\phi$ in the two-point function, which gives the following fourth order differential equation to $O(\lambda^2)$
\begin{equation} \label{TwoBox}
\begin{split}
&\bigg[\xi (1 + \xi) \left( \xi (1 + \xi) \partial_{\xi}^4 + (d + 2)(1 + 2 \xi) \partial_{\xi}^3 \right) + \frac{\left( d(d + 2) + (8 + 6 d (d + 2)) \xi (1 + \xi) \right)}{4} \partial_{\xi}^2  \\
&+ \frac{d^3 (1 + 2 \xi)}{4} \partial_{\xi} + \frac{d^2 (d - 2)^2}{16} \bigg] G_{\phi} (\xi) = D^{(4)} G_{\phi} (\xi) = \lambda_*^2  (N + 2) \left[ (A_{\phi^2}^{N/D})^2 (N + 2) G_{\phi} + 2 G_{\phi}^3 \right].
\end{split}
\end{equation}   
These differential equations can be used to extract some of the BCFT data. To see how to do that, let us recall that the two point function can be expanded into boundary conformal blocks as   
\begin{equation} \label{BoundaryChannelDecomposition}
G_{\phi} (\xi) = \sum_l \mu_l^2 f_{\mathrm{bdry}} (\hat{\Delta}_l; \xi).
\end{equation}
It is easy to see that applying the quadratic and quartic differential operator on the blocks returns the block itself with a coefficient
\begin{equation}
\begin{split}
D^{(2)} f_{\mathrm{bdry}} (\hat{\Delta}_l; \xi) &= \frac{(d - 2 \hat{\Delta}_l)(d - 2 - 2 \hat{\Delta}_l)}{4} f_{\mathrm{bdry}} (\hat{\Delta}_l; \xi), \\
D^{(4)} f_{\mathrm{bdry}} (\hat{\Delta}_l; \xi) &= \frac{(d - 2 \hat{\Delta}_l)^2(d - 2 - 2 \hat{\Delta}_l)^2}{16} f_{\mathrm{bdry}} (\hat{\Delta}_l; \xi).
\end{split}
\end{equation} 
Plugging this decomposition in to eq. \eqref{DiffEqSecondOrder} gives us again the anomalous dimension of the leading boundary operator $\hat{\gamma}_{\phi}^{N/D}$, eq. \eqref{PhiHatAnomPhi4}, which we already found above. At next order, plugging this decomposition in to eq. \eqref{TwoBox} gives us a relation for the boundary operator expansion coefficients for blocks other than the leading block. In Neumann case, it gives  
\begin{equation}
\sum_{l}  \frac{(d - 2 \hat{\Delta}_l)^2(d - 2 - 2 \hat{\Delta}_l)^2}{16} (\mu_l^N)^2 f_{\mathrm{bdry}} (\hat{\Delta}_l; \xi) = \frac{(N + 2) \epsilon^2}{2 (N + 8)^2} \frac{(1 + 2 \xi)^3}{\xi^3 (1 + \xi)^3}
\end{equation}
where the sum does not include the leading operator of dimension $d/2 -1 = 1+O(\epsilon)$. This equation tells us that the first subleading operator has dimension $3$ and appears with a boundary operator expansion coefficient $(\mu_3^N)^2 = \frac{N + 2}{(N + 8)^2} \epsilon^2$. In the Dirichlet case, we get the following equation instead
\begin{equation}
\sum_{l}  \frac{(d - 2 \hat{\Delta}_l)^2(d - 2 - 2 \hat{\Delta}_l)^2}{16} (\mu_l^D)^2 f_{\mathrm{bdry}} (\hat{\Delta}_l; \xi) = \frac{(N + 2) \epsilon^2}{2 (N + 8)^2} \frac{1}{\xi^3 (1 + \xi)^3}.
\end{equation}
Again, the sum does not run over the leading operator of dimension $d/2 = 2+O(\epsilon)$. This tells us that the first subleading operator has dimension $6$ in this case and appears with a coefficient $(\mu_6^D)^2 = \frac{N + 2}{ 800 (N + 8)^2} \epsilon^2$. We can go on and recursively find all other boundary operator expansion coefficients to order $\epsilon^2$.  These results agree with what was found in \cite{Bissi:2018mcq}.  

In fact, in this case, we can do better and fix the full two-point function by solving the differential equations explicitly. Let us start with the second order equation in eq. \eqref{DiffEqSecondOrder}.  We work perturbatively in $\epsilon$ and write the two point function and the differential operator as 
\begin{equation}
\begin{split}
G_{\phi}^{N/D} (\xi) &= G_{0}^{N/D} (\xi) + \epsilon G_{1}^{N/D} (\xi) + \epsilon^2 G_{2}^{N/D} (\xi) + O(\epsilon^3) \\
D^{(2)} &= D^{(2)}_0 + \epsilon D^{(2)}_1 + O(\epsilon^2).
\end{split}
\end{equation}
$G_{0}^{N/D} (\xi)$ is just the solution of the homogeneous differential equation and is given by the free theory two-point function found above in eq. \eqref{EOMSolutionFree}. At next order, $G_{1}^{N/D} (\xi)$ must satisfy the differential equation \eqref{DiffEqSecondOrder} to order $\epsilon$
\begin{equation}
D^{(2)}_0 G_1^{N/D} (\xi) = \pm \frac{(N + 2)}{2 (N + 8)} G_{0}^{N/D}(\xi) - D^{(2)}_1 G_0^{N/D}(\xi).  
\end{equation}
Plugging in the free solution, this can be solved to give
\begin{equation}
G_{1}^{N/D}(\xi) = \frac{c_1}{\xi} + \frac{c_2}{1 + \xi}  +  \frac{\log \xi}{2 \xi} \pm \frac{\log (1 + \xi)}{2 (1 + \xi)} + \frac{N + 2}{2 (N + 8)} \bigg( \frac{\log (1 + \xi)}{\xi} \pm \frac{\log \xi}{1 + \xi} \bigg). 
\end{equation}  
We work in the normalization such that in the bulk OPE limit, $\xi \rightarrow 0$, the leading term in the two-point function goes like $1/\xi^{\Delta_{\phi}}$ which fixes $c_1 = 0$. As we just saw, to order $\epsilon$, we still just have a single boundary block of dimension $d/2 - 1 + \hat{\gamma}_{\phi}^{N} $ or $d/2 + \hat{\gamma}_{\phi}^{D} $. Consistency with this requires setting $c_2 = 0$ for both Neumann and Dirichlet cases. This solution agrees with what was found using a one-loop calculation in \cite{McAvity:1993ue, McAvity:1995zd}. For the fourth order equation in eq. \eqref{TwoBox}, we can again expand $D^{(4)} = D^{(4)}_0 + \epsilon D^{(4)}_1 + \epsilon^2 D^{(4)}_2$ and get a differential equation for the second order correction to the two-point function 
\begin{equation}
D^{(4)}_0 G_2^{N/D} = \frac{(N + 2)}{4 (N + 8)^2} \left[ (N + 2) G_{0}^{N/D} (\xi) + 2 (G_{0}^{N/D})^3 \right] - D^{(4)}_1 G_1^{N/D} - D^{(4)}_2 G_0^{N/D}.
\end{equation} 
This can again be solved to give 
\begin{equation}
\begin{split}
G_2^{N/D}(\xi) &= \frac{d_1}{\xi} + \frac{d_2}{1 + \xi} +   d_3 \frac{  \log \xi}{1 + \xi} +  d_4 \frac{  \log (1 + \xi)}{\xi} \\
& -\frac{N + 2}{4 (N + 8)^2} \bigg( \frac{\log \xi}{\xi} \pm \frac{\log (1 + \xi)}{1 + \xi} \bigg) +  \frac{\log^2 (\xi)}{8  \xi}  \pm \frac{\log^2 (1 + \xi)}{8 (1 + \xi)} \\
& + \frac{(N + 2)^2}{8 (N + 8)^2} \bigg( \frac{\log^2(1 + \xi)}{\xi} \pm \frac{\log^2(\xi)}{1 + \xi}  \bigg) + \frac{N + 2}{4 (N + 8)} \bigg( \frac{1}{\xi} \pm \frac{1}{1 + \xi} \bigg)\log(\xi) \log (1 + \xi).
\end{split}
\end{equation}
Fixing the normalization yields $d_1 = 0$, and consistency with the expansion in the boundary channel spectrum that we just found above fixes $d_2 = 0$ and $d_4^{N/D} = \pm d_3^{N/D}$. The last constant left can be fixed by the bulk OPE behaviour. We know that as $\xi \rightarrow 0$, the correlator must go like 
\begin{equation}
\begin{split}
&G_{\phi}(\xi) = \xi^{-\Delta_{\phi}} + \lambda_{\phi^2} \xi^{\frac{1}{2} ( \Delta_{\phi^2} - 2 \Delta_{\phi})} \  + \  \textrm{higher orders in} \  \xi \\
&= \xi^{-\Delta_{\phi}} + \lambda_{\phi^2}^{(0)} + \epsilon \left[ \lambda_{\phi^2}^{(1)} + \left( \frac{\gamma^{(1)}_{\phi^2}}{2} - \gamma^{(1)}_{\phi} \right) \lambda_{\phi^2}^{(0)} \log \xi  \right] + \\
& \epsilon^2 \left[ \lambda_{\phi^2}^{(2)} + \left[ \lambda_{\phi^2}^{(0)} \left( \frac{\gamma^{(2)}_{\phi^2}}{2} - \gamma^{(2)}_{\phi} \right) + \lambda_{\phi^2}^{(1)} \left( \frac{\gamma^{(1)}_{\phi^2}}{2} - \gamma^{(1)}_{\phi} \right)    \right] \log \xi + \lambda_{\phi^2}^{(0)} \left( \frac{\gamma^{(1)}_{\phi^2}}{2} - \gamma^{(1)}_{\phi} \right)^2 \frac{\log^2 (\xi)}{2}    \right]
\end{split}
\end{equation}
where $\gamma^{(2)}_{\phi^2}$ is the second order correction to the anomalous dimension of bulk operator $\phi^2$ and similarly for all the other notation. Comparing to the correlator we found above at order $\epsilon$ tells us $\lambda_{\phi^2}^{(1)} = (N + 2)/(2 (N + 8))$. Then comparing the $\log \xi$ terms at order $\epsilon^2$ and small $\xi$ gives us the following relation 
\begin{equation}
\lambda_{\phi^2}^{(0)} \left( \frac{\gamma^{(2)}_{\phi^2}}{2} - \gamma^{(2)}_{\phi} \right) + \lambda_{\phi^2}^{(1)} \left( \frac{\gamma^{(1)}_{\phi^2}}{2} - \gamma^{(1)}_{\phi} \right)    = d_3 + \frac{N + 2}{4 (N + 8)}.
\end{equation}
Using the following bulk data (see for instance \cite{Kleinert:2001ax})
\begin{equation}
\gamma^{(2)}_{\phi^2} =\frac{(N + 2)(13N + 44)}{2 (N + 8)^3}, \ \ \gamma^{(2)}_{\phi} = \frac{N + 2}{4 (N + 8)^2}, \ \ \gamma^{(1)}_{\phi^2} =\frac{(N + 2)}{(N + 8)}, \ \ \gamma^{(1)}_{\phi} = 0, \ \ \lambda_{\phi^2}^{(0)} = \pm 1
\end{equation}
fixes the coefficient 
\begin{equation}
d_3^N = \frac{3 (N + 2) (N - 2)}{2 (N + 8)^3}, \ \ \ d_3^D = -\frac{3 (N + 2) (3 N + 14)}{2 (N + 8)^3}. 
\end{equation}
This determines the two-point function completely to order $\epsilon^2$ and agrees exactly with what was found in \cite{Bissi:2018mcq} using analytic bootstrap methods. 

To do a similar analysis in the large $N$ theory, we first decompose the field $\sigma$ into its constant one-point function and fluctuation, $\sigma = \sigma_* + \delta \sigma$. The fluctuation $\delta \sigma$ then only has connected correlators and the equation of motion is $(\nabla^2 - m^2 - \sigma_*) \phi^I (x) =  \delta \sigma \phi^I (x)$ which to leading order in large $N$ gives 
\begin{equation}
(D^{(2)} - \sigma_*) G_{\phi}^{N/D} (\xi) = 0.
\end{equation}
This is the equation of motion of a massive particle and its solution must be the usual bulk-to-bulk propagator, as we have already seen in the previous section. As we expect, expanding this into boundary conformal blocks just tells us that the only allowed blocks at leading order in large $N$ are the ones whose dimensions satisfy the AdS/CFT relation which we used several times above
\begin{equation} \label{AdSCFTAgain}
\frac{(d - 2 \hat{\Delta})(d - 2 - 2 \hat{\Delta})}{4} = \sigma_*, \ \implies \hat{\Delta} = \frac{d - 1}{2} \pm \sqrt{\sigma_* + \frac{1}{4} }.
\end{equation}
If we apply the equation of motion operator on the other $\phi$, we obtain the following quartic differential equation to order $1/N$
\begin{equation}
\left( D^{(4)} - 2 \sigma_* D^{(2)} + (\sigma_*)^2 \right) G_{\phi} (\xi) =  G_{\phi} (\xi) G_{\sigma} (\xi)
\end{equation}
where $G_{\sigma} (\xi)$ is the correlator of $\delta \sigma$ found in subsection \ref{SectionLargeNND} and it appears at order $1/N$ in large $N$ perturbation theory. We can then plug in the boundary channel conformal block decomposition for the correlator (eq. \eqref{BoundaryChannelDecomposition}) into this differential equation. It is easy to see that the leading boundary operator, the one that satisfies eq. \eqref{AdSCFTAgain}, only starts contributing at order $1/N^2$ on the left hand side. So at order $1/N$, we only have subleading operators contributing, which yield 
\begin{equation} \label{EOMSubleadingLargeN}
\sum_{l} \bigg(  \frac{(d - 2 \hat{\Delta}_l)(d - 2 - 2 \hat{\Delta}_l)}{4}  - \sigma_* \bigg)^2 (\mu_l)^2 f_{\mathrm{bdry}} (\hat{\Delta}_l; \xi) = G_{\phi} (\xi) G_{\sigma} (\xi)
\end{equation} 
where the sum runs over all operators other than the leading ones of dimensions $d - 2$  or $d - 3$ for ordinary or special transition respectively. For ordinary transition, plugging in the correlators on the right hand side and comparing powers of $1/\xi$ tells us that the operators appearing must have dimensions $2 d - 2 + 2 k$ with coefficients 
\begin{equation}
(\mu^O_k)^2 = \frac{2^{-d-4 k+2} \sin \left(\frac{\pi  d}{2}\right) \Gamma \left(\frac{d-1}{2}\right) \Gamma \left(\frac{3 (d-1)}{2}+k\right) \Gamma \left(\frac{d}{2}+k\right) \Gamma (d+2 k)}{ N \pi  d (d+2 k) (2 d+2 k-3) \Gamma \left(\frac{d}{2}-2\right) \Gamma \left(\frac{d}{2}\right) \Gamma (k+1) \Gamma \left(d+k-\frac{1}{2}\right) \Gamma \left(\frac{3 (d-1)}{2}+2 k\right)}.
\end{equation}
Near four dimensions, this agrees with the large $N$ limit of what was found using $\epsilon$ expansion in \cite{Bissi:2018mcq}.

Given these OPE coefficients, we can write down the bulk scalar two point function to order $1/N$ as
\begin{equation}
 G_{\phi} (\xi) = (\mu_{d - 2}^O)^2 f_{\textrm{bdry}} (d - 2 + \gamma, \xi) + \sum_{k = 0}^{\infty} \mu^O_k f_{\textrm{bdry}} (2 d - 2 + 2 k, \xi)
\end{equation} 
and in our convention $(\mu_{d - 2}^O)^2 = 1 + O(1/N)$. By crossing symmetry, this expansion must reproduce the bulk OPE expansion in the limit $\xi \rightarrow 0$. Now recall from subsection \ref{SectionLargeNND} that the boundary conformal block has the following expansion in the bulk OPE limit, $\xi \rightarrow 0$
\begin{equation}
\begin{split}
 f_{\textrm{bdry}} (\hat{\Delta}, \xi) = \frac{1}{\xi^{\frac{d}{2} - 1}} &\left( \frac{\Gamma \left(\frac{d}{2}-1\right) \Gamma (-d+2 \hat{\Delta} +2)}{\Gamma (\hat{\Delta} ) \Gamma \left(-\frac{d}{2}+\hat{\Delta} +1\right)} + O(\xi)  \right) \\
 + &\left( \frac{\Gamma \left(1-\frac{d}{2}\right) \Gamma (-d+2 \hat{\Delta} +2)}{\Gamma (-d+\hat{\Delta} +2) \Gamma \left(-\frac{d}{2}+\hat{\Delta} +1\right)} + O(\xi) \right). 
\end{split}
\end{equation} 
The constant term in the second line above corresponds to the presence of $\phi^2$ in the bulk OPE, which is supposed to be absent at the large $N$ fixed point. This term vanishes exactly when $\hat{\Delta} = d - 2$. At order $1/N$, we can allow for an anomalous dimension $\hat{\Delta} = d - 2 + \hat{\gamma}^O/N $, and its contribution to this constant term must be cancelled by the subleading operators of dimensions $2 d - 2 + 2 k$. Plugging in the dimensions, for consistency of bulk and boundary expansions, we must demand 
\begin{equation}
 \frac{\Gamma(1 - \frac{d}{2}) \Gamma(d - 2)}{\Gamma(\frac{d}{2} -1)} \ \hat{\gamma}^O  + \sum_{k = 0}^{\infty} (\mu_k^O)^2 \frac{ \Gamma(1 - \frac{d}{2}) \Gamma(3 d - 2 + 4 k)}{\Gamma(d + 2 k) \Gamma\left( \frac{3 d}{2} - 1 + 2 k \right)} = 0
\end{equation}
which finally gives the result for the $1/N$ correction to the scaling dimension of the leading boundary operator $\hat\phi$ as\footnote{This expression corrects a typo in previous versions of this paper (a missing factor of $\Gamma\left(2-\frac{d}{2}\right)\Gamma\left(d+\frac{1}{2}\right)$ in the last term of the equation). The presence of an issue in eq.~(\ref{eq457}) in the previous versions was recently pointed out in \cite{Ohno:2025qfv}.}
\begin{equation}
\begin{split}
\label{eq457}
\hat{\gamma}^O &
= - \frac{\Gamma(\frac{d}{2} -1)}{\Gamma(d - 2)} \sum_{k = 0}^{\infty} (\mu_k^O)^2 \frac{\Gamma(3 d - 2 + 4 k)}{\Gamma(d + 2 k) \Gamma\left( \frac{3 d}{2} - 1 + 2 k \right)} \\
&= \frac{2^{d-5} (d-4) \sin \left(\frac{\pi  d}{2}\right) \Gamma \left(\frac{3 d}{2}-\frac{1}{2}\right)}{\pi ^2 \Gamma\left(2-\frac{d}{2}\right)\Gamma\left(d+\frac{1}{2}\right)} \Bigg[\frac{\pi^{\frac{3}{2}}  2^d  \Gamma \left(d-\frac{3}{2}\right) \, _3{F}_2\left(1,1-\frac{d}{2},\frac{d}{2}+1;2-\frac{d}{2},d+\frac{1}{2};1\right)}{\Gamma (d) \Gamma (d-2) \sin (\pi d)} \\
&+ \frac{4 (d-2)}{\Gamma (d)} \left(
\frac{\pi^2\, _3{F}_2\left(1-\frac{d}{2},\frac{d-1}{2},\frac{d+2}{2};2-\frac{d}{2},d+\frac{1}{2};1\right)}{ \sin (\pi  d)} \, 
%+\frac{\sqrt{\pi } 2^d \Gamma (3-d) \Gamma \left(d+\frac{1}{2}\right)}{d^2 (2 d-3)}\right)
+\frac{4 \Gamma \left(\frac{3}{2}-\frac{d}{2}\right) \Gamma\left(2-\frac{d}{2}\right)\Gamma\left(d+\frac{1}{2}\right)}{d^2 (2 d-3)}\right)
\Bigg]
\end{split}
\end{equation}
where we performed the sum after plugging in the explicit OPE coefficients of the subleading operators we found above.\footnote{For reader interested in reproducing this result, note that to perform the summation, we had to separate out the $k = 0$ piece and then add it back at the end. Also, the result that Mathematica gives has to be analytically continued using the formula given in eq. 2.12 of \cite{Gopakumar:2018xqi} in order to obtain an expression that is well defined for positive $d$.} The hypergeometric functions appearing in this result are well defined for $d\ge 1$. While we have not been able to find relevant hypergeomtric identities to simplify this formula analytically, we have verified numerically that for all $d\ge 1$ the result precisely agrees with the simpler formula given in \cite{10.1143/PTP.70.1226}, which reads\footnote{Analytic agreement between the infinite series representation in (\ref{eq457}) and the expression in (\ref{gamO-simpler}) was recently shown in \cite{Ohno:2025qfv}.}
\begin{equation}
\hat{\gamma}^O = \frac{(4-d) \Gamma (2 d-3)}{d \Gamma(d-2) \Gamma(d-1)}\,.
\label{gamO-simpler}
\end{equation}
In $d=3$, this gives $\hat\Delta=1+2/(3N)+O(1/N^2)$. We can also verify that in $d=4-\eps$ and $d=2+\eps$ the large $N$ prediction agrees with (\ref{PhiHatAnomPhi4}) and (\ref{AnomalousDimensionNLSM}) for Dirichlet and Neumann conditions respectively. Indeed using (\ref{gamO-simpler}) we find
\begin{equation}
\begin{aligned}
&\hat\Delta = 2-\epsilon+\frac{3\epsilon}{N}+\ldots\,,\qquad d= 4-\epsilon \\
&\hat\Delta = \epsilon+\frac{\epsilon}{N}+\ldots\,,\qquad d=2+\epsilon\,.
\end{aligned}
\end{equation}

For the case of the special transition, eq. \eqref{EOMSubleadingLargeN} tells us that there are two towers of operators that appear at the subleading order: the ones with dimension $d - 1 + 2 p$ and the ones with dimension $2 d - 3 + 2 q$. The coefficients for these can be found recursively using eq. \eqref{EOMSubleadingLargeN}. Carrying out the calculation explicitly is more involved because of the two towers involved, so we leave it for future work. But we expect a similar reasoning as explained above for the ordinary transition to also give anomalous dimension of leading boundary operator at order $1/N$ for the special transition case. The result for this case was also reported previously in \cite{10.1143/PTP.70.1226} using different methods. The explicit result reads
\begin{equation}
\hat\gamma^S = \frac{2 (4-d)}{\Gamma (d-3)} \left(\frac{(6-d) \Gamma (2 d-6)}{d \Gamma (d-3)}+\frac{1}{\Gamma (5-d)}\right)
\label{gamS-simpler}
\end{equation}  
Note that in $d=3$, the anomalous dimension vanishes, consistently with the expectation that this should be lower critical dimension for the special transition (presumably $\hat\Delta=0$ to all orders in the $1/N$ expansion in $d=3$). In $d=4-\epsilon$, (\ref{gamS-simpler}) gives
\begin{equation}
\hat\Delta = 1-\epsilon+\frac{3\epsilon}{N}+\ldots 
\end{equation}
in agreement with the $\epsilon$-expansion result in (\ref{PhiHatAnomPhi4}).

\subsection{Using weakly broken higher spin symmetry} \label{SectionHigherSpin}
We can generalize the equation of motion idea, in a manner similar to \cite{Skvortsov:2015pea, Giombi:2016hkj}, to find the anomalous dimensions of the higher spin displacement operators, which are the operators that appear in the boundary operator expansion of the higher spin currents. The bulk higher spin currents are conserved in the free theory, but are weakly broken in the interacting theory, and hence the corresponding ``higher-spin displacement" operators acquire anomalous dimensions (except of course the spin-2 case, which corresponds to the bulk stress-tensor and displacement operator at the boundary). As usual, it is convenient to package the currents in the index-free notation (see for instance, \cite{Giombi:2016ejx} for a review) 
\begin{equation}
\mathcal{J}_{s} (x,z) = J_{\mu_1 ... \mu_s} (x) z^{\mu_{1}}... z^{\mu_{s}}, \ \ \ z^2 = 0.
\end{equation}
We can free the indices by acting with the Todorov differential operator 
\begin{equation}
D_{z}^{\mu} = \bigg( \frac{d}{2} - 1 + z^{\nu} \frac{\partial}{\partial z^{\nu}}  \bigg) \frac{\partial}{\partial z_{\mu}} - \frac{1}{2} z^{\mu} \frac{\partial}{\partial z^{\nu}} \frac{\partial}{\partial z_{\nu}}.
\end{equation}
Similar tensors can be constructed for the boundary operators and we use the notation $\hat{\mathcal{J}}_l^s$ for the spin $l$ operator appearing in the BOE of a spin $s$ operator. The two-point function of a spin $s$ operator in the bulk and a spin $l$ operator on the boundary is fixed by the conformal symmetry. We will focus on the correlator $\langle \mathcal{J}_{s} (x_1,z_1) \hat{\mathcal{J}}_l^s (\textbf{x}_2, \textbf{z}_2)  \rangle$. The tensor structures appearing in this correlator were found in \cite{Billo:2016cpy} for the case of general defect, and only two of those structures survive in the case of co-dimension one 
\begin{equation}
Q^0_{b \partial} = \mathbf{z}_1 \cdot \mathbf{z}_2 - 2 
\frac{x_{12} \cdot z_1 (\mathbf{x}_{12} \cdot \mathbf{z}_2)}{ x_{1 2}^2 }, \ \ \ Q^2_{b \partial} = z_1^y  - \frac{2 y_1 x_{12} \cdot z_1}{x_{12}^2}. 
\end{equation}
In terms of these structures, the bulk-boundary two point function is 
\begin{equation} \label{SpinSCorrelator}
\langle \mathcal{J}_{s} (x_1,z_1) \hat{\mathcal{J}}_l^s (\textbf{x}_2, \textbf{z}_2)  \rangle = b_{s,l} \frac{(Q^0_{b \partial})^l (Q^2_{b \partial})^{s - l}}{(x_{12}^2)^{\hat{\Delta}_l^s} (y_1)^{\Delta_s - \hat{\Delta}_l^s }}.
\end{equation}

Now consider a CFT with an interaction strength $g$\footnote{Not to be confused with the determinant of the metric. We hope there is no cause for confusion, because this subsection is entirely in flat space and the metric does not appear.}  and suppose we are in the regime where $g$ is small. The dimension of the spin $s$ current then is 
\begin{equation}
\Delta_{s} = d - 2 + s + \gamma_{s}(g).
\end{equation}
The current conservation is weakly broken and its divergence defines a spin $s - 1$ descendant 
\begin{equation} \label{DefinitionDescendant}
\partial_{\mu} D^{\mu}_z \mathcal{J}_{s} (x,z) = g \  \mathcal{K}_{s -1} (x,z).
\end{equation}
Applying this to the correlator gives 
\begin{equation}
\partial_{1 \mu} D^{\mu}_{z_1} \langle \mathcal{J}_{s} (x_1,z_1) \hat{\mathcal{J}}_l^s (\textbf{x}_2, \textbf{z}_2)  \rangle = g \ \langle \mathcal{K}_{s -1} (x_1,z_1) \hat{\mathcal{J}}_l^s (\textbf{x}_2, \textbf{z}_2)  \rangle.
\end{equation}
We can also apply this differential operator on the right hand side of eq. \eqref{SpinSCorrelator}. As a first step, we need
\begin{equation}
\begin{split}
&D^{\mu}_{z_1} ((Q^0_{b \partial})^l (Q^2_{b \partial})^{s - l}) = l \left( \frac{d}{2} + s - 2 \right) (Q^0_{b \partial})^{l - 1} (Q^2_{b \partial})^{s - l} \left( z^i_2 \delta^{\mu}_{i} -  \frac{2 x_{12}^{\mu} (\mathbf{x}_{12} \cdot \mathbf{z}_2)}{ x_{1 2}^2 } \right) \\
& + (s - l) (Q^0_{b \partial})^l (Q^2_{b \partial})^{s - l  - 2}  \bigg[\left( \frac{d}{2} + s - 2 \right) (Q^2_{b \partial}) \left( \delta^{\mu}_{y} - \frac{2 y_1 x_{12}^{\mu}}{x_{12}^2}  \right)  - \frac{(s - l - 1)}{2} z_1^{\mu}  \bigg]
\end{split}
\end{equation}
Using this and after a bit of algebra, we get the following result for the ratio of descendant correlator to the current correlator
\begin{equation} 
\begin{split}
&\frac{ g \ \langle \mathcal{K}_{s -1} (x_1,z_1) \hat{\mathcal{J}}_l^s (\textbf{x}_2, \textbf{z}_2)  \rangle}{\langle \mathcal{J}_{s} (x_1,z_1) \hat{\mathcal{J}}_l^s (\textbf{x}_2, \textbf{z}_2)  \rangle} = \frac{ (  \hat{\Delta}^s_l -\Delta_s ) (s - l) (d + s + l - 3)}{2 y_1 Q^2_{b \partial}} \\
 &+ \frac{\gamma_s(g)}{x_{12}^2} \left[ \frac{(s - l) (s - l -1) x_{12}\cdot z_1}{(Q^2_{b \partial})^2} + (d + 2 s - 4) \left( \frac{y_1 (s - l)}{ Q^2_{b \partial}} + \frac{l ( \mathbf{x}_{12} \cdot \mathbf{z}_2) }{Q^0_{b \partial}} \right) \right].
\end{split}
\end{equation} 
The left hand side of the above equation vanishes if there are no interactions in the bulk ($g = 0$). The bulk anomalous dimension vanishes in that case, $\gamma_s(g) = 0$. The equation above then tells us that in the case of no bulk interactions the dimension of the higher spin displacement operator is fixed to be equal to the dimension of current, $\hat{\Delta}^s_l = \Delta_s = d -2 + s$. It was shown to be true for stress-tensor in \cite{McAvity:1995zd}, but here we get it for higher spin currents as well. This proves the observation made in \cite{Giombi:2019enr} about the higher spin displacements being protected in the presence of interactions localized on the boundary. In the presence of bulk interactions, we parametrize $ \hat{\Delta}^s_l = d -2 + s + \hat{\gamma}^s_l(g)$ and the anomalous dimensions can be obtained from
\begin{equation} 
\begin{split}
&\frac{ g \ \langle \mathcal{K}_{s -1} (x_1,z_1) \hat{\mathcal{J}}_l^s (\textbf{x}_2, \textbf{z}_2)  \rangle}{\langle \mathcal{J}_{s} (x_1,z_1) \hat{\mathcal{J}}_l^s (\textbf{x}_2, \textbf{z}_2)  \rangle} = \frac{ (  \hat{\gamma}^s_l(g) -\gamma_s(g) ) (s - l) (d + s + l - 3)}{2 y_1 Q^2_{b \partial}} \\
&+ \frac{\gamma_s(g)}{x_{12}^2} \left[ \frac{(s - l) (s - l -1) x_{12}\cdot z_1}{(Q^2_{b \partial})^2} + (d + 2 s - 4) \left( \frac{y_1 (s - l)}{ Q^2_{b \partial}} + \frac{l ( \mathbf{x}_{12} \cdot \mathbf{z}_2) }{Q^0_{b \partial}} \right) \right].
\end{split}
\end{equation} 
We can compute the correlators on the left hand side in the free theory, and this will give us the anomalous dimensions to leading order in $g$. We will demonstrate it explicitly in Wilson-Fisher model in $4 - \epsilon$ dimensions. We know that in this model, the anomalous dimension of the bulk current, $\gamma_s(g)$ starts at order $g^2$ (see e.g. \cite{Giombi:2016hkj} and references therein), so we can drop the second line at leading order. This gives us the anomalous dimensions of the boundary operators 
\begin{equation}
\hat{\gamma}^s_l =  \frac{ 2 g }{ (s - l) (d + s + l -3) b_{s,l}} \frac{\langle \mathcal{K}_{s -1} (x_1,z_1) \hat{\mathcal{J}}_l^s (\textbf{x}_2, \textbf{z}_2)  \rangle (x_{12}^2)^{\hat{\Delta}_l^s}  y_1}{(Q^0_{b \partial})^l (Q^2_{b \partial})^{s - l - 1}}.   
\end{equation} 
Let us look at some explicit examples now. Consider the spin $2$ current in the $O(N)$ model near $4$ dimensions, which has a boundary scalar in its BOE. We will find that there is a non-zero anomalous dimension only for the symmetric traceless sector of $O(N)$, which is what we expect because the trace piece is just the stress-tensor. We have the following operators for Dirichlet\footnote{Note that we are using the notation where $\hat{\phi}$ is the leading boundary operator, so in the Dirichlet case, $\hat{\phi} (\mathbf{x}) = \partial_{y}\phi (\mathbf{x}, 0)$} and Neumann case
\begin{equation}
\begin{split}
\mathcal{J}^{I J}_{2} (x_1,z_1) &= \frac{\sqrt{\pi} \Gamma(d - 1) z_1^{\mu_1} z_1^{\mu_2}}{2^{d - 3} \Gamma(\frac{d - 3}{2}) \Gamma(\frac{d}{2} - 1)} \bigg[ \phi^I \partial_{\mu_1} \partial_{\mu_2} \phi^J + I \leftrightarrow J - \frac{2 d }{d - 2} \partial_{\mu_1} \phi^I \partial_{\mu_2} \phi^J  \bigg] \\
\mathcal{K}^{IJ}_{s -1} (x_1,z_1) & = z_1^{\mu_1} \left( - 6 \phi^a \phi^a \partial_{\mu_1} (\phi^I \phi^J) + 6 \phi^I \phi^J \partial_{\mu_1} (\phi^a \phi^a)  \right) \\
(\hat{\mathcal{J}}_0^{2 K L} (\textbf{x}_2))_D &= \hat{\phi}^K \hat{\phi}^L, \ \ (\hat{\mathcal{J}}_0^{2 K L} (\textbf{x}_2))_N = \frac{d - 2}{2} \left( \hat{\phi}^K \mathbf{\partial_i}^2 \hat{\phi}^L + K \leftrightarrow L \right) - \partial_i \hat{\phi}^K \partial^i \hat{\phi}^L.  
\end{split}
\end{equation}
The form of the current is fixed by conservation and condition of being symmetric traceless. The descendant $\mathcal{K}$ can then be figured out using its definition in eq. \eqref{DefinitionDescendant}. The boundary operator $\hat{\mathcal{J}}$ is the boundary limit of the bulk current with both of its Lorentz indices equal to $y$ since we are considering a boundary scalar. Computing the correlator then is just a matter of free-field Wick contractions and it gives
\begin{equation} \label{AnomalousHigherSpinDisplacement}
(\hat{\gamma}^{s=2}_{l=0})^{T, D} = - \frac{N \epsilon}{N + 8}, \hspace{2cm} (\hat{\gamma}^{s=2}_{l=0})^{T, N} = - \frac{N \epsilon}{N + 8}
\end{equation}
where $T$ stands for symmetric traceless. Note that this operator is a composite primary operator on the boundary and also appears in the OPE of $\hat{\phi} \hat{\phi}$ on the boundary. We will check below that the result found here agrees with what we get from a boundary four-point computation in \eqref{AnomDimDisplacementSymTrac}. In principle, this method can be used to calculate anomalous dimensions of all the higher spin displacements, although the algebra gets tedious very soon.   

\subsection{Boundary four-point functions} \label{FourPoint}
In this subsection, we compute the four-point 
function of the leading boundary operator $\hat{\phi}^I$ induced by the bulk scalar $\phi^I$. This will help us compute the 
anomalous dimensions of the boundary composite operators appearing in 
the OPE $\hat{\phi}^I \hat{\phi}^J$. This four-point function can be decomposed into singlet, symmetric traceless and anti-symmetric representations of $O(N)$
\begin{equation} \label{FourPointDecomSTA}
\begin{split}
\langle \hat{\phi}^I(\textbf{x}_1) \hat{\phi}^J(\textbf{x}_2) \hat{\phi}^K(\textbf{x}_3) \hat{\phi}^L(\textbf{x}_4)\rangle &= \delta^{I J } \delta^{K L} \mathcal{G}_S +   \bigg( \frac{\delta^{I K } \delta^{J L} + \delta^{I L } \delta^{J K} }{2} - \frac{\delta^{I J } \delta^{K L}}{N}  \bigg) \mathcal{G}_T  \\
&  +   \frac{\delta^{I K } \delta^{J L} - \delta^{I L } \delta^{J K} }{2}  \mathcal{G}_A. 
\end{split}
\end{equation}
Each of these structures can be decomposed in terms of the conformal blocks 
\begin{equation}
\mathcal{G} = \frac{\hat{C}_{\phi \phi}^2}{(\textbf{x}_{12}^2 \textbf{x}_{34}^2)^{\hat{\Delta}_{\phi}}} \mathcal{F} (u,v), \ \ \ \mathcal{F} (u,v) = \sum_{\hat{\Delta} , l} a_{\hat{\Delta},l} G_{\hat{\Delta},l} (u,v)
\end{equation} 
where 
\begin{equation}
u = \frac{\textbf{x}_{12}^2 \textbf{x}_{34}^2}{\textbf{x}_{13}^2 \textbf{x}_{24}^2}, \ \ \ v = \frac{\textbf{x}_{14}^2 \textbf{x}_{23}^2}{\textbf{x}_{13}^2 \textbf{x}_{24}^2}.
\end{equation}
In the free theory, the four-point function takes the simple form
\begin{equation}
\langle \hat{\phi}^I(\textbf{x}_1) \hat{\phi}^J(\textbf{x}_2) \hat{\phi}^K(\textbf{x}_3) \hat{\phi}^L(\textbf{x}_4)\rangle_0 =  \hat{C}_{\phi \phi}^2 \bigg(\frac{\delta^{I J} \delta^{K L}}{(\textbf{x}_{12}^2)^{\hat{\Delta}_{\phi}} (\textbf{x}_{34}^2)^{\hat{\Delta}_{\phi}} }  + \frac{\delta^{I K} \delta^{J L}}{(\textbf{x}_{13}^2)^{\hat{\Delta}_{\phi}} (\textbf{x}_{24}^2)^{\hat{\Delta}_{\phi}} } + \frac{\delta^{I L} \delta^{J K}}{(\textbf{x}_{14}^2)^{\hat{\Delta}_{\phi}} (\textbf{x}_{23}^2)^{\hat{\Delta}_{\phi}} }   \bigg).
\end{equation}
We will be looking at its decomposition in the s-channel $12 \rightarrow 34$ where the first term is the identity operator and the other two come from the composite operators schematically given by $\hat{\phi}^I (\mathbf{\partial})^{2n} \mathbf{\partial}^l \hat{\phi}^J$ with dimension $ 2 \hat{\Delta}_{\phi} + 2 n + l$ and spin $l$ \cite{Fitzpatrick:2011dm}
\begin{equation}
\frac{1}{(\textbf{x}_{13}^2)^{\hat{\Delta}_{\phi}} (\textbf{x}_{24}^2)^{\hat{\Delta}_{\phi}} } = \frac{(-1)^l}{(\textbf{x}_{14}^2)^{\hat{\Delta}_{\phi}} (\textbf{x}_{23}^2)^{\hat{\Delta}_{\phi}}} = \frac{1}{(\textbf{x}_{12}^2)^{\hat{\Delta}_{\phi}} (\textbf{x}_{34}^2)^{\hat{\Delta}_{\phi}})} \sum_{n,l} a^0_{ n,l} G_{n,l } (u,v)
\end{equation} 
where 
\begin{equation}
a^0_{n,l} = a^0_{2 \hat{\Delta}_{\phi} + 2 n + l,l} = \frac{ (-1)^l[(\hat{\Delta}_{\phi} - \frac{d}{2} + \frac{3}{2})_n (\hat{\Delta}_{\phi})_{l + n}]^2}{l! n! (l + \frac{d -1}{2})_n (2 \hat{\Delta}_{\phi} + n - d + 2)_n (2 \hat{\Delta}_{\phi} + 2 n + l -1)_l (2 \hat{\Delta}_{\phi} + n + l + \frac{1 - d}{2})_n}
\end{equation}
and $G_{n,l} (u,v) = G_{2 \hat{\Delta}_{\phi} + 2 n + l,l} (u,v)$ is the four-point conformal block. In the interacting theory, we expect these dimensions and OPE coefficients to receive corrections. To first order in the expansion parameter, we have 
\begin{equation} \label{4ptChangeGeneral}
\delta \mathcal{F}(u,v) = \sum_{n,l} \left( a^1_{n,l} + \frac{1}{2} a^0_{n,l} \hat{\gamma}_{1} (n,l) \partial_n \right) G_{n,l} 
\end{equation}
We will calculate these corrections in an $\epsilon$ expansion in $d = 4 - \epsilon$ in the Wilson-Fisher model, for both Neumann and Dirichlet boundary conditions with $\hat{\Delta}_{\phi} = d/2 - 1$ and $d/2$ respectively. We will also do an $\epsilon$ expansion calculation in non-linear sigma model in $d = 2 + \epsilon$, for Dirichlet boundary conditions (we leave the case of Neumann boundary conditions to future work). 

\subsubsection{$\epsilon$ expansion in $d = 4 - \epsilon$}
The first order correction to this four-point function in $\phi^4$ theory can be computed using the following contact Witten diagram in AdS
\begin{equation}
\langle \hat{\phi}^I(\textbf{x}_1) \hat{\phi}^J(\textbf{x}_2) \hat{\phi}^K(\textbf{x}_3) \hat{\phi}^L(\textbf{x}_4)\rangle_1   \ = \ \  \vcenter{\hbox{\includegraphics[scale=0.4]{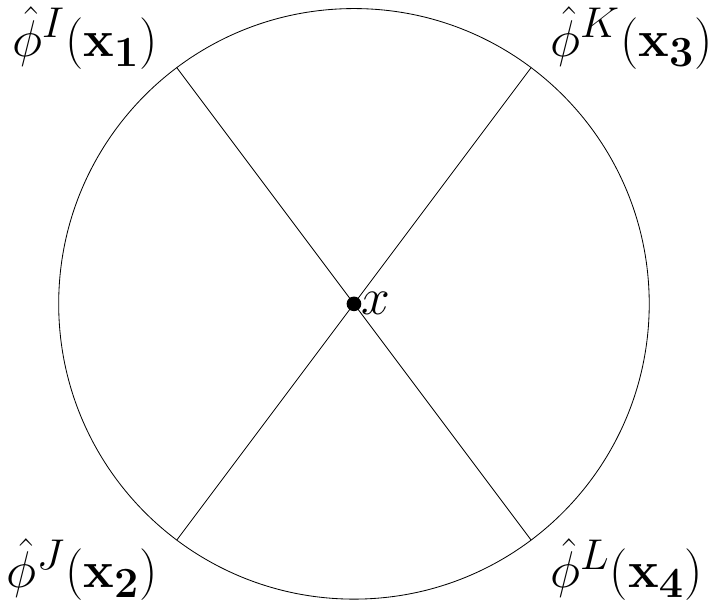}}}
\end{equation}
This contact Witten diagram evaluates to the $D$ function that appears frequently in AdS/CFT literature
\begin{equation}
\begin{split}
\langle \hat{\phi}^I(\textbf{x}_1) \hat{\phi}^J(\textbf{x}_2) \hat{\phi}^K(\textbf{x}_3) \hat{\phi}^L(\textbf{x}_4)\rangle_1  &= -2 \ \hat{C}_{\phi \phi}^4 \ \lambda (\delta^{IJ} \delta^{KL} + \delta^{IK} \delta^{JL} + \delta^{IL} \delta^{JK} ) \int d^d x \sqrt{g} \prod_{i = 1}^4 K_{\hat{\Delta}_{\phi}} (\textbf{x}_i, x)  \\
&  = -2 \ \hat{C}_{\phi \phi}^4 \ \lambda (\delta^{IJ} \delta^{KL} + \delta^{IK} \delta^{JL} + \delta^{IL} \delta^{JK} ) D_{\hat{\Delta}_{\phi} \hat{\Delta}_{\phi} \hat{\Delta}_{\phi} \hat{\Delta}_{\phi}} (\textbf{x}_i).
\end{split}
\end{equation}
Here we are using a normalization where bulk-boundary propagator takes the form

\begin{equation}
K_{\hat{\Delta}_{\phi}} (\textbf{x}_i, x) = \left( \frac{y}{y^2 + (\textbf{x}_i - \textbf{x})^2} \right)^{\hat{\Delta}_{\phi}}.
\end{equation}
The conformal block decomposition of this $D$ function was worked out in \cite{Hijano:2015zsa, Jepsen:2019svc} \footnote{We thank Christian Jepsen for pointing out a normalization typo in eq. 4.8 of \cite{Hijano:2015zsa} which introduces an additional factor of $\mathcal{N}_{\hat{\Delta}_n}$ in our eq. \eqref{CBDD} compared to eq. 4.24 of \cite{Hijano:2015zsa}.} 
\begin{equation} \label{CBDD}
\begin{split}
&D_{\hat{\Delta}_{\phi} \hat{\Delta}_{\phi} \hat{\Delta}_{\phi} \hat{\Delta}_{\phi}} (\textbf{x}_i) = \\
& \frac{1}{(\textbf{x}_{12}^2)^{\hat{\Delta}_{\phi}} (\textbf{x}_{34}^2)^{\hat{\Delta}_{\phi}}} \bigg[ \sum \limits_{\substack{m,n=0 \\ m\neq n}}^{\infty} \frac{2 (-1)^{m + n} (\hat{\Delta}_{\phi})_m^2 (\hat{\Delta}_{\phi})_n^2 \ \ G_{n,0} (u,v)}{\mathcal{N}_{\hat{\Delta}_n} m! n! \left(m_n^2 - m_m^2 \right) \left(2 \hat{\Delta}_{\phi} + m + \frac{1 - d}{2} \right)_m  \left(2 \hat{\Delta}_{\phi} + n + \frac{1 - d}{2} \right)_n} \\
&+ \sum_{n = 0}^{\infty} \frac{4 \ (\hat{\Delta}_{\phi})_n^4 \ \Gamma(\hat{\Delta}_n)^2 }{(n!)^2  \left(2 \hat{\Delta}_{\phi} + n + \frac{1 - d}{2} \right)_n^2 \Gamma \left(\frac{\hat{\Delta}_n}{2} \right)^4    \ \partial_n m_n^2} \partial_n \bigg( \frac{\Gamma\left(\frac{\hat{\Delta}_n}{2} \right)^4}{4 \ \Gamma(\hat{\Delta}_n)^2 \mathcal{N}_{\hat{\Delta}_n} } G_{n,0} (u,v) \bigg)  \bigg] 
\end{split}
\end{equation}
where $m_n^2 = \hat{\Delta}_n (\hat{\Delta}_n - d + 1)$ with $\hat{\Delta}_n = 2 \hat{\Delta}_{\phi} + 2 n $ and 
\begin{equation}
\mathcal{N}_{\Delta} = - \frac{\Gamma (\Delta)}{\pi^{\frac{d-1}{2}} (2 \Delta - d + 1) \Gamma(\Delta + \frac{1 - d}{2})}.
\end{equation}
Comparing this with eq. \eqref{4ptChangeGeneral} tells us that only the spin $0$ operators get anomalous dimension to this order, i.e. $\gamma_1(n, l > 0) = 0$. Moreover, the spin $0$ anomalous dimensions in the singlet sector are given by
\begin{equation}
\frac{N}{2} a^0_{S \ n,0} \hat{\gamma}^S_{1} (n,0) = - \frac{ 2  \hat{C}_{\phi \phi}^2 \ \lambda (N + 2) (\hat{\Delta}_{\phi})_n^4 }{(n!)^2 (2 \hat{\Delta}_{\phi} + n + \frac{1 - d}{2})_n^2 (\partial_n m_n^2) \mathcal{N}_{\hat{\Delta}_n} }
\end{equation}
where $a^0_{S \ n,0} = a^0_{n,l}(1 + (-1)^l)/N$. For Dirichlet boundary condition, we plug in $\hat{C}_{\phi \phi} = \Gamma\left(\frac{d}{2}\right)/\pi^{\frac{d}{2}}$ and $\hat{\Delta}_{\phi} = d/2$ to get 
\begin{equation}
\hat{\gamma}^{S, D}_{1} (n,0) = \frac{\epsilon (N + 2)}{(N + 8)}
\end{equation}
where we used the fixed point value of the coupling $ \lambda =  \lambda^*$. This gives the dimension of the boundary operators up to 1-loop order 
\begin{equation}
\hat{\Delta}^{S, D}_{n,0} = 2 \hat{\Delta}_{\phi} + 2 n + \hat{\gamma}^{S, D}_{1} (n,0) = d + 2 n + 2 \hat{\gamma}^{D}_{\phi} + \hat{\gamma}^{S, D}_{1} (n,0) = d + 2 n +O(\epsilon^2)
\end{equation}
where we used the result for $ \hat{\gamma}^{D}_{\phi}$ from eq. \eqref{PhiHatAnomPhi4}. The $n = 0$ operator is proportional to the displacement operator, which is supposed to be protected to all orders. Similarly, for the Neumann boundary condition, using $\hat{C}_{\phi \phi} = \Gamma \left(\frac{d}{2} - 1 \right)/ (2 \pi^{\frac{d}{2}})$ and $\hat{\Delta}_{\phi} = d/2 - 1$  along with the result for $ \hat{\gamma}^{N}_{\phi}$ from eq. \eqref{PhiHatAnomPhi4} gives 
\begin{equation}
\begin{split}
&\hat{\gamma}^{S, N}_{1} (n,0) = \frac{ \epsilon (N + 2)}{(N + 8)} \implies \hat{\Delta}^{S, N}_{n,0} = d - 2 + 2 n, \ \ \  \forall \ n > 0 \\
&\hat{\gamma}^{S, N}_{1} (0,0) = \frac{ 2 \epsilon (N + 2)}{(N + 8)} \implies \hat{\Delta}^{S, N}_{0,0} = 2 - \frac{6 \epsilon}{N + 8}.
\label{delSN-Neumann}
\end{split}
\end{equation}
Note that the $n = 0$ case has to be treated separately here because first setting $\hat{\Delta} = d/2 - 1$ and then taking $n \rightarrow 0$ gives the wrong result for the OPE coefficient $a^0_{0,0}$. Instead one should directly take $n \rightarrow 0$ which gives $a^0_{0,0} = 1$ for all values of $\hat{\Delta}$. Recall that from the boundary operator expansion  of $\sigma$ at large $N$, we saw leading operators of dimension $ 2 $ and $ d $ to be present in the boundary spectrum, and in (\ref{delSN-Neumann}) we just see the $4 - \epsilon$ description of the same operators.  We can obtain similar results in the symmetric traceless sector, for which we get
\begin{equation} \label{AnomDimDisplacementSymTrac}
\begin{split}
&\hat{\gamma}^{T, D}_{1} (n,0) = \frac{2 \epsilon  }{(N + 8)}, \ \ \ \hat{\gamma}^{T, N}_{1} (n > 0,0) = \frac{ 2 \epsilon }{(N + 8)}, \ \ \ \hat{\gamma}^{T, N}_{1} (0,0) = \frac{ 4 \epsilon }{(N + 8)}  \\
\implies &\hat{\Delta}^{T, D}_{n,0} = d + 2 n - \frac{N \epsilon}{ N + 8}, \ \ \ \hat{\Delta}^{T, N}_{n > 0,0} = d - 2 + 2 n - \frac{N \epsilon}{ N + 8}, \ \ \ \hat{\Delta}^{T, N}_{0,0} = d - 2  + \frac{(2 - N) \epsilon}{ N + 8}.  
\end{split}
\end{equation}
These composite operators also appear in the boundary operator expansion of the bulk higher spin currents. In the previous subsection, we gave another method to calculate the anomalous dimension of such operators and the result above agrees with the example we considered there in eq. \eqref{AnomalousHigherSpinDisplacement}. These results for anomalous dimensions and $\epsilon^2$ corrections to some of these results were also reported in \cite{Bertan:2018afl} by explicitly performing the loop integrals in AdS.

\subsubsection{$\epsilon$ expansion in $d = 2 + \epsilon$}
We can compute a similar four-point function in the non-linear sigma model by evaluating contact Witten diagrams in AdS$_{2+\epsilon}$. We will work to the leading order in $\epsilon$, hence the calculation really reduces to Witten diagrams in AdS$_2$.\footnote{For other recent calculations of Witten diagrams for CFTs in AdS$_2$, see \cite{Beccaria:2019stp, Beccaria:2019mev, Beccaria:2019dju, Beccaria:2020qtk}.} We will restrict ourselves to the simpler case of Dirichlet boundary conditions. The calculation is technically similar to the ones relevant for the defect CFT on BPS Wilson line operators \cite{Giombi:2017cqn}. The case of Neumann boundary conditions is expected to be more subtle, similarly to what was discussed in \cite{Beccaria:2019dws}, due to the presence of zero modes responsible for restoration of the $O(N)$ symmetry, and we will not discuss it in detail here. 

The sigma-model interactions now involves derivatives, so the expression is a little longer to write down
\begin{equation}
\begin{aligned}
&\langle \hat{\varphi}^a (\textbf{x}_1)\hat{\varphi}^b (\textbf{x}_2)\hat{\varphi}^c (\textbf{x}_3)\hat{\varphi}^d (\textbf{x}_4) \rangle_{1} = -t^2 \hat{C}_{\phi \phi}^4 \int d^d x \sqrt{g} g^{\mu \nu} \ \times \\
& \bigg[ \delta^{a b} \delta^{c d} \left(K_{\hat{\Delta}_{\phi}}^1 \partial_{\mu} K_{\hat{\Delta}_{\phi}}^2  + 1 \leftrightarrow 2 \right) \left( K_{\hat{\Delta}_{\phi}}^3 \partial_{\nu} K_{\hat{\Delta}_{\phi}}^4  + 3 \leftrightarrow 4 \right) + \{b,2\} \leftrightarrow \{c,3\} +   \{b,2\} \leftrightarrow \{d,4\} \bigg]
\end{aligned}
\end{equation}  
where we introduced the notation $K_{\hat{\Delta}_{\phi}}^i = K_{\hat{\Delta}_{\phi}}(\textbf{x}_i, x )$. To deal with the derivatives, the following identity is useful, which can be derived just by using the explicit expression of the bulk-to-boundary propagator
\begin{equation}
g^{\mu \nu} \partial_{\mu} K^1_{\Delta_1} \partial_{\nu} K^2_{\Delta_2} = \Delta_1 \Delta_2 \left(K^1_{\Delta_1}K^2_{\Delta_2} - 2 \ \textbf{x}_{12}^2 \ K^1_{\Delta_1 + 1}K^1_{\Delta_2 + 2}  \right). 
\end{equation}
So the integral now produces a linear combination of $D$ functions
\begin{equation}
\begin{split}
&\langle \hat{\varphi}^a (\textbf{x}_1)\hat{\varphi}^b (\textbf{x}_2)\hat{\varphi}^c (\textbf{x}_3)\hat{\varphi}^d (\textbf{x}_4) \rangle_{1} = -t^2 \hat{C}_{\phi \phi}^4 \hat{\Delta}_{\phi}^2 \times \\
\bigg[ &\delta^{a b} \delta^{c d} \bigg( 4 D_{\hat{\Delta}_{\phi}, \hat{\Delta}_{\phi}, \hat{\Delta}_{\phi}, \hat{\Delta}_{\phi}} - 2 \textbf{x}_{2 4}^2  D_{\hat{\Delta}_{\phi}, \hat{\Delta}_{\phi} + 1, \hat{\Delta}_{\phi}, \hat{\Delta}_{\phi} + 1} - 2 \textbf{x}_{1 4}^2  D_{\hat{\Delta}_{\phi} + 1, \hat{\Delta}_{\phi}, \hat{\Delta}_{\phi}, \hat{\Delta}_{\phi} + 1}  \\
&- 2 \textbf{x}_{2 3}^2 D_{\hat{\Delta}_{\phi}, \hat{\Delta}_{\phi} + 1, \hat{\Delta}_{\phi} + 1, \hat{\Delta}_{\phi}} - 2 \textbf{x}_{1 3}^2 D_{\hat{\Delta}_{\phi} + 1, \hat{\Delta}_{\phi}, \hat{\Delta}_{\phi} + 1, \hat{\Delta}_{\phi}}   \bigg) + \{b,2\} \leftrightarrow \{c,3\} +   \{b,2\} \leftrightarrow \{d,4\}\bigg]
\end{split}
\end{equation}
where the permutation also exchanges the subscripts of the $D$ function. For example, under 
$\{b,2\} \leftrightarrow \{c,3\}$ we have $\textbf{x}_{2 4}^2  D_{\hat{\Delta}_{\phi}, \hat{\Delta}_{\phi} + 1, \hat{\Delta}_{\phi}, \hat{\Delta}_{\phi} + 1} \rightarrow \textbf{x}_{3 4}^2  D_{\hat{\Delta}_{\phi}, \hat{\Delta}_{\phi}, \hat{\Delta}_{\phi} + 1, \hat{\Delta}_{\phi} + 1}$. These $D$ functions are well known and all we need here is the particular case with $\hat{\Delta}_{\phi} = 1$. Explicit expressions for this particular case can functions can be explicitly found in, for instance \cite{Giombi:2017cqn, Beccaria:2019dws}. Moreover, since the boundary theory is essentially one-dimensional,\footnote{More precisely, the boundary theory is $1+\epsilon$ dimensional, but to the order we are working we can set $d=2$ everywhere, and hence the boundary is for our purposes one-dimensional.} we only have one cross-ratio which we call $\chi$ with $u = \chi^2$ and $ v = (1 - \chi)^2$. In one dimension, the conformal block is just given by \cite{Dolan:2011dv}
\begin{equation}
G_{n,0} = \chi^{2 \hat{\Delta}_{\phi} + 2 n} {}_2 F_1 (2 \hat{\Delta}_{\phi} + 2 n, 2 \hat{\Delta}_{\phi} + 2 n, 4 \hat{\Delta}_{\phi} + 4 n, \chi)
\end{equation} 
and the derivative with respect to $n$ gives 
\begin{equation}
\partial_{n} G_{n,0} = 2 \chi ^{2 (
\hat{\Delta}_{\phi} +n)} \ \log (\chi ) \, \ _2F_1(2 \hat{\Delta}_{\phi} + 2 n, 2 \hat{\Delta}_{\phi} + 2 n, 4 \hat{\Delta}_{\phi} + 4 n, \chi) + \textrm{ other terms without a} \log .
\end{equation} 
We only focus on the $\log \chi$ term, because that is sufficient for us to extract the anomalous dimensions. Using the explicit expressions for the $D$-functions, we can collect the $\log$ terms and comparing it to the log terms appearing in the boundary operator expansion \eqref{4ptChangeGeneral}, we can read off the anomalous dimensions. For the singlet sector, this gives the following equation 
\begin{equation}
\begin{split}
&\sum_{n} (N - 1) a^0_{S, \ n,0} \hat{\gamma}^{S, D}_1 (n,0) \  \chi^{2 n} \  {}_2 F_1 (2  + 2 n, 2 + 2 n, 4  + 4 n, \chi)  = \\
& - \frac{2 \epsilon}{N - 2} \bigg( - \frac{2 (N + 1)}{1 - \chi} - \frac{N(2 \chi - 3)}{2 (1 - \chi)^2} + \frac{N (3 - \chi)}{2 (1 - \chi)} - \frac{\chi^2}{(1 - \chi)^2} \bigg)
\end{split}
\end{equation}
where we used the coupling at the fixed point $t^2_{*} = 2 \pi \epsilon/(N - 2)$ and the $d = 2$ Dirichlet value of $\hat{C}_{\phi \phi} = 1/\pi$. Using the following orthogonality property of  $F_{\beta} (\chi) = {}_2 F_1 (\beta, \beta, 2 \beta, \chi) $
\begin{equation}
\frac{1}{2 \pi i} \oint_{\chi = 0} \chi^{\beta - \beta' - 1 } F_{\beta} (\chi) F_{1 - \beta'} (\chi) = \delta_{\beta \beta'}
\end{equation}
it is easy to show that the anomalous dimensions are given by 
\begin{equation}
\hat{\gamma}^{S, D}_1 (m,0) = - \frac{2 \epsilon}{(N - 1) a^0_{S, \ m,0}} \ \frac{1}{2 \pi i} \oint_{\chi = 0} \chi^{- 2 m - 1 } \left( \frac{\chi (\chi - 2) + 2}{2 (\chi - 1)^2} \right) \ F_{- 2 m - 1} (\chi).  
\end{equation}
So we just need to calculate the coefficient of $\chi^{2 m}$ in a product of a Hypergeometric function and a polynomial in order to know the anomalous dimension for any $m$. The OPE coefficient $a^0_{S, \ m,0}$ is given by 
\begin{equation}
(N -  1) a^0_{S, \ m,0} = \frac{ 2 \ (\frac{3}{2})_m^2 \Gamma(1 + m)}{(\frac{1}{2})_m (2 + m)_m (\frac{3}{2} + m)_m}. 
\end{equation}
Using the boundary anomalous dimension of $\hat{\phi}$ from eq. \eqref{AnomalousDimensionNLSM}, this gives the following values of dimensions of the composite operators 
\begin{equation}
\hat{\gamma}^{S, D}_1 (n,0) = - \epsilon, \ \ \ \hat{\Delta}^{S, D}_{n,0} = 2\hat{\Delta}^{D}_{\phi} + 2 n + \hat{\gamma}^{S, D}_1 (n,0) = 2 + \epsilon + O(\epsilon^2) =   d + 2 n + O(\epsilon^2).
\end{equation}
Again, the operator with $n = 0$ is proportional to the displacement operator, while the others appear in the boundary operator expansion of the bulk higher spin currents. So far, we dealt with the case of Dirichlet boundary conditions on the unconstrained fields, when the dimension of the leading boundary operator is $ \hat{\Delta}_{\phi} = 1 + \epsilon = d - 1$. This describes extraordinary transition in $d = 2 + \epsilon$. As we mentioned above, we do not discuss the four-point function in the case of Neumann boundary conditions, which would have $\hat{\Delta}_{\phi} = O(\epsilon) $ and hence the propagators would be logarithmic. As we discussed in subsection \ref{SectionLargeNE}, that case describes the ordinary transition in $d = 2 + \epsilon$ dimensions. We leave a more detailed study of that case for future work. 

\section{Conclusion}
\label{sec-Concl}
In this paper, we have explored the idea of placing a CFT in AdS as a way of studying the BCFT problem. Focusing on the concrete example of the large $N$ critical $O(N)$ model, we have explained how to obtain the various boundary critical behaviors of the model in the AdS picture. We have also computed the large $N$ free energies of the model on the hyperbolic ball, and verified consistency of a conjectured $F$-theorem for the behavior of the quantity $\tilde F$ in (\ref{ConjectureDefinition}) under RG flows triggered by boundary relevant operators.  Then, we showed how to use the AdS setup to extract some of the BCFT data in the theory. In particular we suggested that using the bulk equations of motion in a way similar to \cite{Rychkov:2015naa} one can reconstruct in a convenient way the bulk two-point function and the anomalous dimensions of boundary operators encoded in it. We have also presented some calculations of boundary 4-point functions, where one can use the well-known techniques developed for the calculation of Witten diagrams in AdS/CFT. 

It would be interesting to apply the methods used in this paper to other examples of interacting BCFT, in particular explore the idea of using the bulk equations of motion to extract BCFT data as described in subsection \ref{SectionBulkEOM}. For instance, one may consider theories with fermions, like the large $N$ Gross-Neveu model and the related Gross-Neveu-Yukawa model (some results for the Gross-Neveu BCFT at large $N$ from the AdS approach were recently obtained in \cite{Carmi:2018qzm}). It would be also quite interesting to study bosonic and fermionic vector models coupled to Chern-Simons gauge theory in $d=3$ \cite{Giombi:2011kc, Aharony:2011jz} by placing them in AdS. One may then compute their free energy and other BCFT data, and perhaps provide further evidence for the 3d boson/fermion duality.  

Another technically interesting direction to pursue would be to develop more methods for the $1/N$ perturbation theory in the $O(N)$ model BCFT. Analytic functionals have been developed in \cite{Mazac:2018biw} to do perturbative expansions around a mean field solution in BCFT. In a mean field theory, there is an elementary field with dimension $\Delta_{\varphi}$ and all its higher point correlators factorize into products of two-point functions.  In the presence of a boundary, there are two possibilities for the boundary spectrum: Neumann with boundary dimensions being $\Delta_{\phi} + 2 n$ or Dirichlet with boundary dimensions being $\Delta_{\phi} + 2 n + 1$. However, as we saw in subsection \ref{SectionLargeNND}, in the large $N$ solution the elementary bulk field has dimensions $d/2 - 1$, while the leading boundary operator for ordinary or special transition has dimension $d -2$ or $d - 3$. So the large $N$ theory is quite different from a mean field solution, and the analytic functionals developed so far cannot be used for the $1/N$ perturbation theory. It would be interesting to see if a suitable set of functionals may be developed for the large $N$ expansion. Improving our knowledge of the $O(N)$ model BCFT at large $N$ may also help in better understanding the holographic description of the model, which should presumably be related to Vasiliev higher-spin theory in AdS$_{d+1}$ in the presence of an AdS$_d$ ``boundary brane".\footnote{To avoid possible confusion, let us stress here that this AdS$_d$ ``brane" is not the same as the AdS$_d$ on which we placed the CFT in this paper.}  

\section*{Acknowledgments}
We thank Christian Jepsen for useful discussions, and Xinan Zhou for discussions and collaboration on related topics. This research was supported in part by the US NSF under Grants No.~PHY-1620542 and PHY-1914860.

\appendix
\section{Bulk OPE coefficients at large $N$} \label{BulkOPEAppendix}
In this appendix, we show how to obtain the bulk OPE coefficients for the two-point function of $\phi$ in the case of special and ordinary transtions. As we saw in the main text, the two-point function for any scalar operator in the flat half-space can be written as (the AdS expression is the same with the conformal factors stripped off)
\begin{equation}
\langle O(x) O(x') \rangle = \frac{A}{(4 y y')^{\Delta_O}} \xi^{- \Delta_O} G(\xi) = \frac{A}{(4 y y')^{\Delta_O}} \mathcal{G}(z), \ \ \ z = \frac{1}{1 + \xi}.
\end{equation} 
We introduced a new variable $z$ because that is more convenient to work with for us in this appendix. In terms of $z$, we have
\begin{equation}
\begin{split}
&\mathcal{G}(z) =  \frac{z^{\Delta_O}}{(1 - z)^{\Delta_O}} \sum_{k} \lambda_k f_{\textrm{bulk}} \left( \Delta_k; 1 - z \right) = \sum_l \mu_l^2 f_{\textrm{bdy}} (\hat{\Delta}_l; z) \\
&f_{\textrm{bulk}} (\Delta_k; z) = z^{\frac{\Delta_k}{2}} \ {}_2 F_1 \left(\frac{\Delta_k  -d}{2} + 1, \frac{\Delta_k}{2}; \Delta_k + 1 - \frac{d}{2}; z \right) \\ 
&f_{\textrm{bdy}} (\hat{\Delta}_l; z)  =  z^{\hat{\Delta}_l} \ {}_2 F_1 \bigg(\hat{\Delta}_l, \hat{\Delta}_l + 1 - \frac{d}{2} ; 2 \hat{\Delta}_l + 2 - d;z \bigg) . 
\end{split}
\end{equation}  
The bulk and boundary block expansions are obtained respectively in the limit $\xi \rightarrow 0 \ (z \rightarrow 1)$ and $\xi \rightarrow \infty \ (z \rightarrow 0)$ limit of the two point function. For the special transition, we have 
\begin{equation}
G(\xi) = \frac{1 + 2 \xi}{(1 + \xi)^{\frac{d}{2} - 1}} \implies \mathcal{G}(z) = \frac{z^{d - 3} (2 - z)}{(1 - z)^{\frac{d}{2} - 1}}. 
\end{equation}
By expanding this for small $\xi$, it is easy to see that the operators appearing in bulk have dimensions $\Delta_n = 2 n + 2$. To obtain the OPE coefficients, we can use the Euclidean inversion formula \cite{Mazac:2018biw} for BCFT, which gives the bulk coefficient function as 
\begin{equation}
I_{\Delta} =  \int_0^1 d y \ y^{-2} (1 - y)^{1 - \frac{d}{2}}  \
{}_2 F_1 \left( \frac{\Delta}{2}, \frac{d - \Delta}{2},1,1 - 
\frac{1}{y} \right) \mathcal{G}(1 - y).
\end{equation}
To actually do this integral, we need to transform the parameters of this hypergeometric function, so that it becomes a combination of hypergeometric functions at $y$ instead of $1 - 1/y$, and the integral can be then performed after that. The bulk coefficients can be determined from the residues of this coefficient function. The coefficient function near behaves near the poles as
\begin{equation}
I_{\Delta} \frac{\Gamma(\frac{\Delta}{2}) \Gamma(\frac{\Delta - d + 2}{2})}{2 \Gamma(\Delta - \frac{d}{2})} \sim \frac{(\lambda )_n}{\Delta - \Delta_n} , \ \ \textrm{as} \ \ \Delta \rightarrow \Delta_n = 2n + 2.  
\end{equation}
Using this, we can find the bulk coefficients 
\begin{equation}
(\lambda )^S_n = \frac{(d^2 - 4 d (n + 2) + 8 (1 + n)^2 + 4) \Gamma(1 - \frac{d}{2} + n) \Gamma(2 - \frac{d}{2} + n)^2}{4 \Gamma(2 - \frac{d}{2})^2 \Gamma(n + 2) \Gamma(2 - \frac{d}{2} + 2 n)}. 
\end{equation}
This agrees with the result that was found in \cite{Liendo:2012hy}. For the ordinary transition, we have
\begin{equation}
G(\xi) = \frac{1 }{(1 + \xi)^{\frac{d}{2} -1}} \implies \mathcal{G}(z) = \frac{z^{d - 2} }{(1 - z)^{\frac{d}{2} - 1}}.
\end{equation}
The Euclidean inversion formula can again be used to get a coefficient function which gives the bulk coefficients
\begin{equation}
(\lambda)_n^O = -\frac{ (-1)^n \Gamma \left(\frac{d}{2}\right) \, _2F_1\left(-n-1,-n;\frac{1}{2} (d-4 n-2);1\right)}{\Gamma (n+2) \Gamma \left(\frac{d}{2}-n-1\right)}.
\end{equation}
This formula, to our knowledge, has not appeared before. 

\section{Details on $\sigma$ propagator} 
\label{SigmaCorrelatorAppendix}
In this appendix, we give some details on deriving the $\sigma$ propagator in position space in the large $N$ theory. In order to do that, we first show how to invert functions of chordal distance in AdS space, following a similar derivation on half-space in \cite{McAvity:1995zd}. We start with the following equation
\begin{equation}
\int d^d x \sqrt{g_x} \ G(\xi_{x_1,x}) H(\xi_{x, x_2}) = \int d^d x \ y^{-d} \ G(\xi_{x_1,x}) H(\xi_{x, x_2}) =  \frac{\delta^d(x_1 - x_2)}{\sqrt{g_{x_1}}}
\end{equation}
and the problem is to find $H$ given $G$. First we note that we can integrate over the boundary directions using the following formulae
\begin{equation}
\int d^{d - 1} \textbf{x} \ G(\xi_{x,x'}) = (4 y y')^{\frac{d - 1}{2}} g (\rho_{y,y'}), \ \ \ \rho_{y,y'} = \frac{(y - y')^2}{4 y y'}, \\
g (\rho) = \frac{\pi^{\frac{d - 1}{2}}}{\Gamma(\frac{d - 1}{2})} \int_0^{\infty} du u^{\frac{d - 3}{2}} G(u + \rho)
\end{equation}
and the above transform can be inverted as 
\begin{equation}
G(\xi) = \frac{\pi^{-\frac{d - 1}{2}}}{\Gamma(-\frac{d - 1}{2})} \int_0^{\infty} d \rho \ u^{\frac{-d - 1}{2}} g(\rho + \xi).
\end{equation}
Using this, we can integrate both sides over the boundary coordinates $\textbf{x}_1$
\begin{equation}
\int_0^{\infty} \frac{d y }{y} g(\rho_{y_1,y}) h(\rho_{y,y_2}) = \frac{y_1 \delta(y_1 - y_2)}{4^{d - 1}}.
\end{equation}
We can make a change of variable to $y = e^{2 \theta}$ to simplify the integral to
\begin{equation}
\int d \theta \ g(\sinh^2(\theta_1 - \theta)) h (\sinh^2(\theta - \theta_2)) = \frac{\delta(\theta_1 - \theta_2)}{4^d}.
\end{equation}
This can then be solved by a Fourier transform 
\begin{equation}
\hat{g}(k) = \int d \theta e^{i k \theta} g (\sinh^2 \theta)
\end{equation}
which gives 
\begin{equation}
\hat{g}(k) \hat{h}(k) = \frac{1}{4^d}.
\end{equation}
So we need to be able to go from $\hat{g}(k)$ to $G(\xi)$. It was shown in \cite{McAvity:1995zd} that $\hat{g}(k)$ of the form 
\begin{equation}
\hat{g}_{a, b}(k) = \frac{\Gamma\left(a - \frac{i k }{4} \right) \Gamma\left(a + \frac{i k }{4} \right)}{\Gamma\left(b - \frac{i k }{4} \right) \Gamma\left(b + \frac{i k }{4} \right)}
\end{equation}
corresponds to 
\begin{equation}
G_{a,b}(\xi) = \frac{\Gamma\left( 2 a + \frac{d - 1}{2} \right)}{4^{2 a - 1} \pi^{\frac{d - 1}{2}} \Gamma\left( b - a \right) \Gamma\left(b + a \right)} \frac{1}{\xi^{2 a + \frac{d - 1}{2}}} \, _2F_1 (2 a + \frac{d - 1}{2}, a + b - \frac{1}{2}; 2 a + 2 b - 1; - \frac{1}{\xi}).
\end{equation}
For the ordinary transition in section \ref{SectionLargeNND}, we have
\begin{equation}
G(\xi) = - \frac{N}{2} G_{\phi}(\xi)^2 = - \frac{N \Gamma \left( \frac{d}{2} - 1 \right)^2}{2 (4 \pi)^{d}} \frac{1}{(\xi( 1 + \xi))^{d - 2}} = - \frac{N \Gamma \left( \frac{d}{2} - 1 \right)^2 \Gamma \left(- \frac{d}{2} + 2 \right)}{2^{d + 5} \Gamma(d - 2) \pi^{d/2}} G_{\frac{3 d - 7}{4}, \frac{d + 1}{4}}
\end{equation} 
which immediately gives us 
\begin{equation}
G_{\sigma}(\xi) = H(\xi) = - \frac{2^{d + 5} \Gamma(d - 2) \pi^{d/2}}{N 4^d \Gamma \left( \frac{d}{2} - 1 \right)^2 \Gamma \left(- \frac{d}{2} + 2 \right)} G_{\frac{d + 1}{4}, \frac{3 d - 7}{4}}.
\end{equation}
For application to the extraordinary transition in section \ref{SectionLargeNE} we need to use
\begin{equation}
G(\xi) = \tilde{\Pi}(\xi) = - \frac{(N - 1) \Gamma\left( \frac{d}{2} \right)^2}{(4 \pi)^d} \frac{1}{(\xi( 1 + \xi))^{d - 1}} = - \frac{ (N - 1) \pi^{\frac{3 - d}{2}} \csc \left(\frac{\pi d}{2} \right) }{ 4^d \ \Gamma\left( \frac{d - 1}{2} \right)} G_{\frac{3 d - 3}{4}, \frac{d + 1}{4}}.  
\end{equation}
For the special transition, getting the $\sigma$ propagator needs a little more work. We do not do it here and refer the reader to \cite{McAvity:1995zd} for details. 

\section{Displacement operator and the $b$-anomaly coefficient} \label{AppendixDisplacement}
In this appendix we collect some comments on the displacement operator, which appears in every CFT with a boundary or a defect. In the case of a $d-$ dimensional CFT with a boundary this operator is a scalar of conformal dimension $d$ appearing in the spectrum of the $d - 1$ dimensional boundary theory. In the flat half-space space setup, the displacement operator $D$ can be defined by 
\begin{equation}
\partial^{\mu} T_{\mu y} (\mathbf{x}, y) = D (\mathbf{x}) \delta(y)
\end{equation}
and it can be thought of as related to the broken translational symmetry perpendicular to the boundary. This equation can be integrated in a Gaussian pill box located at the boundary which implies $D = T_{y y}|_{y \rightarrow 0} $. The coefficient of the two-point function of the displacement operator is a piece of BCFT data, and in $d=3$ it is related to the trace anomaly coefficient $b$ in (\ref{TraceAnomaly3d}). We will first discuss the calculation of this two-point function in the theory of a single free scalar field, and then move to the interacting case. The improved stress tensor in flat half-space can be written as 
\begin{equation}
T_{\mu \nu} = \partial_{\mu} \phi \partial_{\nu}\phi - \frac{1}{2}\delta_{\mu \nu} \partial \phi \cdot \partial \phi -  \frac{(d - 2)}{4 (d -1)} \big( \partial_{\mu} \partial_{\nu} - \delta_{\mu \nu} \partial^2 \big) \phi^2.
\end{equation}
We can use this to explicitly write the displacement operator. With Neumann boundary condition, we have $\partial_y \phi(\mathbf{x}, y = 0) = 0$, which gives after using equations of motion 
\begin{equation}
D^N = T_{y y} |_{y \rightarrow 0} =  - \frac{d -2}{2 (d - 1)} \phi \partial_y^2 \phi - \frac{1}{2 (d - 1)} \partial_i \phi \partial^i \phi\,.
\end{equation}
In the Dirichlet case, $\phi(\mathbf{x}, y = 0) = 0 $, which gives 
\begin{equation}
D^D = \frac{1}{2} (\partial_y \phi)^2.
\end{equation}  
Computing the correlator then is just a matter of Wick contractions and taking derivatives, and yields the result
\begin{equation}
\langle D^D (\mathbf{x}) D^D (\mathbf{x}') \rangle = \langle D^N (\mathbf{x}) D^N (\mathbf{x}') \rangle = \frac{2}{S_d^2} \frac{1}{|\mathbf{x} - \mathbf{x}'|^{2 d}} 
\end{equation}
where $S_d = 2 \pi^{\frac{d}{2}} / \Gamma\left( \frac{d}{2} \right)$ is the volume of a $d - 1$ dimensional sphere. This equation gives the coefficient of the two-point function of the displacement operator for a free scalar, $\hat{C}_{DD} =  2 / S_d^2$ for both Neumann and Dirichlet cases. These results agree with what was found in \cite{McAvity:1995zd}.

In the large $N$ interacting theory, one way to define the displacement operator is through its appearance in the BOE of the $\sigma$ operator. In hyperbolic space, this means that if we look at the propagator of $\sigma$ near the boundary, the contribution of the displacement operator should look like
\begin{equation}
\langle \sigma(x) \sigma (x') \rangle_{y,y' \rightarrow 0} \sim ({B_{\sigma}}^D)^2 (y y')^d  \langle D(\mathbf{x}) D (\mathbf{x}') \rangle. 
\end{equation}
Using the boundary limit of the bulk correlator of $\sigma$ from eq. \eqref{SigmaBulkBoundaryDirichlet}, \eqref{SigmaBulkBoundaryNeumann} and eq. \eqref{SigmaBulkBoundaryEx} for special, ordinary and extraordinary transition gives 
\begin{equation}
\begin{split}
&({B_{\sigma}}^{D})^2 \hat{C}_{DD}|_{\textrm{Special}} = \frac{4^{d}}{3 N} \frac{(4 - d) \Gamma(d - 2)^2 }{\Gamma(2 - \frac{d}{2}) \Gamma(\frac{d}{2} - 1)^3} \frac{(6 - d) \Gamma(d) }{(d -2) \Gamma( 2d - 5)}\\
&({B_{\sigma}}^{D})^2 \hat{C}_{DD}|_{\textrm{Ordinary}} = \frac{4^{d}}{N} \frac{(4 - d)\Gamma(d - 2)^2 }{\Gamma(2 - \frac{d}{2}) \Gamma(\frac{d}{2} - 1)^3} \frac{\Gamma(d)}{\Gamma(2 d - 4)}\\
&({B_{\sigma}}^{D})^2 \hat{C}_{DD}|_{\textrm{Extraordinary}} = -\frac{2^{d+1} \sin \left(\frac{\pi  d}{2}\right) \Gamma \left(\frac{d-1}{2}\right) \Gamma (d+1)}{\pi  (N - 1)\Gamma \left(\frac{d}{2}-1\right) \Gamma \left(d-\frac{1}{2}\right)}.
\end{split}
\end{equation}
We can combine these equations with the constraint coming from the Ward identity relating the two point function of a bulk operator $O$ with the displacement to its one-point function \cite{McAvity:1995zd, Billo:2016cpy} \footnote{Note that the conventions for the OPE coefficients in \cite{McAvity:1995zd} and \cite{Billo:2016cpy} are different, which makes their Ward identities look different by factors of 2 etc., but the physical content is the same. We use the conventions of \cite{McAvity:1995zd} suitably adapted to hyperbolic space.}
\begin{equation} \label{WardIdentity}
{B_{O}}^{D} \hat{C}_{DD} = - \Delta_O \frac{ 2^d A_{O}} {S_d} 
\end{equation}
where $A_O$ is the one-point function coefficient and $\Delta_O$ is the conformal dimension of $O$ . Using this relation for the $\sigma$ operator with dimension $2$ gives us
\begin{equation} \label{DisplacementTwoPointCoeff}
\begin{split}
&\hat{C}_{DD}|_{\textrm{Special}} = \frac{24 N}{S_d^2} \frac{\Gamma\left( \frac{d}{2} \right)^3 \Gamma\left( 4 - \frac{d}{2} \right) \Gamma\left( 2 d - 5\right)}{\Gamma\left( d \right) \Gamma\left( d - 1 \right)^2}\\
&\hat{C}_{DD}|_{\textrm{Ordinary}} = \frac{2 N}{S_d^2} \frac{\Gamma\left( \frac{d}{2} \right)^3 \Gamma\left( 3 - \frac{d}{2} \right) \Gamma\left( 2 d - 3\right)}{\Gamma\left( d \right) \Gamma\left( d - 1 \right)^2} \\
&\hat{C}_{DD}|_{\textrm{Extraordinary}} =-\frac{ (N - 1) \pi^{3/2} (d-2) d \  \Gamma \left(d-\frac{1}{2}\right)}{2 \ S_d^2 \ \Gamma \left(\frac{d-1}{2}\right) \Gamma \left(\frac{d+1}{2}\right) \sin \left(\frac{\pi  d}{2}\right)}.
\end{split}
\end{equation}
The Neumann and Dirichlet results agree with \cite{McAvity:1995zd}. Note that in $3<d<4$ we have $\hat{C}_{DD}|_{\textrm{Special}}>\hat{C}_{DD}|_{\textrm{Ordinary}}$, but the value for the extraordinary transition (which is supposed to be in the IR of both special and ordinary) is not consistent with a possible ``$\hat{C}_{DD}$-theorem" for boundary RG flows. 

As a check of the large $N$ result, for extraordinary transition, we can also easily calculate this quantity in $d = 4 - \epsilon$. In that case, $\phi^N$ or its fluctuation $\chi$ is the operator that contains the displacement in its BOE. From the BOE of $\chi$ in eq. \eqref{BulkBoundaryChi}, we get 
\begin{equation}
({B_{\phi^N}}^D)^2 \hat{C}_{DD} = \frac{4^4}{160 \pi^2}
\end{equation}
which when combined with the Ward identity \eqref{WardIdentity} for $O = \phi^N$ gives 
\begin{equation}
\hat{C}_{DD}|_{\textrm{Extraordinary}} = \frac{40 (N + 8)}{S_d^2 \epsilon}
\end{equation}
in agreement with the large $N$ result. 

As we discussed in subsection \eqref{SectionTraceAnomaly}, in $d = 3$ boundary CFT, the trace anomaly contains two terms as in eq. \eqref{TraceAnomaly3d}. The coefficient of one of the terms called as  $a_{3d}$ is related to the logarithmic term of the free energy with spherical boundary, and it was discussed at length in the main text. 
The other anomaly coefficient $b$ is related to the two- point function coefficient of the displacement operator as $b = \pi^2 \hat{C}_{DD}/8$ \cite{Herzog:2017kkj}. Plugging in $d = 3$ in the results from above \eqref{DisplacementTwoPointCoeff}, we get 
\begin{equation} 
b^{\textrm{Special}}_{3d} = \frac{9 N \pi^2}{1024}, \ \ \ b^{\textrm{Ordinary}}_{3d} =  \frac{ N \pi^2}{1024}, \ \ \ b^{\textrm{Extraordinary}}_{3d} = \frac{9 N \pi^2}{1024}    .
\end{equation} 
There was a conjectured bound in \cite{Herzog:2017kkj} implying $a_{3d}/b \ge - 2/3$. It does not seem to hold true in the case of the extraordinary transition. The bound was based on a conjectured relation between $a_{3d}$ and coefficients appearing in the stress tensor two-point function, which was also found to not hold true in case of a $\phi^6$ theory with a $\phi^4$ boundary interaction \cite{Herzog:2020lel}. 
%So extraordinary transition discussed here provides another counterexample to that conjectured relationship. 

\bibliographystyle{ssg}
\bibliography{InteractingO_N_BCFT-bib}

\end{document}